\documentclass[symmetry,review,accept,oneauthor, pdftex]{Definitions/mdpi}
\usepackage{graphicx,amsmath,amssymb,amsfonts,latexsym,color,dcolumn,bm}
\usepackage{revsymb}
\usepackage{gensymb}
\usepackage[compat=1.1.0]{tikz-feynman}
\newcommand{\be}{\begin{equation}}
\newcommand{\ee}{\end{equation}}
\newcommand{\bea}{\begin{eqnarray}}
\newcommand{\eea}{\end{eqnarray}}
\newcommand{\ra}{\rangle}
\newcommand{\la}{\langle}
\newcommand{\om}{\omega}

\usepackage[utf8]{inputenc}
\usepackage{gensymb}
\usepackage{braket}
\usepackage{caption}
\usepackage{subcaption}
\usepackage{geometry}
\usepackage{array}
\usepackage{comment}
\usepackage{lineno}
\usepackage{url}
\usepackage{wrapfig}
\usepackage{url}
\usepackage{enumitem}

\usepackage[colorinlistoftodos]{todonotes}

\usepackage{caption}
\firstpage{1} 
\pubvolume{00}
\issuenum{0}
\articlenumber{0}
\pubyear{2023}
\copyrightyear{2023}

\history{Received: 2020; Published: 2020; Updated 2023}
\updates{yes} 

\Title{Dynamical Symmetries of the H Atom, One of the Most Important Tools Of Modern Physics: SO(4) to SO(4,2), Background, Theory, and Use in Calculating Radiative Shifts}

\Author{{G. Jordan Maclay}}

\AuthorNames{G. Jordan Maclay}
\address{%
 \quad Quantum Fields LLC, St. Charles, IL 60174 USA; Emeritus, University of Illinois at Chicago, Chicago, IL 60607 \\ jordanmaclay@quantumfields.com}

\corres{Correspondence: jordanmaclay@quantumfields.com}

\abstract{Understanding the hydrogen atom has been at the heart of modern physics. Exploring the symmetry of the most fundamental two body system has led to advances in atomic physics, quantum mechanics, quantum electrodynamics, and elementary particle physics. In this pedagogic review we present an integrated treatment of the symmetries of the Schrodinger hydrogen atom, including the classical atom, the SO(4) degeneracy group, the non-invariance group or spectrum generating group SO(4,1) and the expanded group SO(4,2).  After giving a brief history of these discoveries, most of which took place from 1935-1975, we focus on the physics of the hydrogen atom, providing a background discussion of the symmetries, providing explicit expressions for all the manifestly Hermitian generators in terms of  position and momenta operators in a Cartesian space, explaining the action of the generators on the basis states, and giving a unified treatment of the bound and continuum states in terms of eigenfunctions that have the same quantum numbers as the ordinary bound states.  We present some new results from SO(4,2) group theory that are useful in a practical application, the computation of the first order Lamb shift in the hydrogen atom. By using SO(4,2) methods, we are able to obtain a generating function for the radiative shift for all levels. Students, non-experts and the new generation of scientists may find the clearer, integrated presentation of the symmetries of the hydrogen atom helpful and illuminating. Experts will find new perspectives, even some surprises.}  

\keyword{symmetry, H atom, hydrogen atom, group theory, SO(4), SO(4,2), dynamical symmetry, non-invariance group, spectrum generating algebra, Runge-Lenz, Coulomb potential, Lie algebra, radiative shift, Lamb shift}

\begin{document}

\tableofcontents

\section{Introduction}

\subsection{Objective of this paper}
This pedagogic review is focused on the symmetries of the Schrodinger nonrelativistic hydrogen atom exclusively to give it the attention we believe it deserves. The fundamental results of the early work are known and do not need to be derived again. But having this knowledge permits us to use the modern language of group theory to do a clearer, more focused presentation, and to use arguments from physics to develop the proper forms for the generators, rather than dealing with detailed, mathematical derivations to prove results we know are correct. 

There are numerous articles about the symmetry of the Schrodinger hydrogen atom, particularly the SO(4) group of the degenerate energy eigenstates, including discussions from classical perspectives.  The spectrum generating group SO(4,1) and the non-invariance group SO(4,2) have been discussed but in many fewer articles, often in appendices, with different bases for the representations.  For example, numerous papers employ Schrodinger wave functions in parabolic coordinates, not with the familiar $nlm$ quantum numbers, often with the emphasis on the details of the mathematical structures, not the physics, and with the emphasis on the potential role of the symmetry in elementary particle physics.  Generators may be expressed in complex and unfamiliar terms, for example, in terms of the raising and lowering operators for quantum numbers characteristic of parabolic coordinates.  Other approaches involve regularizing Schrodinger's equation by, for example, multiplying by $r$.  The approach results in generators that are not always Hermitian or manifestly Hermitian, and the need for nonstandard inner products. Indeed, most of the seminal articles do not have the words "hydrogen atom" in the title but are focused on the prize down the road, understanding what was called at the time "the elementary particle zoo." 

Yet these foundational articles and books taken together present the information we have about the symmetry of the H atom, of which experts in the field are aware.  On the other hand, to the non-expert, the student, or a researcher new to the field, it does not appear that the relevant information is in a form that is conveniently accessible.  Since the hydrogen atom is the most fundamental physical system with an interaction, whose exploration and understanding has led to much of the progress in atomic physics, quantum physics, and quantum electrodynamics, we believe a comprehensive treatment is  warranted and, since most of the  relevant papers were published four decades ago, is timely.  Many younger physicists may not be acquainted with these results.

Unlike in a number of the foundational papers, here the operators are all Hermitian, and given in terms of the canonical position and momentum variables in the simplest forms. The transformations they generate are clearly explained, and we provide brief explanations for the group theory used in the derivations. 

In most papers, a separate treatment for bound and scattering states in needed. In contrast we are able to clarify and simplify the exposition since we use a set of basis states which are eigenfunctions of the inverse of the coupling constant \cite {brow}, that include both the bound states and the scattering states in a uniform way, and that employ the usual Cartesian position and momenta, with the usual inner product, with the exact same quantum numbers $nlm$ as the ordinary bound states; a separate treatment for bound and scattering states is not required. In addition, we have two equivalent varieties of this uniform basis, one more suitable for momentum space calculations and one more suitable for configuration space calculations. This advantage again allows us to simplify the exposition.

We focus on the utility of group theoretic methods using our representation and derive expressions for the unitary transformation of group elements and some new results that allow us to readily compute the first order radiative shift (Lamb shift) of a spinless electron, which accounts for about 95\% of the total shift. This approach allows us to obtain a generating function for the shifts for all energy levels. For comparison, we derive an expression for the Bethe log.

In summary, we present  a unified treatment of the symmetries of the Schrodinger hydrogen atom, from the classical atom to SO(4,2) that focuses on the physics of the hydrogen atom, that gives explicit expressions for all the manifestly Hermitian generators in terms of  position and momenta operators in a Cartesian space, that explains the action of the generators on the basis states, that evaluates the Casimir operators characterizing the group representations, and that gives a unified treatment of the bound and continuum states in terms of wave functions that have the same quantum numbers as the ordinary bound states.  We give an example of the use of SO(4,2) in a practical application, the computation of the first order radiative shift in the hydrogen atom. 

Hopefully students and non-experts and the new generation of scientists will find this review helpful and illuminating, perhaps motivating some to use these methods in various new contexts. Senior researchers will find new perspectives, even some surprises and encouragements. 

\subsection{Outline of this Paper}
In the remainder of Section 1 we give a brief historical account of the role of symmetry in quantum mechanics and of the work done to explore the symmetries of the Schrodinger hydrogen atom.

In Section 2, we provide some general background observations about symmetry groups and non-invariance groups and discuss the degeneracy groups for the Schrodinger, Dirac, and Klein-Gordon equations. We also introduce the  uncommon $(Z \alpha)^{-1}$ eigenstates that allow us to treat the bound and scattering states in a uniform way, using the usual quantum numbers.  In Section 3, the classical equations of motion of the nonrelativistic hydrogen atom in configuration and momentum space are derived from symmetry
considerations. The physical meaning of the symmetry
transformations and the structure of the degeneracy group SO(4)
is discussed. In Section 4, we discuss the symmetries using the language of quantum mechanics.  In order to display the symmetries in quantum
mechanics in the most elegant and uniform way we use a
basis of eigenstates of the inverse of the coupling constant $(Z \alpha)^{-1}$.  In Section 5 we discuss these wave functions in momentum and configuration space, how they transform and their classical limit for Rydberg states.

In Section 6 we discuss the noninvariance or spectrum
generating group of the hydrogen atom SO(4,l) and relate it to the conformal group in momentum space. In Section 7 the enlarged spectrum generating group SO(4,2) is introduced, with a discussion of the physical meaning of the generators. All physical states together form a basis for a unitary irreducible representation of these noninvariance groups. We derive manifestly Hermitian expressions in terms of the momentum and position canonical variables for the generators of the group transformations and obtain the values possible for the Casimir operators. We discuss the important subgroups of SO(4,2).

In Section 8 we use the group theory of SO(4,2) to determine the radiative shifts in energy levels due
to the interaction of a spinless electron with its own radiation
field, or equivalently with the quantum vacuum.  In the nonrelativistic or dipole
approximation the level shift contains
a matrix element of a
rotation operator of an O(1,2) subgroup of the group
SO(4,2). We can sum over this all states, obtaining the character of the representation, yielding a single integral which is a generating
function for the radiative shift for any level in the nonrelativistic
or dipole approximation.  A brief conclusion follows.

\subsection{Brief History of Symmetry in Quantum Mechanics and its Role in Understanding the Schrodinger Hydrogen Atom}
The hydrogen atom is the fundamental two-body system and perhaps the most important tool of atomic physics and the continual challenge is to continually improve our understanding of the hydrogen atom and to calculate its properties to the highest accuracy possible.  The current QED theory is the most precise of any physical theory\cite{beyer}: 
\begin{quotation}
The study of the hydrogen atom has been at the heart of the development of modern physics...theoretical calculations reach precision up to the 12th decimal place...high resolution laser spectroscopy experiments...reach to the 15th decimal place for the 1S--2S transition...The Rydberg constant is known to 6 parts in $10^{12}$ \cite{beyer}\cite{moh}. Today the precision is so great that measurement of the energy levels in the H atom has been used to determine the radius of the proton.
\end{quotation}
Continual progress in understanding the properties of the hydrogen atom has been central to progress in quantum physics\cite{rigden}.  Understanding the atomic spectra of the hydrogen atom drove the discovery of quantum mechanics in the 1920's. The measurement of the Lamb shift in 1947 and its explanation by Bethe in terms of atom's interaction with the quantum vacuum fluctuations ushered in a revolution in quantum electrodynamics\cite{lamb}\cite{bethe}\cite{maclayrad}. Exploring the symmetries of the hydrogen atom has been an essential part of this progress. Symmetry is a concept that has played a broader role in physics in general, for example, in understanding the dynamics of the planets, atomic and molecular spectra, and the masses of elementary particles.

When applied to an isolated system, Newton's equations of motion imply the conservation of momentum, angular momentum and energy.  But the significance of these conservation laws was not really understood until 1911 when Emily Nother established the connection between symmetry and conservation laws\cite{nother}.  Rotational invariance in a system results in the conservation of angular momentum; translational invariance in space results in conservation of momentum; and translational invariance in time results in the conservation of energy.  We will discuss Nother's Theorem in more detail in Section 2.  

Another critical ingredient of knowledge, on which Nother based her proof, was the idea of an infinitesimal transformation, such as a infinitesimal rotation generated by the angular momentum operators in quantum mechanics. These ideas of infinitesimal transformations originated with the Norwegian mathematician Sophus Lie who was studying differential equations in the latter half of the nineteenth century. He studied the collection of infinitesimal transformations that would leave a differential equation invariant\cite{lie}. In 1918 German physicist and mathematician Hermann Weyl, in his classic book with the translated title "The Theory of Groups and Quantum Mechanics,"  would refer to this collection of differential generators leaving an operator invariant as a linear algebra, ushering in a little of the terminology of modern group theory\cite{weyl}. Still this was a very early stage in understanding the role of symmetry in the language of quantum theory. When he introduced the new idea of a commutator on page 264, he put the word "commutator" in quotes. In the preface Weyl made a prescient observation: ".. the essence of the new Heisenberg-Schrodinger-Dirac quantum mechanics is to be found in the fact that there is associated with each physical system a set of quantities, constituting a non-commutative algebra in the technical mathematical sense, the elements of which are the physical quantities themselves."

A few years later Eugene Wigner published in German, "Group Theory and Its Application to the Quantum Mechanics of Atomic Spectra\cite{wigner}." One might ask why was this classic not translated into English until 1959.  In the preface to the English edition, Prof. Wigner recalled: "When the first edition was published in 1931, there was a great reluctance among physicists toward accepting group theoretical arguments and the group theoretical point of view. It pleases the author that this reluctance has virtually vanished.."   It was the application of group theory in particle physics in the early sixties, such as SU(3) and chiral symmetry,  that reinvigorated interest in Wigner's book and the field in general.  In the 1940's Wigner and Bargmann developed the representation theory of the Poincare group that later provided an infrastructure for the development of relativistic quantum mechanics\cite{wigner}\cite{barg}.

The progress in understanding the symmetries of the hydrogen atom in particular has some parallels to the history of symmetry in general:  there were some decades of interest but after the 1930's interest waned for about three decades in both fields, until stimulated by the work on symmetry in particle physics.

Probably the first major advance in understanding the role of symmetry in the classical treatment of the Kepler problem after Newton's discovery of universal gravitation, elliptical orbits, and Kepler's Laws, was made two centuries ago by Laplace 
when he discovered the existence of three new constants of the motion in addition to the components of
the angular momentum\cite{lapl}. These additional conserved
quantities are the components of a vector which determines
the direction of the perihelion of the motion (point closest to the focus)
and whose magnitude is the eccentricity of the orbit.
The Laplace vector was later rediscovered by Jacobi
and has since been rediscovered numerous times under different names. Today it is generally referred to as the Runge-Lenz vector. But the significance of this conserved quantity was not well understood until the nineteen thirties.

In 1924 Pauli made the next major step forward in understanding the role of symmetry in the hydrogen atom\cite{pauli}. He used the conserved Runge-Lenz vector \textbf{A} and the conserved angular momentum vector \textbf{L}
to solve for the energy spectrum of the hydrogen atom
by purely algebraic means, a beautiful result, yet he did not explicitly identify that \textbf{L} and \textbf{A} formed the symmetry group SO(4) corresponding to the degeneracy. At this time, the degree of degeneracy in the hydrogen energy levels was believed to be $n^2$ for a state with principal quantum number $n$, clearly greater than the degeneracy due to rotational symmetry which is $(2\textit{l}+1)$. The $n^2$ degeneracy arises from the possible values of the angular momentum $l=0, 1, 2, \dots n-1,$ and the $2l+1$ values of the angular momentum along the azimuthal axis $m=-l,-l+1,..0,1,2, l+1$. The additional degeneracy was referred to as "accidental degeneracy\cite{maci}."

Six years after Pauli's paper, Hulthen used the new Heisenberg matrix mechanics to simplify the derivation of the energy eigenvalues of Pauli by showing that the sum of the squares $\textbf{L}^2 + \textbf{A}^2$ could be used to express the Hamiltonian and so could be used to find the energy eigenvalues\cite{hult}.  In a one sentence footnote in this 3 page paper, Hulthen gives probably the most important information in his paper: Prof. Otto Klein, who had collaborated for years with Sophus Lie, had noticed that the two conserved vectors formed the generators of the Lorentz group, which we can describe as rotations in four dimensions, the fourth dimension being time. This is the non-compact group SO(3,1), the special orthogonal group in four dimension whose transformations leave the magnitude $g_{\mu\nu}z^{\mu}z^{\nu}=-t^2 + x^2 + y^2 + z^2$ unchanged\cite{units}.  Klein's perceptive observation triggered the introduction of group theory to understanding the hydrogen atom.

About a decade later, in 1935, the Russian physicist Vladimir Fock published a major article in Zeitshrift fur Physik, the journal in which all the key articles about the hydrogen atom cited were published\cite{fock}.  He transformed Schrodinger's equation for a given energy eigenvalue from configuration space to momentum space, and did a stereographic projection onto a unit sphere, and showed that the bound state momentum space wave functions were spherical harmonics in four dimensions. He stated that this showed that rotations in four dimensions corresponded to the symmetry of the degenerate bound state energy levels in momentum space, realizing the group SO(4), the group of special orthogonal transformations which leaves the norm of a four-vector $U_0^2+U_1^2 + U_2^2 + U_3^2 $ constant. By counting the number of four-dimensional spherical harmonics $Y_{nlm}$ in momentum space $(m=-l, -l+1...0, 1,...l,$ where the angular momentum $l$ can equal $l=n-1, n-2,..0)$ he determined that the degree of degeneracy for the energy level characterized by the principal quantum number $n$ was $n^2$.   It is interesting that Fock did not cite the work by Pauli implying the four dimensional rotational symmetry in configuration space.  Fock also presented some ideas about using this symmetry in calculating form factors for atoms.

A year later the German-American mathematician and physicist Valentine Bargmann showed that for bound states (E<0) Pauli's conserved operators, the angular momentum \textbf{L} and the Runge-Lenz vector \textbf{A}, obeyed the commutation rules of the SO(4)\cite{barg}.  His  use of commutators was so early in the field of quantum mechanics, that Bargmann explained the square bracket notation he used for a commutator in a footnote\cite{dirac}. He gave differential expression for the operators, adapting the approach of Lie generators in the calculation of the commutators.  He linked solutions to Schrodinger's equation in parabolic coordinates to the existence of the conserved Runge-Lenz vector and was thereby able to establish the relationship of Fock's results to the algebraic representation of SO(4) for bound states implied by Fock and Pauli \cite{barg}.  He also pointed out that the scattering states (E>0) could provide a representation of the group SO(3,1).  In a note at the end of the paper, Bargmann, who was at the University in Zurich, thanked Pauli for pointing out the paper of Hulthen and the observation by Klein that the Lie algebra of \textbf{L} and \textbf{A} was the same as the infinitesimal Lorentz group, which is how he referred to a Lie algebra. Bargmann's work was a milestone demonstrating the relationship of symmetry to conserved quantities and it clearly showed that to fully understand a physical system one needed to go beyond the usual ideas of geometrical symmetry.  This work was the birth, in 1936, without much fanfare, of the idea of dynamical symmetry.  

Little attention was paid to these developments until the 1960's when interest arose primarily because of the applications of group theory in particle physics, particularly modeling the mass spectra of hadrons.
Particle physicists were faced with the challenge of achieving a quantitative description of hadron properties, particularly the mass spectra and form factors, in terms of quark models. Since little was know about quark dynamics they turned to group-theoretical arguments, exploring groups like SU(3), chiral U(3)xU(3), U(6)xU(6) etc. 
The success of the eight-fold way of SU(3) (special unitary group in three dimensions) of American physicist Murray Gell-Mann in 1962 brought attention to the use of symmetry considerations and group theory as tools for exploring systems in which one was unsure of the exact dynamics\cite{gell}.

In 1964, three decades after Fock's work, American physicist Julian Schwinger published a paper using SO(4) symmetry to construct a Green function for the Coulomb potential, which he noted was based on a class he taught at Harvard in 1949\cite{schw1}.  The publication was a response to the then current emphasis on group theory and symmetry which led to, as Israeli physicist Yuval Ne'eman described it,  " 'the great-leap-forward' in particle physics during the years 1961-66\cite{neem}." Some of the principal researchers leading this effort were Ne'eman\cite{neem2}, Gell-Mann\cite{gell}, \cite{gell2} \cite{eight}, Israeli physicist Y. Dothan\cite{doth}, Japanese physicist Yochiri Nambu\cite{nambu}, and English-American Freeman Dyson\cite{dyso1}. Advantage was taken of the mathematical infrastructures of group theory developed years earlier\cite{weyl}\cite{wigner}\cite{barg}\cite{thom}\cite{hari}.

Interest was particularly strong in systems with wave equations with an infinite number of components, which characterize non-compact groups. In about 1965, this interest in particle physics gave birth to the identification of SO(4,1) and SO(4,2) as Spectrum Generating Algebras that might serve as models for hadronic masses. The hydrogen atom was seen as a model to explore the infinite dimensional representations of non-compact groups.  The first mention of SO(4,1) was was by Barut, Budini, and Fronsdal \cite{baru6} where the H atom was presented as an illustration of a system characterized by non-compact representation, and so comprising an infinite number of states. 
The first mention of a six dimensional symmetry, referred to as the "non-compact group O(6)", appears to be by the Russian physicists I. Malin and V. Man'ko of the Moscow Physico-technical Institute\cite{malk}. In a careful three page paper, they showed that all the bound states of the H atom energy spectrum in Fock coordinates provided a representation of this group, and they calculated the Casimir operators for their symmetric tensor representation in parabolic coordinates.

Very shortly thereafter Turkish-American theoretical physicist Asim Barut and his student at University of Colorado, German theoretical physicist Hagen Kleinert, showed that including the dipole operator  $ er$   as a generator led to the expansion of SO(4,1) to SO(4,2), and that all the bound states of the H atom formed a representation of SO(4,2)\cite{baru0}.  This allowed them to calculate dipole transition matrix elements algebraically. They give a position representation of the generators based on the use of parabolic coordinates. The generators of the transformations are given in terms of the raising and lowering operators for the quantum numbers for solutions to the H atom in parabolic coordinates.  The dilation operator is used to go from one SO(4) subspace with one energy to a SO(4) subspace with different energy and it has a  rather complicated form.  They also used SO(4,2) symmetry to compute form factors \cite{baru4}.

The papers of the Polish-American physicist Myron Bander and French physicist Claude Itzakson published in 1966, when both were working at SLAC (Stanford Linear Accelerator in California) provide the first mathematically rigorous and "succinct" review of the O(4) symmetry of the H atom and provide an introduction to SO(4,1)\cite{band1}\cite{band2}, which is referred to as a spectrum generating algebra SGA, meaning that it includes generators that take the basis states from one energy level to another. They use two approaches in their mathematical analysis, the first is referred to as "the infinitesimal method," based on the two symmetry operators, \textbf{L} and \textbf{A} and the O(4) group they form, and the other, referred to as the "global method," first done by Fock, converts the Schrodinger equation to an integral equation with a manifest four dimensional symmetry in momentum space. They establish the equivalence of the two approaches by appealing to the solutions of the H atom in parabolic coordinates, and demonstrate that the symmetry operators in the momentum space correspond to the symmetry operators in the configuration space.  As they note, the stereographic projection depends on the energy, so the statements for a SO(4) subgroup are valid only in a subspace of constant energy.  They then explore the expansion of the SO(4) group to include scale changes so the energy can be changed, transforming between states of different principal quantum number, which correspond to different subspaces of SO(4). To insure that this expansion results in a group, they include other transformations which results in the the generators forming the conformal group O(4,1). Their mathematical analysis introducing SO(4,1) is based on the projection of a p dimensional space (4 in the case of interest) on a parabaloid in p + 1 dimensions (5 dimensions).  In their derivation they treat bound states in their first paper \cite{band1} and scattering states in the second paper \cite{band2}.

As we have indicated the interest in the SO(4,2) symmetry of the Schrodinger equation was driven by a program focused on developing equations for composite systems that had infinite multiplets of energy solutions and ultimately could lead to equations that could be used to predict masses of elementary particles, perhaps using other than 4 dimensions \cite{nambu}\cite{band1}\cite{band2}\cite{frons3}\cite{frons2}\cite{baru7}  \cite{fron1}. In 1969 Jordan and Pratt showed that one could add spin to the generators $\bm{A}$ and $\bm{L}$, and still form a SO(4) degeneracy group.  By defining $\bm{J}=\frac{1}{2}(\bm{L}+\bm{A}) + \bm{S}$, they showed one could obtain a representation of O(4,1) for any spin $s$\cite{prat}. 

In their review of the symmetry properties of the hydrogen atom, Bander and Itzakson emphasize this purpose for exploring the group theory of the hydrogen atom\cite{band1} :
\begin{quotation}
The construction of unitary representations of non-compact groups which have the property that the irreducible representations of their maximal subgroup appear at most with multiplicity one is of certain interest for physical applications. The method of construction used here in the Coulomb potential case can be extended to various other cases. The geometrical emphasis may help visualize things and provide a global form of the transformations.  
\end{quotation}  

Special attention was also given to solutions for the hydrogen atom from the two body Bethe-Salpeter equation for a proton and electron interacting by a Coulomb potential, since the symmetry was that of a relativistic non-compact group \cite{nambu}\cite{fron1}\cite{frons8}\cite{kyri2}.

Finally in 1969, 5 years after it was published, Schwinger's form of the Coulomb Green's function based on the SO(4) symmetry was used to calculate the Lamb shift by Michael Lieber, one of Schwinger's students at Harvard \cite{schw1}\cite{lieber}.  A year later  
 Robert Huff, a student of Christian Fronsdal at UCLA, focused on the use of the results from SO(4,2) group theory to compute the Lamb shift\cite{huff}.  He converted the conventional expression for the Lamb shift into a matrix element containing generators of SO(4,2), and was able to perform  rotations and scale changes to simplify and evaluate the matrix elements. After clever mathematical manipulation, he obtained an expression for the Bethe log in terms of a rapidly terminating series for the level shifts. He provided an appendix with a brief discussion of the fundamental of SO(4,2) representations for the H atom, showing the expressions for the three generators needed to express the Schrodinger equation.

In the next few years, researchers published a few mathematically oriented papers\cite{prat}\cite{must1}\cite{baruMag} \cite{baru9}\cite{mack} \cite{deco}, a short book \cite{engl} dealing with the symmetries of the Coulomb problem, and a paper by Barut presenting a SO(4,2) formulation of symmetry breaking in relativistic Kepler problems, with a 1 page summary of the application of SO(4,2) for the non-relativistic hydrogen atom\cite{baru4}. Bednar published a paper applying group theory to a variety of modified Coulomb potentials which included some matrix elements of SO(4,2) using hydrogen atom basis states with quantum numbers $nlm$\cite{bednar}. There also was interest in application of the symmetry methods and dynamical groups in molecular chemistry \cite{wulf7} and atomic spectroscopy\cite{wybou}.

In the 1970's researchers focused on developing methods of group theory and on understanding dynamical symmetries in diverse systems\cite{mari}\cite{akyi}
\cite{fron4}. A book on group theory and its applications appeared in 1971\cite{wulf1}. Barut and his collaborators published a series of papers dealing with the hydrogen atom as a relativistic elementary particle, leading to an infinite component wave equation and mass formula\cite{baru2}\cite{baru3}\cite{barut1}\cite{baru5}. 

 Papers on the classical Kepler problem, the Runge-Lenz vector, and SO(4) for the hydrogen atom  have continued to appear over the years, from 1959 to today. Many were published in the 1970's  \cite{shib}\cite{dahl}\cite{coll}\cite{rodg}\cite{maju}\cite{stic}\cite{ligo} and some since 1980, including \cite{laks}\cite{ross}\cite{vale}\cite{more}\cite{lee}.  Papers dealing with SO(4,2) are much less frequent. In 1986 Barut, A. Bohm, and Ne'eman published a book on dynamical symmetries that included some material on the hydrogen atom\cite{barut2}.  In 1986, Greiner and Muller published the second edition of Quantum Mechanics Symmetries, which had 6 pages on the Hydrogen atom, covering only the SO(4) symmetry\cite{grei}. The 2005 book by Gilmore on Lie algebras has 4 pages of homework problems on the H atom to duplicate results in early papers \cite{gilm}.  The last papers I am aware of that used SO(4,2) were applications in molecular  physics\cite{kibl1}\cite{hamm} and more general in scope\cite{lev}. Carl Wulfman published a book on dynamical symmetries in 2011, which provides a helpful discussion of dynamical symmetries for the hydrogen atom\cite{wulf2}. He regularizes the Schrodinger equation, essentially multiplying by $r$, obtaining Sturmian wave functions in parabolic coordinates.  This approach allows him to treat bound and scattering states for SO(4,2) at one time, but requires redefining the inner product, and leads to a non-Hermitian position operator.

\subsection{The Dirac Hydrogen Atom}
We have focused our discussion on the symmetries of the non-relativistic hydrogen atom described by the Schrodinger equation.  Quantum mechanics also describes the hydrogen atom in terms of the relativistic Dirac equation, which we will discuss only briefly in this paper. 

The gradual understanding of the dynamical symmetry of the Dirac atom parallels that of the Schrodinger atom, but it has received much less attention, probably because the system has less relevance for particle physics and for other applications.  It was know that the rotational symmetry was present and that the equation predicted that the energy depended on the principal quantum number and the quantum number for the total angular momentum $j$, but not the spin $s$ or orbital angular momentum \textit{l} separately. This remarkable fact meant that in some sense angular momentum contributed the same to the total energy no matter whether it was intrinsic or orbital in origin.  This degeneracy is lifted if we include the radiative interactions which leads to the Lamb shift. 

 To understand the symmetry group for the Dirac equation consider that for a given total angular momentum quantum number j>0 there are two degenerate levels for each energy level of the Dirac hydrogen atom: one level has \textit{l}=j+1/2 and the other has \textit{l}=j-1/2.  Since the \textit{l} values differ by unity, the two levels have opposite parity.  Dirac described a generalized parity operator $K$, which was conserved.  For an operator $\Lambda$ to transform one degenerate state into the other,  it follows that the operator has to commute with j and have parity -1. This means it has to anticommute with $K$, and so it is a conserved pseudoscalar operator. 
                           
The parity  $(-1)^{\textit{l}+j-1/2}$ is conserved in time, so the states are parity eigenstates.  Using the two symmetry operators $\Lambda$ and K, one can build a SU(2) algebra.  If we include the O(3) symmetry due to the conservation of angular momentum, we obtain the full symmetry group $SU(2)xO(3)$ which is isomorphic to $SO(4)$ for the degeneracy of the Dirac hydrogen atom.

In 1950, M.Johnson and B. Lippman discovered the operator $\Lambda$\cite{john}. Further work was done on understanding $\Lambda$ by Biedenharn\cite{bied1}.  The Johnson-Lippman operator has been rediscovered and reviewed several times over the decades\cite{lanik}\cite{stahl}\cite{chen4}. It has been interpreted in the non-relativistic limit as the projection of the Runge-Lenz vector onto the spin angular momentum\cite{chen4}\cite{khac}\cite{zhan}. The SO(4) group can be expanded to include all states, and then the spectrum generating group is SO(4,1) or SO(4,2) depending on the assumptions regarding relativistic properties and the charges present\cite{baru0}\cite{frons2}. 
We will not discuss the symmetries of the Dirac H atom further.

\section{Background}
\subsection{The Relationship between Symmetry and Conserved Quantities}
The nature of the relationship between symmetry,
degeneracy, and conserved operators is implicit in the
equation
\be
[H,S] = 0
\ee
where H is the Hamiltonian of our system, S is a
Hermitian operator, and the brackets signify a commutator
if we are discussing a quantum mechanical system, or i times a Poisson
bracket if we are discussing a classical system. 
If S is viewed as the generator of a transformation
on H, then Eq. 1 says the transformation leaves H
unchanged. We therefore say S is a symmetry operator of
H and leaves the energy invariant. The fact that a non-
trivial S exists means that there is a degeneracy. To
show this, consider the action of the commutator on an
energy eigenstate $|E\ra$:
\be
[H,S] |E\ra= 0
\ee
or
\be
 H(S |E\ra) = E (S |E\ra)  .
 \ee
If $S$ is nontrivial then $S|E\ra$ is a different state
from $|E\ra$ but has the same energy eigenvalue. If we label
all such degenerate states by
\be
|E,m\ra , m = 1,... ,N
\ee
then clearly $S |E,m\ra$ is a linear combination of degenerate
states:
\be
S|E,m\ra = S_{ ms}|E,s\ra  .
\ee
$S_{ms}$ is a matrix representation of S in the subspace 
of degenerate states. In a classical Kepler system S generates
an orbit deformation that leaves H invariant. The
existence of a nontrivial S therefore implies degeneracy in which multiple states have the same energy eigenvalue.
We can show that the complete set of symmetry
operators for H forms a Lie algebra by applying Jacobi's identity to our set of
Hermitian operators $S_i$ :
\be
\left[\mathrm{H}, \mathrm{S}_{\mathrm{i}}\right]=0, \quad \mathrm{i}=1, \ldots, \mathrm{L}.
\ee
\be
[S_j, [H, S_i]] + [S_i, [S_j, H]] + [H, [S_i, S_j]] = 0
\ee
so
\be
[H, [S_i, S_j]]=0   .
\ee

The commutator of $S_i$ and $S_j$ is therefore a symmetry operator of H.  Either the commutator is a linear combination of all the symmetry operators $S_i, i=1,...,L$:
\be
[S_i, S_j]=C_{ij}^k S_k
\ee
or the commutator defines a new symmetry operator which we label $S_{L+1}$.  We repeat this procedure until the Lie algebra closes as in Eq. 9.

By exponentiation we assume we can locally associate a group of unitary transformations 
\be
e^{iS_ia^i}
\ee
for real $a^i$ with our Lie algebra and so conclude that a
group of transformations exists under which the Hamiltonian
is invariant\cite{heine}. We call this the symmetry or degeneracy
group of H. Our energy eigenstates states form a realization of this group.

It is possible to form scalar operators, called Casimir operators, from the generators of the group that commute with all the generators of the group, and therefore have numerical values. The values of the Casimir operators characterize the particular representation of the group. 
For example, for the rotation group in three dimensions, the generators are $\bm{L}= (L_1, L_2, L_3)$ and the quantity $\bm{L}^2=L(L+1)$ commutes with all the generators. $L$ can have any positive integer value for a particular representation. The Casimir operator for O(3) is   $\bm{L}^2$.  The number of Casimir operators that characterize a group is called the rank of the group. O(3) is rank 1 and SO(4,2) is rank 3. 

Now let us consider Eq. 1 in a different way. If we
view H as the generator of translations in time, then we
recall that the total time derivative of an operator $S_i$ is
\be
\frac{\mathrm{d} S_i}{\mathrm{d} t} =\frac{i}{\hbar}\left[\mathrm{H}, \hspace{4pt} \mathrm{S}_{\dot{\mathrm{i}}}\right]+\frac{\partial \mathrm{S_i}}{\partial t}
\ee
where the commutator and the partial derivative give the
implicit and explicit time dependence respectively. Provided that the symmetry operators have no explicit time dependence $(\frac{\partial S_i}{dt}=0)$, then Eq. 11 implies that Eq. 1 means that the symmetry operators $S$ are conserved in time and $\frac{dS_i}{dt}=0$. Conversely we can say that conserved Hermitian operators with no explicit time dependence are symmetry operators of H. This very important relationship between conserved Hermitian operators and symmetry was first discovered by German mathematician Emmy Nother in 1917, and is called Nother's Theorem \cite{nother} \cite{noth}\cite{neue}\cite{hanc}\cite{byer}\cite{history}.

\subsection{Non-Invariance Groups and Spectrum Generating Group}

As we have discussed, the symmetry algebra contains conserved generators $S_i$ that transform one energy eigenstate into a linear combination of eigenstates all with the same energy. To illustrate with hydrogen atom eigenstates:
\be
S_i|nlm\ra = S_{nlm}^{nl'm'}|nl'm'\ra  
\ee
where $|nlm\ra $ refers to a state with energy $E_n$, angular momentum $l(l+1)$ and $l_z=m$.

A non-invariance algebra contains generators $D_{i}$ that can be used to transform one energy 
eigenstate $|nlm \ra $ into a linear combination of other eigenstates, with the same or a different energy, different angular momentum $l$ and different azimuthal angular momentum $m$:

\be
D |nlm\ra = D_{nlm}^{n'l'm'}
|n'l'm'\rangle.
\ee
Since the set of energy eigenstates is complete, the action of the most general operator would be identical to that shown to Eq 13. Therefore this requirement alone is not sufficient to select the generators needed.

The goal is to expand the degeneracy group with its generators $S_i$ into a larger group, so that some or all of the eigenstates form a representation of the larger group with the degeneracy group as a subgroup. Thus we need to add generators $G_i$ such that the combined set of generators 
$$\{S_i, G_j; \text{ for all } i, j\}\equiv \{D_k; \text{ for all } k\}$$ forms an algebra that closes
\be
[D_i, D_j] = i \epsilon_{ijk}D_k  .
\ee
This is the Lie algebra for the expanded group.  To illustrate with a specific example, consider the O(4) degeneracy group with 6 generators. One can expand the group to O(5) or O(4,1) which has ten generators by adding a 4-vector of generators. One component might be a scalar and the other 3 a three-vector.  The question then is can some or all of the energy eigenstates provide a representation of O(5)? If so, then this would be considered a non-invariance group. The group might be expanded further in order to obtain generators of a certain type or to include all states in the representation. For the H atom the generators $D_i$ can transform between different energy eigenvalues meaning between eigenfunctions with different principal quantum numbers.  
 
Another way to view the expansion of the Lie algebra of the symmetry group is to consider additional generators $D_i$ that are constants in time\cite{doth2time} but do not commute with the Hamiltonian so 
$$
\frac{\mathrm{d} D_i}{\mathrm{d} t}=0 =\frac{i}{\hbar}\left[\mathrm{H}, \quad \mathrm{D}_{\dot{\mathrm{i}}}\right]+\frac{\partial \mathrm{D_i}}{\partial t} .
$$
If we make the additional assumption that the time dependence of the generators is harmonic
$$
\frac{\partial^2 D_i(t)}{\partial t^2} = \omega_{in} D_n(t)  .
$$
then generators $D_i$, and the first and second partial derivatives with respect to time could close under commutation, forming an algebra. This approach does not tell us what generators to add, but as we demonstrate in Section 7.5 it does reflect the behavior of the generators that have been added to form the spectrum generating group in the case of the hydrogen atom.

We may look for the largest set of generators $D_i$ which can transform the set of solutions into itself in an irreducible fashion (meaning no more generators than necessary). These generators form the Lie algebra for the non-invariance or spectrum generating algebra\cite{malk}\cite{sudar}. If the generators for the spectrum generating algebra can be exponentiated, then we have a group of transformations for the spectrum generating group. The corresponding wave functions form the basis for a single irreducible representation of this group.  This group generates transformations among all the solutions for all energy eigenvalues and is called the Spectrum Generating Group \cite{kyri}. For the H atom, SO(4,1) is a spectrum generating group or non-invariance group, which can be reduced to contain one separate SO(4) subgroup for each value of $n$. 

To get a representation of SO(4,1) we need an infinite number of states, which we have for the H atom.  This  group has been expanded by adding a five vector to form SO(4,2) because the additional generators can be used to express the Hamiltonian and the dipole transition operator. 
The group SO(p,q) is the group of orthogonal transformations that preserve the quantity $X=x_1^2 + x_2^2 + ...+x_p^2 -...-x_{p+q}^2$, which may be viewed as the norm or a p+q-dimensional vector in a space that has a metric with p plus signs and q minus signs.  The letters SO stand for special orthogonal, meaning the orthogonal transformations have determinant equal to +1.

In terms of group theory there is a significant difference between a group like SO(4) and SO(4,1).  SO(4) and SO(3) are both compact groups, while SO(4,1) and SO(4,2) are non-compact groups. A continuous group G is compact if each function f(g), continuous for all elements g of the group G, is bounded.  The rotation group in three dimensions O(3), which conserves the quantity $r^2=x_1^2 + x_2^2 + x_3^2$ is an example of a compact group.  

For a non-compact group consider the Lorentz group O(3,1) of transformations to a coordinate system moving with a velocity $v$.  The transformations preserve the quantity $r^2 - c^2t^2$. The matrix elements of the Lorentz transformations are proportional to $1/\sqrt{1-\beta^2}$, where $\beta=v/c$, and are not bounded as $\beta\rightarrow 1$. Therefore $r$ and $ct$ may increase without bound while the difference of the squares remains constant. Unitary representations of non-compact groups are infinite dimensional, for example the representation of the non-invariance group SO(4,1) has an infinite number of states. Unitary representations of compact groups can be finite dimensional, for example our representation of SO(4) for an energy level $E_n$ has dimension $n^2$.     
 
In the nineteen sixties and later, the spectrum generating group was of special interest in particle physics, because it was believed it could provide guidance where the precise particle dynamics were not known.  The hydrogen   atom provided a physical system as a model. Because the application was in particle physics,  there was less interest in exploring representations in terms of the dynamical variables for position and momentum.

The expansion of the group from SO(4,1) to SO(4,2) was motivated by the fact that the additional generators could be used to write Schrodinger's equation entirely in terms of the generators, and to express the dipole transition operator.  This allowed algebraic techniques and group theoretical methods to be used to obtain solutions, calculate matrix elements, and other quantities\cite{baru0}\cite{frons2}. 

\subsection[Basic Idea of Eigenstates of Z alpha-1]{Basic Idea of Eigenstates of $(Z\alpha)^{-1}$}

We briefly introduce the idea behind these states since they are unfamiliar\cite{brow}.  The full derivation is given in Section 4.
Schrodinger's equation in momentum space for bound states can be written as
$$
 \left[p^2 + a^2 - \frac{2mZ\alpha}{r}\right] |a\ra=0. 
$$
where $a^2=-2mE>0$ and $Z\alpha$ is the coupling constant, which we will now view as a parameter. This equation has well behaved solutions for certain discrete eigenvalues of the energy or $a^2$, namely 
$$a_n^2=-2mE_n$$
or equivalently
$$
( \frac{a_n}{mZ\alpha}) =\frac{1}{n}.
$$
This last equation shows that solutions exist for certain values $a_n$ of the RMS momentum $a$.  To introduce eigenstates of $(Z\alpha)^{-1}$ we simply take a different view of this last equation and say that instead of quantizing $a$ and obtaining $a_n$, we imagine that we quantize $(Z\alpha)^{-1}$, let $a$ remain unchanged, obtaining
$$\frac{a}{m(Z\alpha)_n}=\frac{1}{n}.$$
So now we can interpret Schrodinger's equation as an eigenvalue equation that has solutions for certain values of $(Z\alpha)^{-1}$ namely $$(Z\alpha)^{-1}_n = \frac{m}{an}.$$  We have the same equation but can view the eigenvalues differently but equivalently. Instead of quantizing $a$ we quantize $(Z\alpha)^{-1}.$

This roughly conveys the basic idea of eigenstates of the inverse of $(Z\alpha)$, but this simplified version does at all reveal the advantages of our reformulation because we have left the Hamiltonian unchanged.  In Section 4, we transform Schrodinger's equation to an eigenvalue equation in $a$ so that the kernel is bounded, which means that there are no states with E>0, no scattering states, and all states have the usual quantum numbers.  Other important advantages to this approach will also be discussed.

\subsection{Degeneracy Groups for Schrodinger, Dirac and Klein-Gordon Equations}

The degeneracy groups for the bound states described by the different equations of the hydrogen atom are summarized in Table 1.  The degeneracy (column 2) is due to the presence of conserved operators which are also symmetry operators (column 3), forming a degeneracy symmetry group (column 4). For example, 
The symmetry operators for the degeneracy group in the Schrodinger hydrogen atom are the angular momentum $\textbf{L}$ and the Runge-Lenz vector $\textbf{A}$. In Section 3, it will be shown that together these are the generators for the direct product SO(3)xSO(3) which is isomorphic to SO(4). 
Column 5 gives the particular representations present. These numbers are the allowed values of the Casimir operators for the group and they determine the degree of degeneracy (last column) and the corresponding allowed values of the quantum numbers for the degenerate states.

\begin{table}[ht]
\centering

\setlength{\arrayrulewidth}{.5mm}
\setlength{\tabcolsep}{16pt}
\renewcommand{\arraystretch}{1.5}
\newcolumntype{s}{>{\columncolor[HTML]{AAACED}} p{3cm}}

\begin{tabular}[l]{ | p{1.4cm} | p{1.2cm} | p{1.0cm} |p{1.1cm}| p{1.6cm} | p{1.2cm}|}
\hline
\multicolumn{6}{|c|}{Degeneracy Groups for Bound States in A Coulomb Potential} \\
 \hline
 
 Equation&Degeneracy&Conserved Quantities&Degeneracy Group& Representation&Dimension\\
 \hline
 Schrodinger & E indep. of $\textit{l}$, $\textit{l}_z$ & $\textbf{A}$, $\textbf{L}$&$SO(4)$&$(\frac{n-1}{2}$,$\frac{n-1}{2})$&$n^2$\\
Klein-Gordon&E indep. of $\textit{l}_z$&$\textbf{L}$&$O(3)$&Casimir op. is $\textit{l}(\textit{l}+1)$&$2\textit{l}+1$\\
Klein-Gordon without $V^2$ term& E indep. of $\textit{l}$, $\textit{l}_z$ & $\textbf{A}$, $\textbf{L}$&$SO(4)$&$(\frac{n-1}{2}$,$\frac{n-1}{2})$&$n^2$\\
Dirac&E depends on J,n only&$\Lambda,K, \textbf{J}$&$SO(4)$&$(1/2, J)$&2(2J+1)\\
\hline
\end{tabular}
\caption{  In the table \textbf{L}$=$\textbf{r}x\textbf{p} is the orbital angular momentum; \textbf{A} is the Runge-Lenz vector; \textbf{J}$=$\textbf{L}+\textbf{$\sigma$}/2 \hspace{3pt} is the total angular momentum$;$ K is the generalized parity operator;  $\Lambda$ is the conserved pseudoscalar operator.}
\end{table}

The Casimir operators, which are made from generators of the group, have to commute with all the members of the group, and the only way this can happen is if they are actually constants for the representation.  The generators are formed from the dynamical variables of the H atom, so the Casimir operators are invariants under the group composed of the generators, and their allowed numerical values reflect the underlying physics of the system and determine the appropriate representations of the group\cite{lie}\cite{wulf2}\cite{lipk}. For example, $\bm{L}^2$ is the Casimir operator for the group O(3) and can have the values $\textit{l}(\textit{l} +1)$.  The relationship between Casimir operators and group representations is true for all irreducible group representations, including the SO(4) degeneracy group, as well as the spectrum generating group SO(4,2)\cite{bednar}\cite{kyri2}.

For the Schrodinger equation there are $n^2$ states $|n l m \ra$ that form a representation of the degeneracy group SO(4) formed by $\bm{L}$ and $\bm{A}$.  These states correspond to the principal quantum number $n$, the $n$ different values of the angular momentum quantum number $\textit{l}$, and $2l +1$ different values of the $z$ component of the angular momentum $\textit{l}_Z=m$.

For the Dirac equation, the $2(2J+1)$ dimensional degeneracy group for bound states is realized by the total angular momentum operator $\textbf{J}$, the generalized parity operator K, and the Johnson-Lippman operator $\Lambda$, which together form the Lie algebra for SO(4).  

For the fully relativistic Klein-Gordon equation, only the symmetry from rotational symmetry survives, leading to the degeneracy group O(3).  If the $V^2$ term, the four-potential term squared is dropped in a semi-relativistic approximation as we describe in Section 4.3, then the equation can be rewritten in the same form as the non-relativistic Schrodinger equation, so a Runge-Lenz vector can be defined and the degeneracy group is again SO(4).

\section{Classical Theory of the H Atom}
In order to discuss orbital motion and the continuous deformation or orbits we give this discussion in terms of classical mechanics, but much of it is valid in terms of the Heisenberg representation of quantum mechanics if the Poisson brackets are converted to commutators, as will be discussed in Section 4.

For a charged particle in a Coulomb potential there
are two classical conserved vectors: the angular momentum
\textbf{L}, which is perpendicular to the plane of the orbit, and
the Runge-Lenz vector \textbf{A}, which goes from the focus corresponding to the center of mass and force along the semi-major axis to the perihelion (closest point) of the elliptical orbit. The conservation of \textbf{A} is related to the fact that non-relativistically the orbits do not precess. The Hamiltonian of our bound state classical system with an energy E<0 is\cite{units}

\be
H=\frac{p^{2}}{2 m}-\frac{Z \alpha}{r}=E
\ee
where m=mass of the electron, r is the location of the electron, p is it's momentum, $\alpha$ is the fine structure constant, E is the total non-relativistic energy.

The Runge-Lenz vector is 
\be
\textbf{A}=\frac{1}{\sqrt{-2 m H}}(\textbf{p} \times \textbf{L}-m Z \alpha \frac{\textbf{r}}{r})
\ee
where \textbf{L} is the angular momentum. From Hamilton's equation, $H=E$ so  
\be
\textbf{A}=\frac{\textbf{p} \times \textbf{L}}{a}-\frac{m Z \alpha}{a} \frac{\textbf{r}}{r}
\ee
where  \hspace{5pt}$a$\hspace{5pt}   is defined by 
\be
a = \sqrt{-2mE}    .
\ee
From the virial theorem, the average momentum $\la p^2 \ra=-2mE$ so \hspace{3pt}$a$\hspace{3pt} is the root mean square momentum. We are  discussing bound states so E<0.
It is straightforward to verify that \textbf{A} is conserved in time:
\be
[\textbf{A},H]=\frac{d\textbf{A}}{dt}=0  .
\ee
From the definition of \textbf{A} and the definition of angular momentum
\be
\textbf{L}=\textbf{r}\hspace{3pt}\times\hspace{3pt}\textbf{p}
\ee
if follows that \textbf{A} is orthogonal to the angular momentum vector 
\be
\textbf{A} \cdot \textbf{L} = 0  .
\ee
Using the fact that \textbf{A} and \textbf{L} are conserved, we can
easily obtain equations for the orbits in configuration and momentum space and the eccentricity,
and other quantities, all usually derived by solving the
equations of motion directly.

$\bm{A}$ and $\bm{L}$ are the generators of the group O(4).  If we introduce the linear combinations $
\bm{N}=\textstyle\frac{1}{2}(\bm{L} + \bm{A})$ and $
\bm{M}=\frac{1}{2}(\bm{L} - \bm{A})$, we find that $\bm{N}$ and $\bm{M}$ commute reducing the nonsimple group O(4) to O(3)x O(3), which we will discuss in Section 4.2 in the language of quantum mechanics.  

\subsection{Orbit in Configuration Space}

To obtain the equation of the orbit one computes
\be
\textbf{r} \cdot \textbf{A} = r A cos{\phi_r}=-r \frac{mZ\alpha}{a}+\textbf{r}\cdot\textbf{p}\times\textbf{L}  .
\ee
Noting that $\textbf{r}\cdot\textbf{p}\times \textbf{l}=L^2$ we can solve for r
\be
r=\frac{L^2/mZ\alpha}{(a/mZ\alpha)A\hspace{3pt}cos{\phi_r} + 1}  \hspace{7pt}.
\ee
This is the equation of an ellipse with eccentricity $e=(a/mZ\alpha)A$ and a focus at the origin (Fig. 1).  To find $\hspace{3pt} e \hspace{3pt}$ we calculate $\textbf{A}\cdot \textbf{A}$ using the identity $\textbf{p} \times \textbf{L} \cdot \textbf{p} \times \textbf{L} = p^2L^2$ and obtain
\be
A^{2}=\frac{p^{2} L^{2}}{a^{2}}-\frac{2 m Z \alpha}{a^{2}} \frac{L^{2}}{r}+\left(\frac{m Z \alpha}{a}\right)^{2}   .
\ee
Substituting E for the Hamiltonian Eq. 15 gives the usual result
\be
\frac{a}{mZ \alpha} A=e=\sqrt{\frac{2 {EL}^{2}}{m(Z\alpha)^{2}}+1} \hspace{7pt}.
\ee
The length $r_c$ of the semi-major axis is the average of the radii at the turning points
\be
r_{c}=\frac{r_{1}+r_{2}}{2}  .
\ee
Using the orbit equation we find
\be
r_{c}=\frac{L^{2}}{m Z \alpha} \frac{1}{1-e^{2}}
\ee
or
\be
r_c = -\frac{Z \alpha}{2 E}=\frac{m Z \alpha}{a^{2}}   .
\ee
The energy depends only on the length of the semi-major
axis $r_c$, not on the eccentricity. This important result is a consequence of the symmetry of the problem.
It is convenient to parameterize the eccentricity
in terms of the angle $\nu$ (see Fig. 1) where
\be e = \sin{\nu}
\ee
From this definition and from Eqs. 25, 27 and 28 follow the
useful results
\be
L = r_{c} a \cos{\nu} ,\hspace{13pt} A  = r_{c} a \sin{\nu} 
\ee
which immediately imply
\be
L^2  +  A^2 = \left(r_{c} a\right)^{2}=\left(\frac{m Z \alpha}{a}\right)^{2}  .
\ee
This equation is the classical analogue of an important
quantum mechanical result first obtained by Pauli and Hulthen allowing us to determine the
energy levels from symmetry properties alone \cite{pauli}\cite{hult}. From Fig. 1, it is apparent that this equation is a statement of Pythagoras's theorem for right triangles.
 \begin{figure}[H] 
\centering
\includegraphics[scale =.6]{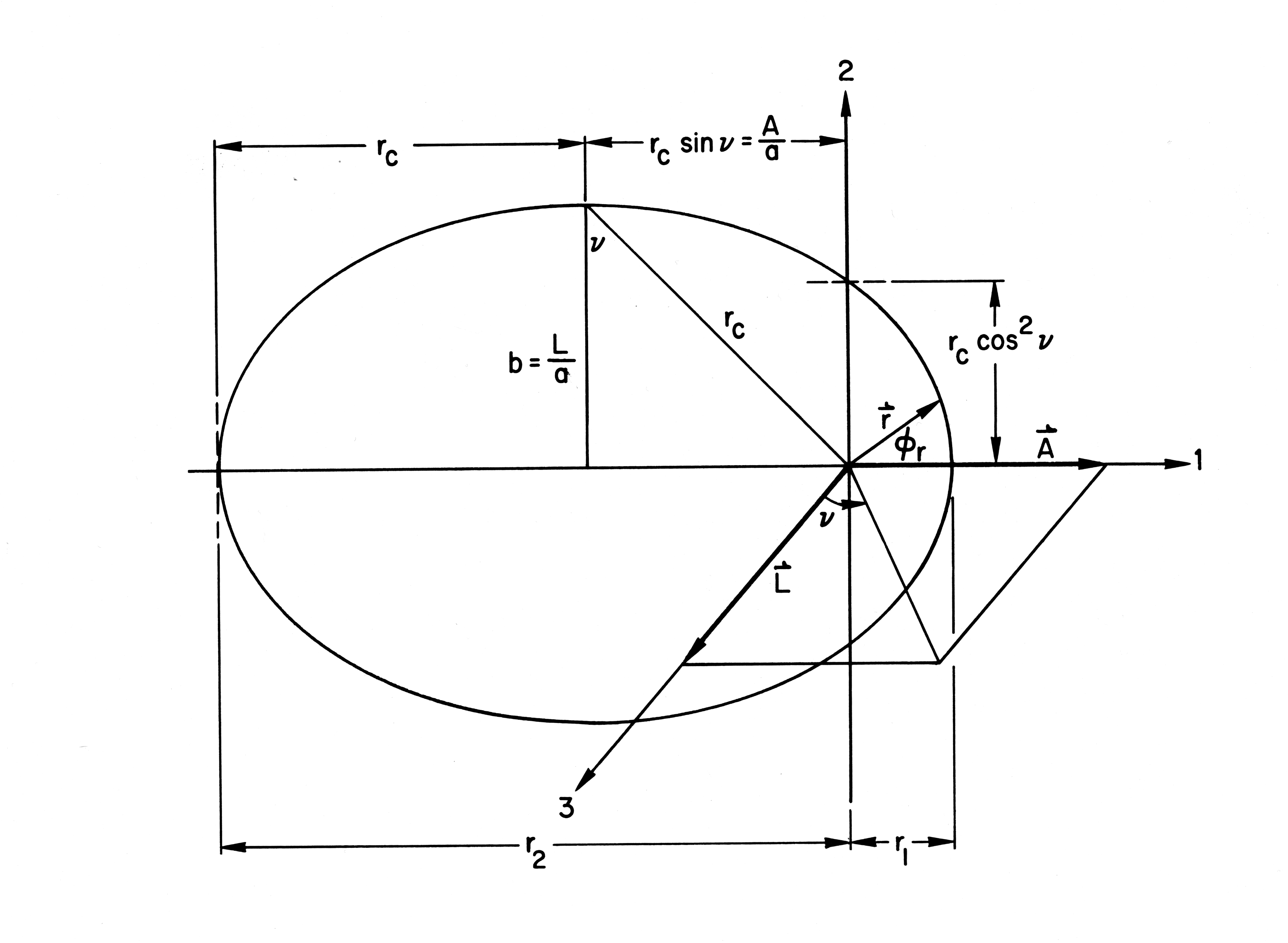}
 \caption{{Classical} Kepler orbit in configuration space. The orbit is in the 1-2 plane (plane of the paper). One focus, where the proton charge is located, is the origin. The semi-minor axis is $b=r_c \sin {\nu}$. The semi-major axis is $r_c$. }
 \label{Fig1}
\end{figure}

The energy
equation (Eq. 15) and the orbit equation (Eq. 23)
respectively may be rewritten in terms of $\hspace{4pt} a\hspace{2pt}$, $r$ , and $\nu$ :
\be
\frac{r_{c}}{r}=\frac{p^{2}+a^{2}}{2 a^{2}}
\ee
\be
r=\frac{r_{c} \cos ^{2} \nu}{1+\sin \nu \cos \phi_{r}}  .
\ee
\subsection{The Period}
To obtain the period we 
use the geometrical definition of the eccentricity
\be
e=\sqrt{1-(b/r_c)^2}
\ee
where $b$ is the semi-minor axis. Using $e=\sin{\nu}$ we find
\be
b=r_c \cos{\nu}
\ee 
so from Eq.30 we obtain
\be
L=a b   .
\ee
From classical mechanics we know the magnitude of the angular momentum is equal to twice the mass times the area
swept out by the radius vector per unit time. The area of the ellipse is $\pi b r_c$. It the period of the classical motion is $T$, then $L=2m \pi b r_c / T$.  Therefore the classical period is
\be
T=2 \pi \frac{m r_c}{a}=2 \pi \sqrt{\frac{m(Z \alpha)^{2}}{-8 E^{3}}}   .
\ee
and the classical frequency $\omega_{cl}=2\pi/T$ is
\be
\omega_{cl}=\frac{a}{m r_c}    .
\ee
\subsection{Group Structure SO(4)}
The generators of our symmetry operations form the closed Poisson bracket algebra of $O(4)$:
\be
[L_i , L_j ] = i \epsilon_{ijk} L_k\hspace{3pt}, \hspace{7pt} [L_i, A_j] = i\epsilon_{ijk} A_k \hspace{3pt}, \hspace{7pt} [A_i, A_j]= i \epsilon_{ijk} L_k\hspace{4pt}.
\ee
The brackets mean   i  times the Poisson bracket, which is the classical limit of a commutator.  The first bracket says that the angular momentum generates rotations and forms a closed Lie algebra corresponding to O(3). The second bracket says that the Runge-Lenz vector transforms as a vector under rotations generated by the angular momentum. The last commutator says that the multiple transformations generated by the Runge-Lenz vector are equivalent to a rotation. Taken together the commutators form the Lie algebra of O(4). The connected symmetry group for the classical bound state Kepler problem is obtained by exponentiating our algebra giving the symmetry group SO(4).  The scattering states with $E<0$ form a representation of the non-compact group SO(3,1). 

We now want to determine the nature of the transformations generated by $A_i$ and $L_i$. Clearly $\bm{L}\cdot\bm{\delta\omega}$
generates
a rotation of the elliptical orbit about the axis $\bm{\delta\om}$ by an amount
$\delta\om$. To investigate the transformations generated by $\bm{A}\cdot\bm{\delta\nu}$
we assume a particular orientation of the orbit, namely that
it is in the 1-2 (or x-y) plane and that $\bm{A}$ is along
the 1-axis (see Fig 1). The more general problem is
obtained by a rotation generated by $L_i$. For an example, 
we choose a transformation with $\bm{\delta\nu}$	pointing along the
2-axis so that $\bm{A}\cdot\bm{\delta\nu}=A_2 \delta\nu$ 	. The change in $\bm{A}$ is defined by $\bm{\delta A}$ where
\be
\bm{\delta A} = i [\bm{A}\cdot\bm{\delta \nu}, \bm{A}]  .
\ee
From the Poisson bracket relations we find for this particular case:
\be
\delta A_1 = L_3 \delta \nu \hspace{3pt}, \hspace{7pt} \delta A_2 = 0 \hspace{3PT},\hspace{7pt}  \delta A_3 = -L_1 \delta \nu
\ee
For our orbit, $L_1=0$ so $\delta A_3=0$. We perform a similar computation to find $\bm{\delta L}$ .  We find we can characterize the transformation by 
\be
\begin{array}{ll}\delta {A_{1}}=L_{3} \delta v & \text { or } \quad \delta{A}=L \delta \nu \\ \delta{L_{3}}=-A_{1} \delta \nu & \text { or } \quad \delta L=-A \delta \nu \\ \delta e=\sqrt{1-e^{2}} \delta \nu & \text { or } \quad \delta(\sin \nu)=\cos \nu \delta\nu\end{array}  .
\ee 
Recalling $e=\sin{\nu}$ and Eq. 30 we see that these transformations are equivalent to the substitution
\be
\nu \longrightarrow \nu + \delta \nu   .
\ee
In other words the eccentricity of the orbit, and therefore $\bm{A}$ and $\bm{L}$ are all changed in such a way that the energy, $\hspace{3pt} a \hspace{3pt}$ and $r_c$ (length of the semi-major axis) remain constant. In our example, both $\bm{L}$ and $\bm{A}$ are changed in length but not direction, so the plane and orientation of the orbit are unchanged.  The general transformation $\bm{A}\cdot \bm{\delta \mu}$ will also rotate the plane of the orbit or the semi-major axis.

Figure 2 shows a set of orbits in configurations space with different values of the eccentricity $e=\sin{\nu}$ but the same total energy and the same semimajor axis $r_c$, which is the bold hypotenuse. The bold vertical and horizontal legs are $A/a$ and $L/a$ and are related to the hypotenuse $r_c$ by Pythagoras's theorem.  The generator $A_2 \nu$ produces a deformation of the circular orbit into the various elliptical orbits shown. This classical degeneracy corresponds to the quantum mechanical degeneracy in 
energy levels that occurs for different eigenvalues of the angular momentum with a fixed principal quantum number.
 \begin{figure}[H] 
\centering
 \includegraphics[scale = 0.55 ]{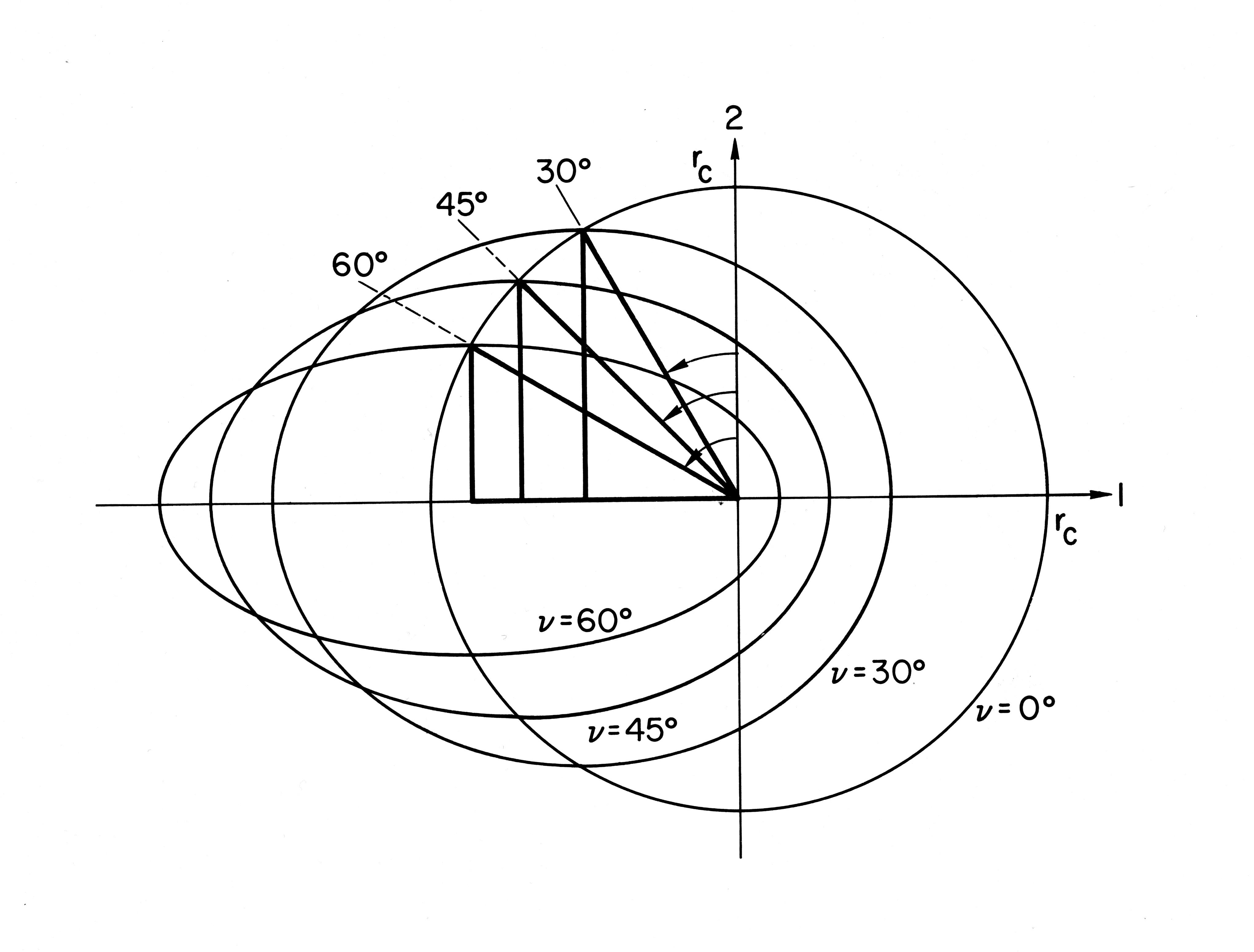}
 \caption{{Kepler} bound state orbits in configuration space for a fixed energy and different values of the eccentricity $e = \sin{\nu}$. The bold hypotenuse is the semi-major axis $r_c$ which makes an angle $\nu$ with the vertical 2-axis.}
 \label{Fig2}
\end{figure}
We can visualize all possible elliptical orbits for
a fixed total energy or semi-major axis through a simple
device. It is possible to produce an elliptical orbit with
eccentricity $\sin {\nu}$ as the shadow of a circle of radius $r_c$
which is rotated an amount $\nu$ about an axis perpendicular
to the illuminating light. With a complete rotation of the
circle we will see all possible classical elliptical
orbits corresponding to a given total energy. In quantum
mechanics only certain angles of rotation would be possible
corresponding to the quantized values of $L$. As the circle
is rotated we must imagine that the force center shifts as the sine of the angle of rotation so that it always remains at the focus\cite{center}.

\subsection{The Classical Hydrogen Atom in Momentum Space}
We can derive the equation for the classical orbit
in momentum space of a particle bound in a Coulomb potential using the conserved operators $\bm{L}$ and $\bm{A}$. For convenience we assume that we have rotated our
axes so that $\bm{L}$ lies along the 3-axis and $\bm{A}$
the 1-axis as shown in Fig 1. We compute
\be
\bm{p}\cdot \bm{A} = p_1 A =\frac{-mZ\alpha}{a}\bm{p}\cdot \frac{\bm{r}}{r}\equiv\frac{-mZ\alpha}{a} p_r
\ee
and we employ Eq. 1.30, $A =\frac{mZ\alpha}{a}\sin{\nu}$, to obtain\cite{cordinate}
\be
p_r=-\sin{\nu}p_1
\ee
which we substitute into the identity
\be
p_r^2 + \frac{\bm{L}^2}{r^2} = p^2=p_1^2 + p_2^2
\ee
Using Eqs.30 and 32 we find \be
1=\left(\frac{2 a p_{1}}{p^{2}+a^{2}}\right)^{2}+\left(\frac{2 a p_{2}}{p^{2}+a^{2}}\right)^{2} \frac{1}{\cos ^{2} v}
\ee and
\be 
p^{2}-a^{2}=2 a p_{2} \tan{\nu}
\ee
which may also be written as
\be
p_{1}^{2}+\left(p_{2}-a\tan{\nu} \right)^{2}=\frac{a^{2}}{\cos^2{\nu}}  .
\ee
From Eq. 49 we see the orbit in momentum space is a circle of radius $a/ \cos{\nu}$	with its center displaced from the
origin a distance $a\tan{\nu}$ along the 2-axis. Fig. 3 shows the momentum space orbit that corresponds to the configuration space orbit in Fig. 1.
 \begin{figure}[ht] 
\centering
 \includegraphics[scale=.6]{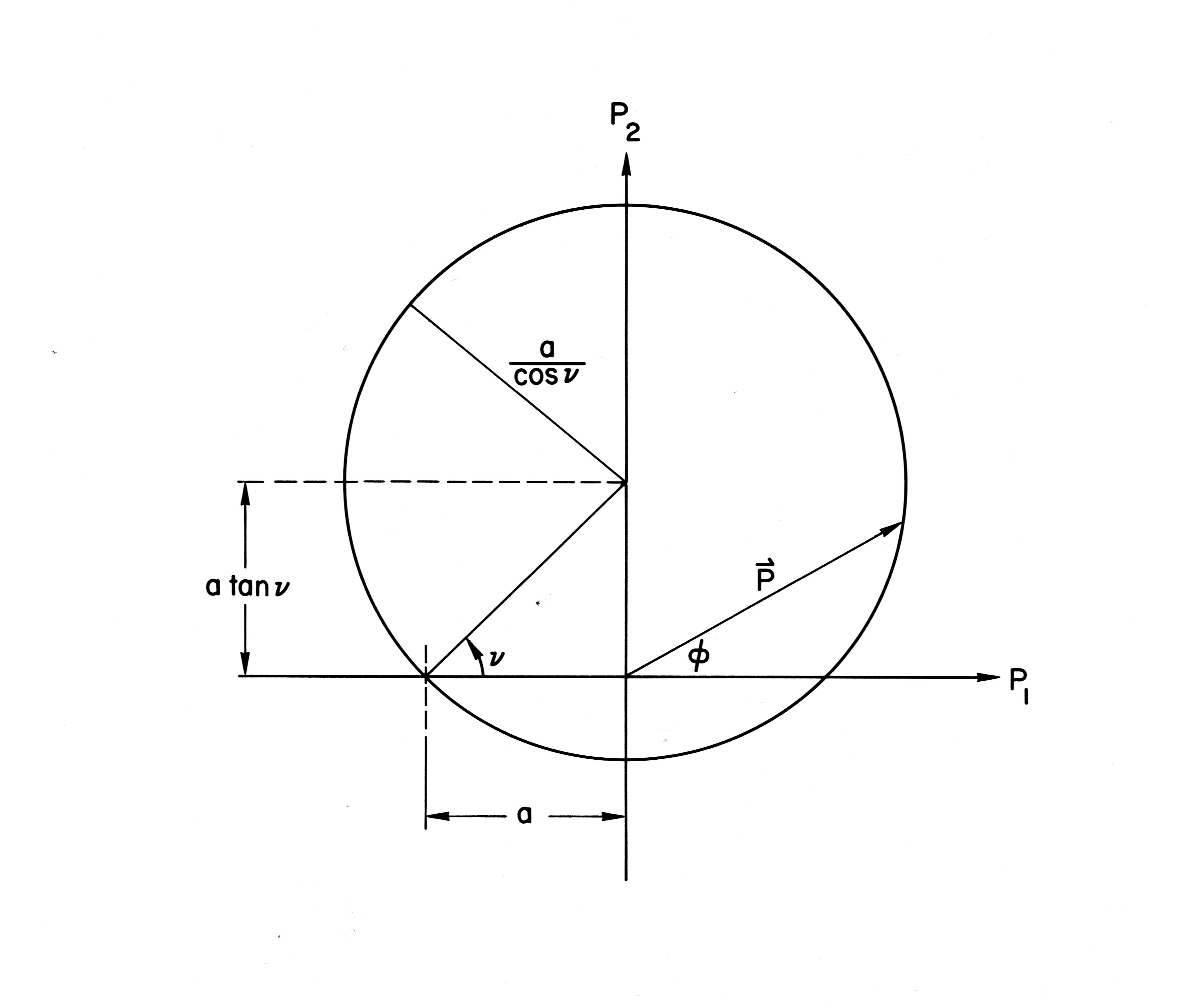}
 \caption{{Kepler} orbit in momentum space of radius $a/\cos \nu$, with its center at $p_2=a\tan\nu$, corresponding to the orbit in configuration space shown in Fig. 1. A circular orbit in configuration space corresponds to a circular orbit in momentum space centered on the origin with radius $a$. }
 \label{Fig3}
\end{figure}

As an alternative method of showing the momentum space orbits are circular we can compute\cite{brown}
\be
\left(\bm{p}-a\frac{\bm{L} \times \bm{A}}{\bm{L}^{2}}\right)^{2}=C^{2}  .
\ee 
Using the lemma
\be
\bm{p}\times\bm{A} =  -\frac{\bm{L}}{2a}(p^2 - a^2)   ,
\ee
the fact that $\bm{L}\cdot \bm{A}=0$, and Eq. 30, we find $C=\frac{a}{\cos{\nu}}$. The orbit is a circle of radius $\frac{a}{cos{\nu}}$ whose center lies at $a\frac{\bm{L} \times \bm{A}}{\bm{L}^{2}}$, in agreement with the previous result.

We now consider what the generators $A_i$ and $L_i$	do to the orbit in momentum space. Clearly $\bm{L}$ generates a rotation of the axes. For an $A_i$ transformation consider the same situation we considered in our discussion
of the configuration space orbit (see Figs. 1 and 3). Since the generator $A_2 \delta \nu$ changes $\nu$ to $\nu + \delta \nu$, we conclude that in momentum space this shifts the center of the orbit along the 2-axis and changes the radius of the circle.
However,the distance  $a$  from the 2-axis to the intersection of
the orbit with the 1-axis remains unchanged. Fig. 4 shows a set of
momentum space orbits for a fixed energy which correspond to the set of orbits in configuration space shown in Fig. 2.
 \begin{figure}[H] 
\centering
 \includegraphics[scale=0.65]{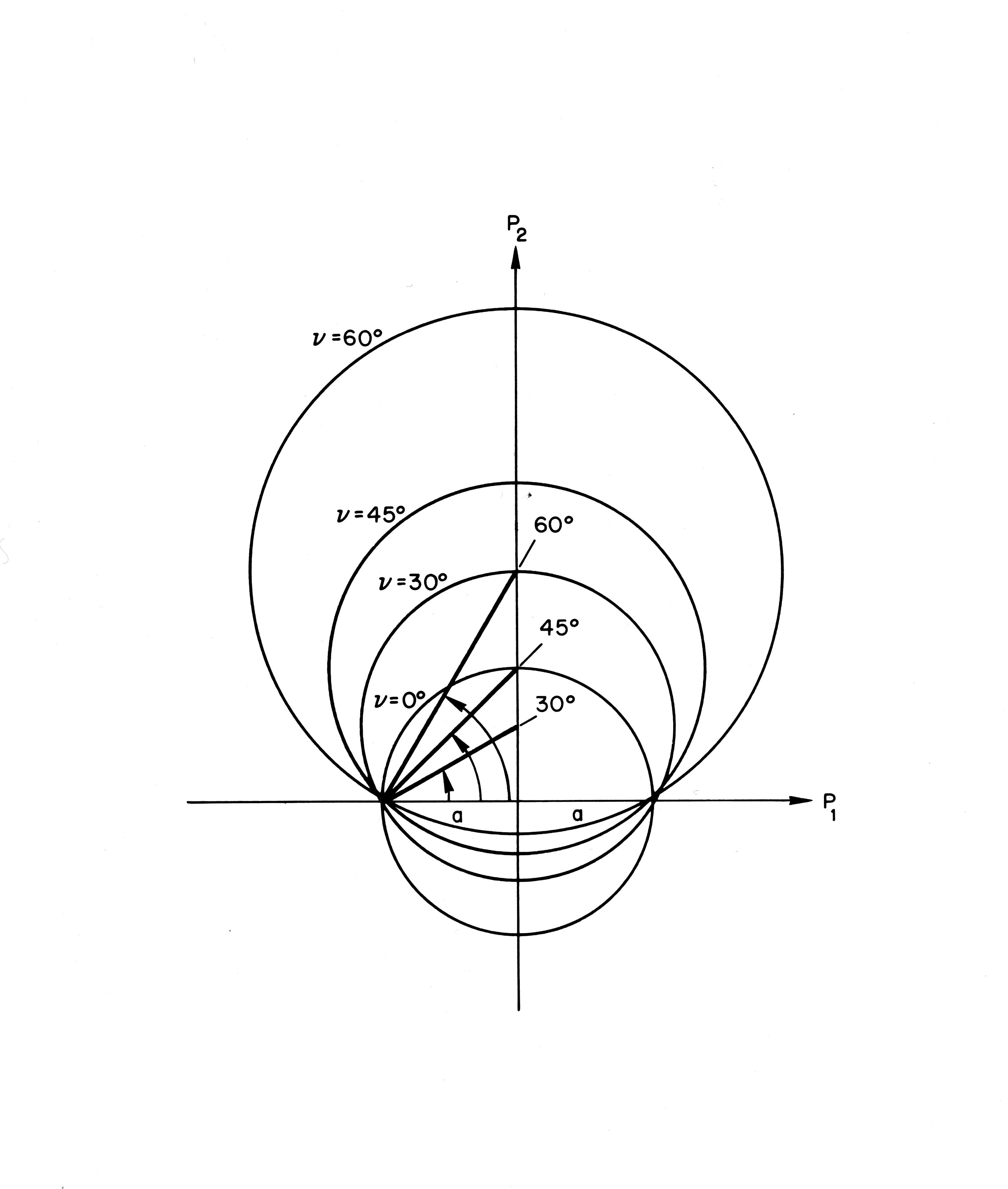}
 \caption{{Kepler} orbits in momentum space for a fixed energy and RMS momentum $a$ with different values of the eccentricity $e=\sin{\nu}$ corresponding to the orbits in configuration space shown in Fig. 2. }
 \label{Fig4}
\end{figure}
\subsection{Four-dimensional Stereographic Projection in Momentum Space}
It is interesting that in classical mechanics the bound state orbits in a Coulomb potential are simpler in momentum space than in configuration space. In quantum mechanics the momentum space wave functions become simply four-dimensional
spherical harmonics if one normalizes the momentum $\bm{p}$ by dividing by the RMS momentum $a = \sqrt{-2mE}$ and
performs a stereographic projection onto a unit hyper-sphere in a four-dimensional space\cite{fock}\cite{band1}. We will do the analogous projection procedure for the classical orbits. As shown in Fig. 5, the three-dimensional momentum space hyperplane passes through the center of the four-dimensional hypersphere. The unit vector in the fourth direction is $\hat{n} = (1,0,0,0)$. The unit vector $\hat{U}$ goes from the center of the sphere to the surface of the
hypersphere where it is intersected by the line connecting the vector $\bm{p}/a$ to the north pole of the sphere. 
 \begin{figure}[ht] 
\centering
 \includegraphics[scale=0.45]{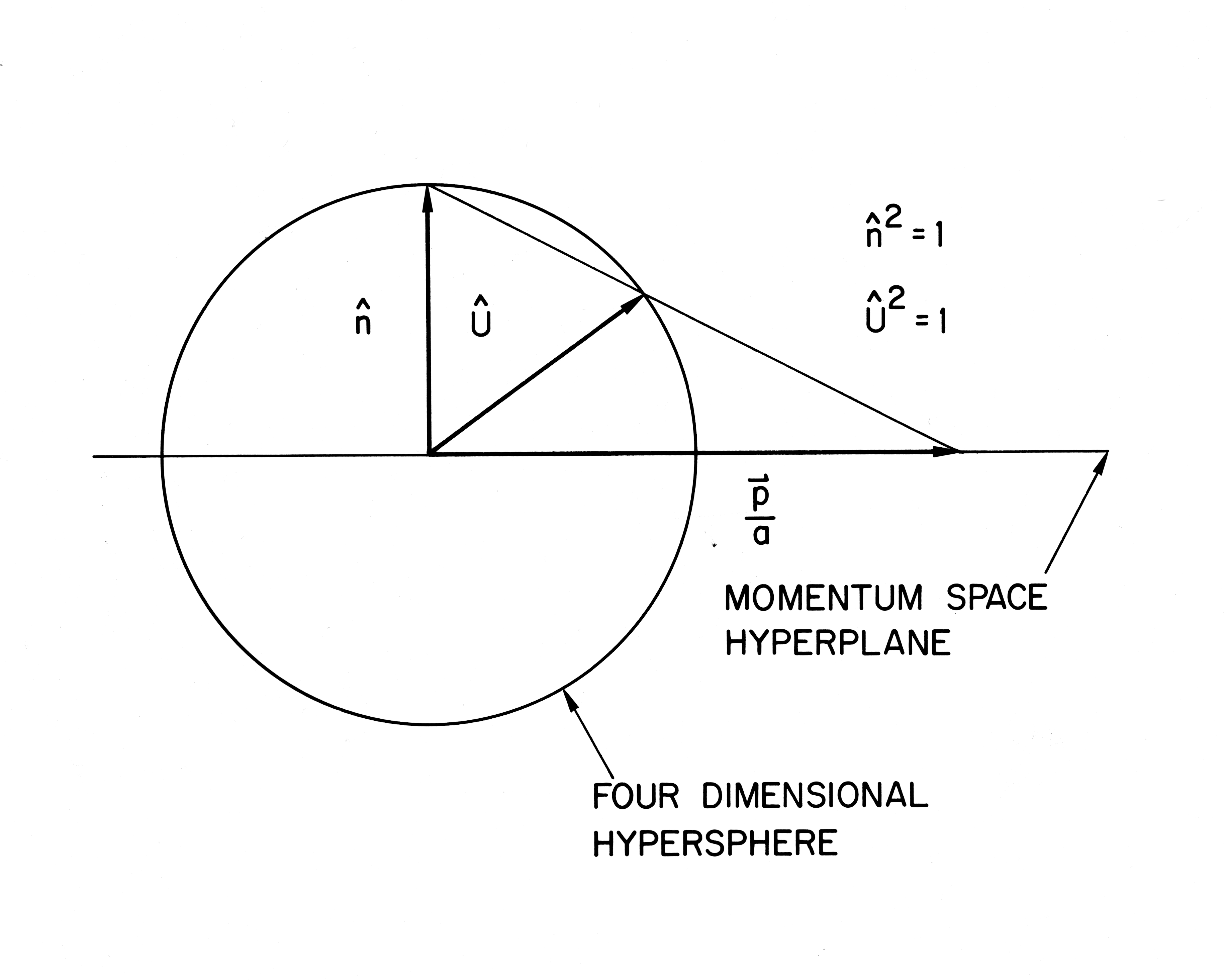}
 \caption{Stereographic projection in momentum space for a fixed energy, mapping $\bm{p}/a$ into $\hat{U}$. The unit vector in the 4 direction is $\hat{n}$ and $\hat{n} \cdot \hat{U}=\cos\Theta_4$. }
 \label{Fig5}
\end{figure}
We find

\be
U_{i}=\frac{2 ap_{i} }{p^{2}+a^{2}}\hspace{3pt} \hspace{4pt} i=1,2,3. \hspace{12pt} U_{4}=\frac{p^{2}-a^{2}}{p^{2}+a^{2}}
\ee
or inverting, 
\be
p_{i}=\frac{aU_{i} }{1-U_{4}}\hspace{16pt}  p^{2}=a^{2} \frac{1+U_{4}}{1-U_{4}}  .
\ee
Momentum space vectors for which $p/a<1$ are mapped onto
the lower hyperhemisphere. The advantage of this projection over one in which the hypersphere is tangent to the hyperplane is that we may have $|\hat{n}| = |\hat{U}|=1$. At times
it is convenient to describe $\hat{U}$ in terms of spherical polar coordinates in four dimensions. Since $\hat{U}$ is a unit
vector we define
\be
\begin{array}{l}U_{4}=\cos \theta_{4} \\ U_{3}=\sin \theta_{4} \cos \theta \\ U_{2}=\sin \theta_{4} \sin \theta \sin \phi \\U_{1}=\sin \theta_{4} \sin \theta \cos \phi\end{array}
\ee
where $\theta$ and $\phi$ are the usual coordinates in three dimensions. By comparison to Eq. 52, we have
\be
\theta_{4}=2 \cot ^{-1} \frac{p}{a} \qquad \theta=\cos ^{-1} \frac{p_{3}}{p} \qquad \phi=\tan ^{-1} \frac{p_{2}}{p_{1}}
\ee
\subsection{Orbit in U space}
We want to find the trajectory of the particle on the
surface of the hypersphere corresponding to the Kepler 
orbits in configuration space or the displaced circles in
momentum space. We assume we have rotated the axes in
configuration space so that $\bm{L}$ is along the 3-axis and
$\bm{A}$ is along the 1-axis as shown in Fig. 1. The equation
for the orbit in three-dimensional momentum space is
given by Eqs. 48 or 50. Dividing Eq. 48 by $p^2 + a^2$ immediately gives a parametric equation for the projected orbit in U space:
\be
U_4 = U_2 \tan{\nu}  .
\ee

\begin{figure}[ht] 
\centering
 \includegraphics[scale=0.61]{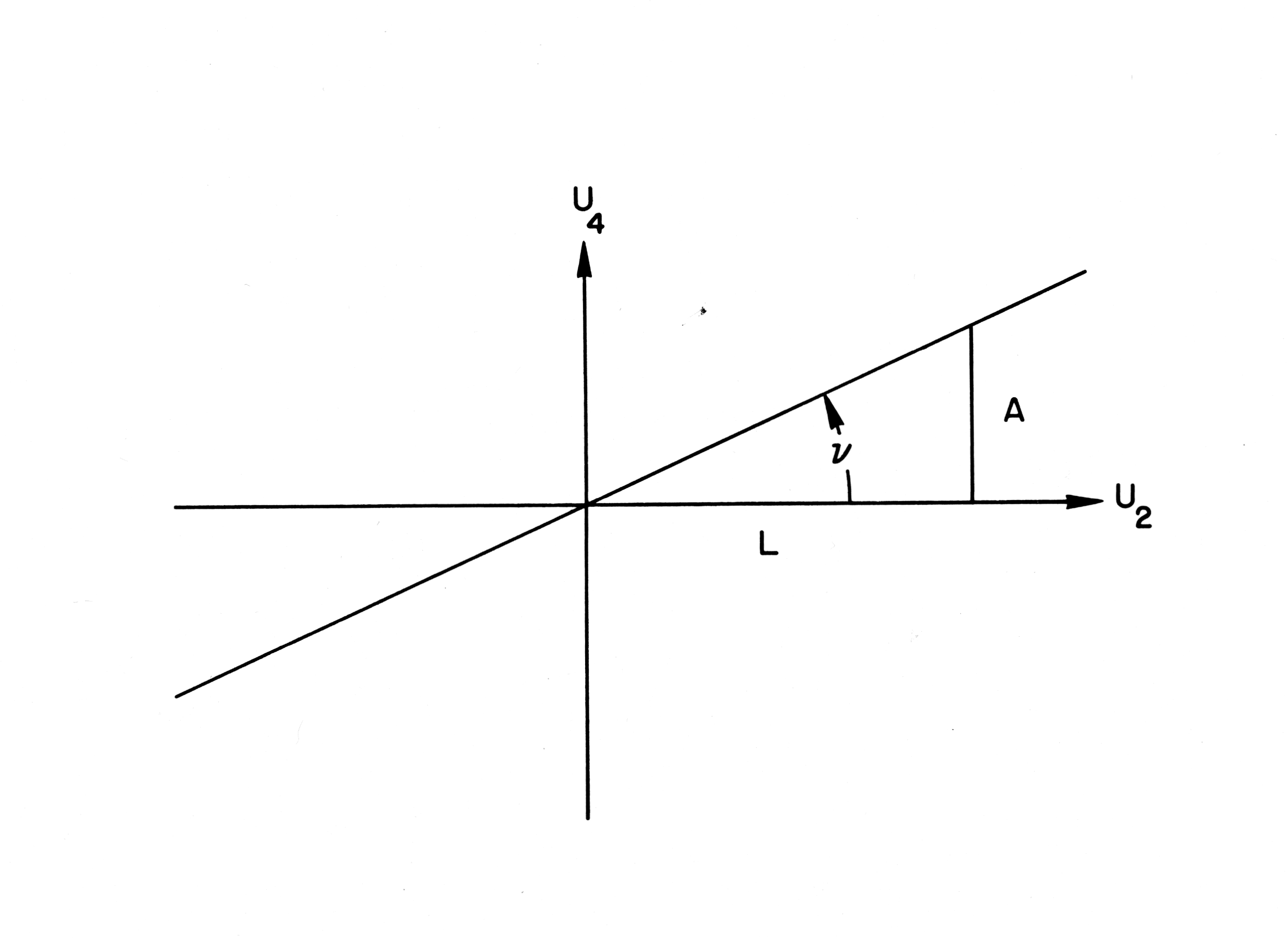}
 \caption{Shows the hyperplane which contains the orbit and makes an angle $\nu$ with the $U_2$ plane. Notice $\tan{\nu} = A/L$ as required by Eq. 30. }
 \label{Fig6}
\end{figure}
Since the orbit is in the 1-2 plane in configurations space, $U_3=0$. The orbit lies in a hyperplane perpendicular to 
the 2-4 plane that goes through the origin and makes an angle $\pi/2 - \nu$ with the 4-axis as shown in Fig. 6 \cite{angle}.  
The orbit is the intersection of this plane with the hypersphere and is therefore a great circle. To derive the exact equation for
the projected orbit we express p in Cartesian components $p_1$ and $p_2$ 
in Eq. 48 and substitute Eq. 52 obtaining
\be
U_l^2 + U_2^2 - 2\tan{\nu}\hspace{3pt} U_2(l - U_4) = (1 - U_4)^2  .
\ee
To interpret this equation we consider it in a rotated
coordinate system. If we perform a rotation by an amount
$\delta \nu$ about the 1-3 plane ($A_2 \delta \nu$ is the generator of this rotation), the equations of transformation
may be written\cite{later}
\be
\begin{array}{l}U_{2}=U_{2}` \cos {\delta \nu}+U_{4}` \sin{ \delta \nu}, \quad U_{3}=U_{3}` \\ U_{4}=U_{4}` \cos \delta \nu-U_{2}^{\prime} \sin {\delta \nu}, \quad U_{1}=U_{1}`   .\end{array}
\ee
This transformation is equivalent to making the substitution $\nu \longrightarrow \nu + \delta \nu$
in the equations relating to the orbit.  For example, Eq. 56 becomes
\be
U_4'  =  U_2' \tan{(\nu + \delta \nu)}   .
\ee
We choose $\delta \nu = -\nu$  , which means the orbital plane becomes $U_4' = 0$.  Writing Eq. 57 in terms of the primed coordinates, we find
\be
U_1'^2  +  U_2'^2  =  1  
\ee
which in the original system is
\be
U_1^2 + (U_2 \cos{\nu} + U_4 \sin{\nu} )^2  = 1  .
\ee
This is the equation of a great hypercircle $(\nu, 0)$
centered at the origin and lying in a hyperplane making an angle $\pi/2 - \nu$ with the 4-axis and an angle $\pi/2 - 0$ with the
3-axis. If $\bm{L}$ did not lie along the 3-axis but, for
example, was in the 1-3 plane, at an angle $\Theta$ from the
3-axis, then Eq. 61 would be modified by the substitution
\be
U_1 \longrightarrow U_1 cos{\Theta} + U_3 \sin{\Theta}
\ee
which follows since $U_i$ transforms as a three-vector. The corresponding great circle $(\nu, \Theta)$ lies in a hyperplane making an angle $\pi/2 - \nu$ with the 4-axis and $\pi/2 - \Theta$ with the 3-axis.

The motion of the orbiting particle corresponds to a dot moving along the great circle $(\nu , 0$ or $\Theta)$
with a period T given by the classical period Eq. 37.  The velocity in configuration space can be expressed in terms of $U_4$
by using its definition in terms of $p^2$ Eq. 53 or in terms of $\theta_4$ Eq. 55.  The particle is moving at maximum velocity when $\theta_4$ is a minimum, which occurs at the perihelion when $\theta _4 = \pi/2 - \nu$: 
\be
max(\frac{p}{a}) = \sqrt{\frac{1 + e}{1 - e}} = \sqrt {\frac{r_2}{r_1}}
\ee
and at a minimum velocity when $\theta_4 = \pi/2 + \nu$:
\be
min(\frac{p}{a}) = \sqrt{\frac{1 - e}{1 + e}} .
\ee
These values of $\theta_4$ correspond to turning points at which  $r$ and $p$ have extreme values.  This is apparent when we use Eq. 32 for the total energy to show
\be
U_4 = \frac{r_c - r}{r_c}  .
\ee
When $r>r_c$ then $p^2<a^2$, so the particle is moving more slowly than the RMS velocity.  Applying the virial theorem to any orbit we find ${\la p^2 \ra}= a^2$ so as expected $a$ is the RMS momentum and $\la \frac{1}{1 - U_4} \ra = 1 = \la r_c/r \ra$.

Fig. 7 is a picture of a simple device illustrating
the stereographic projection of the orbit in p/a-space onto the four-dimensional hypersphere in U-space. We assume that the orbit is in the 1-2 plane and that $\bm{A}$ lies along
the 1-axis so $p_3 = 0$, $U_3 = 0$. Because of this trivial
dependence on $p_3$ we have omitted the 3-axis. The
vertical pin or rod represents the unit vector $\hat{n}$ lying along the 4-axis. The circumference of the larger circle perpendicular to the 4-axis represents the orbit in $p/a$-space.
One can see it is displaced from the origin along the
2-axis. Centered at the origin we must imagine a hypersphere of unit radius $\hat{U}^2 = 1$. The stereographic projection $\hat{U}$ of the vector $\bm{p}/a$ is obtained by placing the
string coming from the top of n directly at the head of
the vector $\bm{p}/a$. The intersection of the string with the
unit hypersphere defines $\hat{U}$. As the string is moved along
the orbit in $p/a$-space, it intersects the hypersphere
along a great circle shown by the circumference of the unit
circle making an angle $\nu$ with the 1-2 plane. We can see
for example, that at the closest approach, $\theta_4$ is a minimum and U4 is a
maximum, $\hat{U} \cdot \bm{A} = U_1$ is a minimum, and $p/a$ is a maximum. 
 \begin{figure}[ht] 
\centering
 \includegraphics[scale=0.45]{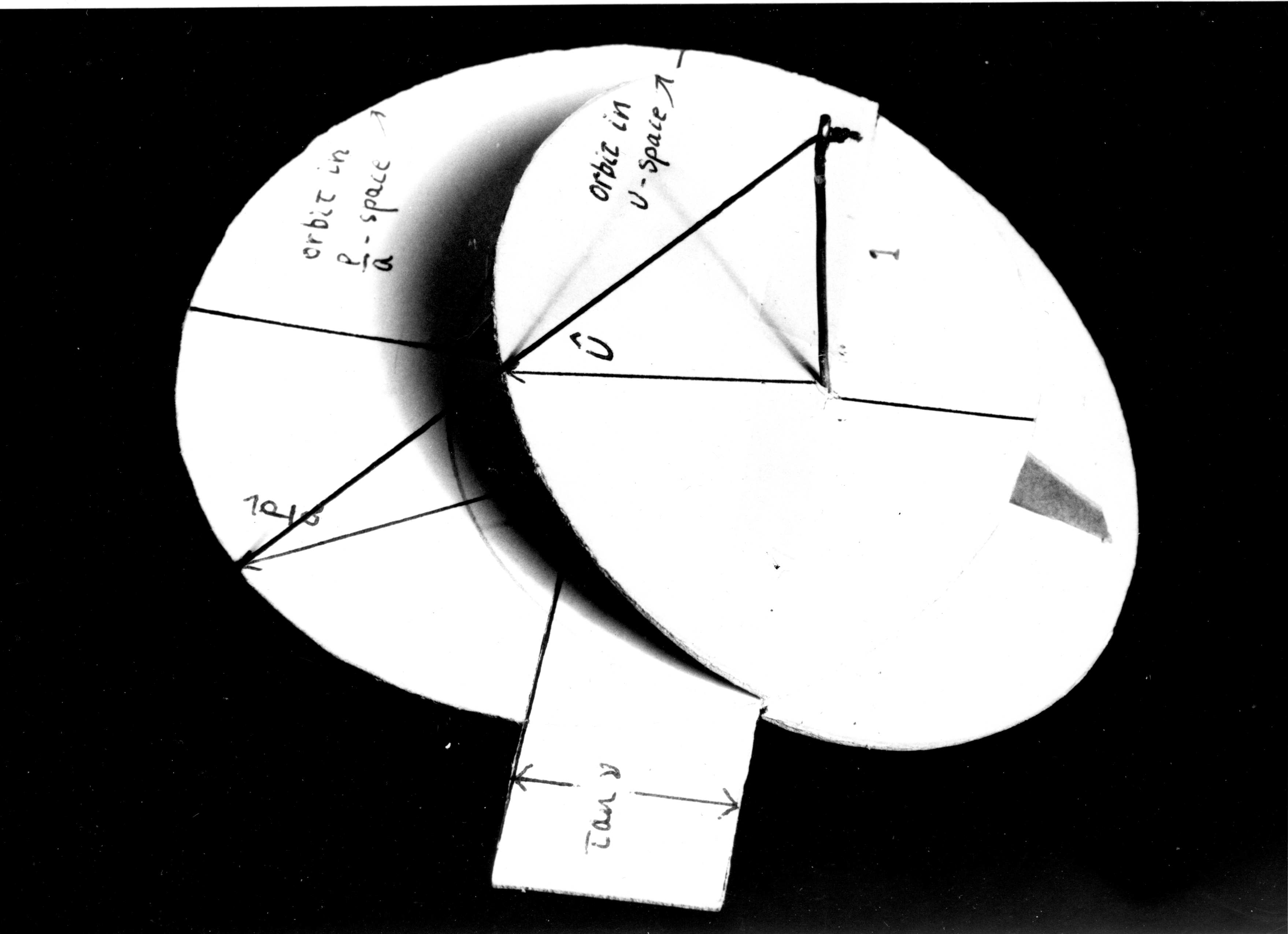}
 \caption{{Model} illustrating the stereographic projection from the 1-2 plane to a 4-d hypersphere. The pin represents the unit vector $\hat{n}$ along the 4-axis, normal to the 1-2 plane.}
 \label{Fig7}
\end{figure}

\subsection{Classical Time Dependence of Orbital Motion}
We can determine the time dependence of the orbital motion by integrating the expression for the angular momentum $L = mr^2 d\phi_r/dt$
\be
\int {dt}=\int \frac{{mr}^{2}}{{L}}{d} \phi_{r}
\ee
where r is given by the orbit equation Eq. 33 and we are assuming the orbit is in the $1-2$ plane. After integrating, we can
use the equations relating the momentum space and
configuration space variables to obtain the time dependence
in p-space and U-space.
We obtain
\be
\frac{1}{r_{c}^2} \frac{1}{\cos \nu} \frac{L}{m} \int_0^t d t=\cos ^{3}\nu \int_0^{\phi_r(t)} \frac{d \phi_r}{\left(1+\sin \nu \cos \phi_{r}\right)^{2}}  .
\ee
 
The left-hand side of this equation is equal $\omega_{cl}t$, where $\omega_{cl}$ is the classical frequency. This follows by substituting Eqs. 30 and 38    
\be
L=a r_c \cos{\nu} \qquad \omega_{cl}=\frac{a}{m r_c}  .
\ee
The integral on the right side gives\cite{petit}
\be
\omega_{cl} t=-\frac{\sin {\nu}\cos{\nu}\sin{\phi_{r}}}{1+\sin v \cos \phi_{r}}-\tan ^{-1} \frac{\cos{\nu} \sin{\phi _r}}{\sin{\nu}+\cos{\phi_r}}
\ee
which may be simplified as
\be
\omega_{cl}t=-\frac{A}{L} \frac{y}{r_{c}}-\tan ^{-1} \frac{A}{L} \frac{y}{r_{c}-r}
\ee
where  $ y=r \sin{\phi_r}$  .  

The relationship between the angle $\phi\equiv \phi_p$ in momentum
space and $\phi_r$ in configuration space follows by either
differentiating the orbit equation Eq. 33 with respect to time and using $L=mr^2\dot{\phi_r}$ or by solving simultaneously the configuration space orbit, the momentum space orbit equation Eq. 48 and the energy equation Eq. 32. We find
\be
\sin \phi_{r}=-p \cos \phi \dfrac{\cos{\nu}}{a} \hspace{15pt} \cos \phi_{r}=p \sin{\phi} \dfrac{\cos v}{a}-\sin{\nu} .
\ee
From these equations, the definitions of the $U_i$, Eq. 52, and the orbit and energy equations, it follows that
for the classical orbit in the $1-2$ plane
\be
\begin{array}{l}U_{1}=-\dfrac{r \sin{\phi_r}}{r_{c} \cos{\nu}}\vspace{5pt} \\  U_{2}=\dfrac{r}{r_{c}} \dfrac{\sin{\nu}+\cos{\phi_r}}{\cos v}=U_{4} \cot \nu \vspace{4pt} \\  U_{3}=0 \vspace{3pt} \\ U_4=\dfrac{r_c - r}{r_c} .\end{array}
\ee
Using these results in Eq. 69 gives
\be
\omega_{cl}t=U_{1} \sin\nu+\tan ^{-1}\left(\frac{U_{2}}{U_{1} \cos \nu}\right)
\ee
which gives the time dependence in U space, and agrees with the results of \cite{doth2}\cite{doth2time}.  We could do rotations to generalize this result\cite{cordinate}.  We can also rewrite the inverse tangent as $\cos^{-1}{U_1}$ using 
\be
U_1^2 + \frac{U_2^2}{cos^2\nu} = 1  .
\ee
If we consider the last two equation for circular orbits with $e=\sin\nu=0$ we obtain $\tan^{-1}(U_2/U_1)=\phi(t) = \omega_{cl}t$  $=\phi_r(t)+\pi/2$,  and our familiar circle $U_1^2 + U_2^2 = 1$ .

\subsubsection{Remark on Harmonic Oscillator}  We can find a conserved Runge-Lenz vector for the non-relativistic hydrogen atom because the elliptical orbit does not precess, as it does for the relativistic atom.  The only central force laws which yield classical elliptical orbits that do not precess are the inverse Kepler force and the linear harmonic oscillator force\cite{brownNon}. Thus it seems reasonable that one could construct a constant vector similar to $\bm{A}$ for the oscillator, although the force center for the atom is at a focus and for the oscillator it is at the center of the ellipse.  However,it is not readily possible\cite{bacry}.  Instead one can construct a constant Hermitian second rank tensor $T_{ij}$:
\be
T_{ij} = \frac{1}{m\omega_0}p_ip_j + m \omega_0 x_i x_j  .
\ee
This constant tensor is analogous to the moment of inertia
tensor for rigid body motion. The eigenvectors of the
tensor will be constant vectors along the principal axes for the particular orbit being considered. The existence of the conserved tensor leads to the U(3) symmetry algebra of the oscillator. The generators are $\lambda^a_{ij} T_{ij}$ where the $\lambda$'s are are the usual $U(3)$ matrices\cite{neem}. The spectrum generating algebra is SU(3,1).

In another approach, the Schrodinger equation for the hydrogen atom has been transformed into an equation for a four dimensional harmonic oscillator or two two dimensional harmonic oscillators.  This approach which fits well with parabolic coordinates was used especially in the 1980's to analyze the group structure of the atom and relate it to SU(3)\cite{barutSHO} \cite{boit}\cite{hugh}\cite{chen1}\cite{chen2} \cite{kibl2}\cite{chen3} \cite{gerr} \cite{chen3a}\cite{meer}.  We will not discuss this approach further.

\section{The Hydrogenlike Atom in Quantum Mechanics; Eigenstates of the Inverse of the Coupling Constant}
In this section we switch from classical dynamics to quantum mechanics and discuss the group structure and exploit it to determine the bound state energy spectrum directly, as Pauli and followers did almost a century ago \cite{pauli} \cite{hult}. In Section 4.3 we introduce a new set of basis states for the hydrogenlike atom, eigenstates of the coupling constant. Using these states allows us to display the symmetries in the most convenient manner and to treat bound and scattering states uniformly. 
\subsection{The Degeneracy Group SO(4)}
The quantum mechanical Hamiltonian is \be
H=\frac{p^{2}}{2 m}-\frac{Z \alpha}{r}=E   .
\ee
The classical expression for the Runge-Lenz vector needs to be symmetrized to insure the corresponding quantum mechanical operator $\bm{A}$ is Hermitian:
\be
\bm{A}=\frac{1}{\sqrt{-2 mH}}\left(\frac{\bm{p} \times \bm{L}-\bm{L} \times \bm{p}}{2}-m Z\alpha \frac{\bm{r}}{r}\right)  .
\ee
We may verify that $\bm{A}$ and $\bm{L} = \bm{r} \times \bm{p}$ both commute with
the Hamiltonian $H$. The commutation relations of $L_i$
and $A_i$ are the same as the corresponding classical
Poisson bracket relations for bound states:
\be
[L_i , L_j ] = i \epsilon_{ijk} L_k\hspace{3pt}, \hspace{10pt} [L_i, A_j] = i\epsilon_{ijk} A_k \hspace{3pt}, \hspace{10pt} [A_i, A_j]= i \epsilon_{ijk} L_k\hspace{4pt}
\ee
and form the algebra of O(4)\cite{band1}.
We can write the commutation relations in a single equation
which makes the 0(4) symmetry explicit. If we define
\be
S_{ij}= \epsilon _{ijk} L_k \hspace{17pt} S_{i4} = A_i    
\ee
then
\be
[S_{ab} , S_{cd}] = i(\delta_{ac} S_{bd} +\delta_{bd}S_{ac} - \delta_{ad}S_{bc} - \delta_{bc} S_{ad})\hspace{11pt} a,b = 1,2,3,4 .
\ee
The Kronecker delta function $\delta_{ab}$ acts like a metric tensor.
\subsection{Derivation of the Energy Levels}
We can obtain the energy levels by determining which representations of the group SO(4) are realized by the degenerate eigenstates of the hydrogenlike atom\cite{barg}\cite{pauli}\cite{band1}.  The representations of SO(4) may be characterized by the numerical values of the two Casimir operators for SO(4):
\be
C_1 = \bm{L}\cdot\bm{A}    \hspace{15pt}C_2 = \bm{L}^2 + \bm{A}^2
\ee
 Once we know the value of $C_2$, then the eigenvalues of H follow from the quantum mechanical form of Eq. 31, namely
\be
\bm{L}^2 +\bm{A}^2 +1 = \frac{(mZ\alpha)^2}{-2mH} .
\ee
In order to determine the possible values of $C_2$ we
factor the 0(4) algebra into two disjoint SU(2)
algebras\cite{bacr1}, each of which has the same commutation relations as the ordinary angular momentum operators, 
\be
\bm{N}=\textstyle\frac{1}{2}(\bm{L} + \bm{A})\hspace{18pt}
\bm{M}=\frac{1}{2}(\bm{L} - \bm{A})  .
\ee
The commutation relations are
\be
[M_i, N_j] = 0 \hspace{12pt}[M_i, M_j] = i \epsilon_{ijk} M_k \hspace{12pt} [N_i, N_j] = i \epsilon_{ijk} N_k   .
\ee
In analogy with the results for the ordinary angular momentum operators, the Casimir operators are
\be
\begin{array}{l}{M}^{2}={j}_{1}\left({j}_{2}+{1}\right), \quad{j}_{1}=0, \frac{1}{2}, 1, \ldots \\ {N}^{2}=j_{2}\left({j}_{2}+1\right), \quad {j}_{2}=0, \frac{1}{2}, 1, \ldots\end{array}
\ee
The numbers $j_1$ and $j_2$, which may have half-integral values for SU(2) but not O(3), define the $(j_1, j_2)$ representation of SO(4). From the definitions of A and
L in terms of the canonical variables it follows that $C_1 = \bm{L}\cdot\bm{A} = 0$ which means $j_1=j_2=j$ as in the classical case. For our representations we find
\be
\bm{M}^2 = \bm{N}^2 =\textstyle \frac{1}{4}(\bm{L}^2 + \bm{A}^2) = j(j+1),   j=0, \frac{1}{2}, 1, ..
\ee
and therefore
\be
\bm{L}^2 + \bm{A}^2 + 1 = (2j + 1)^2 .
\ee
Substituting this result into Eq. 82 gives the usual formula for the bound state energy levels of the hydrogen atom:
\be
H^{\prime}=-\frac{m(Z \alpha)^{2}}{2 n^{2}}=E_{n}
\ee
where the principal quantum number n = 2j+1 = 1,2,.. and the prime on $H$ signifies an eigenvalue of the operator $H$.

Within a subspace of energy $E_n$, the Runge-Lenz vector is
\be
\bm{A}=\frac{1}{a_n}\left(\frac{\bm{p} \times \bm{L}-\bm{L} \times \bm{p}}{2}-m Z\alpha \frac{\bm{r}}{r}\right)  .
\ee
where $a_n=\sqrt{-2mE_n}=\frac{mZ\alpha}{n}$.

Our considerations of the Casimir operators have
shown that the hydrogen atom provides completely symmetrical tensor representations of SO(4), namely, $(j, j) = (\textstyle\frac{n-1}{2},\frac{n-1}{2} ),\hspace{10pt} n= 1,2,\dots$ The dimensionality is $n^2$ , corresponding to the $n^2$
degenerate states. The appearance of only symmetrical
tensor representations $(j_1=j_2)$ may be traced to $\bm{L}\cdot\bm{A}$ vanishing,
which is a consequence of the structure of $\bm{L}$ and $\bm{A}$ in
terms of the dynamical variables for the hydrogenlike
atom. For systems other than
the hydrogenlike atom it is not generally possible to
find the expression for the energy levels in terms of all
the different quantum numbers alone. It worked here
since we could express the Hamiltonian as a function of
the Casimir operators which contained all quantum numbers
explicitly.

There are a variety of possible basis states.  We could choose basis states for the SO(4) representation that reflect the SU(2) decomposition, namely eigenstates of $\bm{M}^2$, $\bm{N}^2$, $M_3$ and $N_3$ \cite{bacr1}.   Another possibility is to have a basis with eigenstates of the Casimir operator $C_2$, and $A_3$, and $M_3$. This choice fits well with the use of parabolic coordinates\cite{baru2}. A more physically understandable choice  is to choose the common basis states $|n l m\ra$ that are eigenstates of $C_2$,  $\bm{L}^2$, and $L_3$.
For this set of basis states, we have
\be
\sqrt{(\bm{L}^2+\bm{A}^2+1)}|nlm\ra=n|nlm\ra \hspace{15pt}\bm{L}^2|nlm\ra=l(l+1)|nlm\ra \hspace{10pt}L_3|nlm\ra = m|nlm\ra
\ee
We can define raising and lowering operators for $m$:
\be
L_{\pm}=L_{1}\pm i L_2
\ee
which obeys the commutation relations
\be
[\bm{L}^2,L_{\pm}]=0]\hspace{20pt}[L_3, L_{\pm}]=\pm L_{\pm}   .
\ee
Therefore we can use $L_{\pm}$ to change the value of $m$ for the basis states
\be
L_{\pm} |nlm \ra = \sqrt{(l(l+1)-m(m\pm 1)}|n l\hspace{3pt} m\pm 1 \ra
\ee
for $l \ge 1$.
We can also use the generators $\bm{A}$ to change the angular momentum. A general SO(4)  transformation can be expressed as a rotation induced by $\bm{L}$, followed by a rotation induced by $A_3$, followed by another rotation generated by $\bm{L}$ \cite{bied2}. Our interest is primarily in changing the angular momentum $l$,  which is most directly done using $A_3$, which commutes with $L_3$ and $C_2$, and so only changes $l$:
\be
A_3|n l m\ra=\left(\frac{(n^2 - (l+1)^2)((l+1)^2-m^2)}{4(l+1)^2-1}\right)^{\frac{1}{2}} |n\hspace{2pt} l+1 \hspace{2pt}m\ra+\left(\frac{(n^2 -l^2)(l^2-m^2)}{4l^2 -1}\right)^{\frac{1}{2}}|n \hspace{3pt}l-1\hspace{3pt}m\ra   .
\ee
for $l\ge 1$. 

\subsection{Relativistic and Semi-relativistic Spinless Particles
in the Coulomb Potential and Klein-Gordon Equation}

The Klein-Gordon equation
$$
\left(p^{2}-(\tilde{E}-V)^{2}+m^{2}\right) \tilde{\psi}=0
$$
where $\tilde{E}$ is the relativistic total energy, may be solved
exactly for a Coulomb potential, $V=-(Z\alpha)/r $ \cite{shif}.  The energy levels depend on a principal quantum number and on the magnitude of the angular momentum but not its direction. The only degeneracy present is associated with the O(3) symmetry of the Hamilton.  For a relativistic scalar particle there is no degeneracy to be lifted by a Lamb shift.

If we neglect the $V^2$ term the resulting equation can be written in the form 
$$
\left(\frac{{p}^{2}}{2 \tilde{{E}}}+{V}-\frac{\tilde{{E}}^{2}-{m}^{2}}{2 \tilde{{E}}}\right) \tilde{\psi}=0
$$
This is exactly the same as the nonrelativistic Schrodinger equation with the substitutions
$$
m\rightarrow \tilde{E} \hspace{15pt}{E}\rightarrow \frac{\tilde{E}^2-m^2}{2\tilde{E}}  .
$$
Thus we regain the O(4) symmetry of the nonrelativistic hydrogen atom, and can define two conserved vectors, as indicated in Table 1. It is possible to take the "square root" of this approximate Klein-Gordon equation (in the same sense that the Dirac equation is the square root of the Klein-Gordon equation) and get an approximate Dirac equation whose energy eigenvalues are independent of the orbital angular momentum \cite{biden}.

\subsection[Eigenstates of the Inverse Coupling Constant Z alpha -1]{Eigenstates of the Inverse Coupling Constant $(Z\alpha)^{-1}$}
Solutions to Schrodinger's equation for a particle of energy $E=-\frac{a^2}{2m}$
in a Coulomb potential 
\be
\left[p^{2}+a^{2}-\frac{2 m Z \alpha}{r}\right]|a\rangle=0
\ee
may be found for certain critical values of the energy $E_n= -\frac{a_n^2}{2m}$ where $a_n = \frac{mZ\alpha}{n}$.  The corresponding eigenstates of the Hamiltonian are $|n \textit{l} m\ra$
which satisfy Eq. 95 with $a$ replaced by $a_n$.
In addition
to the bound states, because there is no upper bound on $p^2$ in the Hamiltonian, we also have the continuum of scattering states that have E>0.

Since the quantity which must have discrete values for a solution to exist is actually $\frac{a}{mZ\alpha}$, as noted in Section 2.3, we might ask if eigensolutions to Eq. 95 exist for certain critical values of $Z\alpha$ while
keeping  $a$  and the energy fixed\cite{brow}. To investigate such solutions it is convenient to algebraically transform Eq. 95:
$$
\left[\frac{1}{\sqrt{\rho(a)}}(\frac{p^2+a^2}{a})\frac{1}{\sqrt{\rho(a)}}-\frac{1}{\sqrt{\rho(a)}}(\frac{2mZ\alpha}{ar})\frac{1}{\sqrt{\rho(a)}}\right]\sqrt{\rho(a)}| a\ra=0
$$ where
\be
\rho(a)=\frac{p^2+a^2}{2a^2}.
\ee 
Since $\rho(a)$ commutes with $p^2$ we obtain the eigenvalue equation
\be
\left[\left(\frac{a}{m Z \alpha}\right)-K(a)\right] | a) = 0
\ee
where the totally symmetric and real kernel is
\be
K(a)=\sqrt{\frac{2 a^{2}}{p^{2}+a^{2}}} \frac{1}{a r} \sqrt{\frac{2 a^{2}}{p^{2}+a^{2}}}
\ee
and
\be
| a)=(\rho(a))^{\frac{1}{2}}|a\rangle.
\ee
As before, solutions to this transformed equation may found for the eigenvalues
\be
K'(a) = (\frac{a}{mZ\alpha})' = \frac{1}{n}  .
\ee
If we hold $Z\alpha$ constant and let  $a$  vary, we obtain the
usual spectrum $a_n =\sqrt{-2mE_n}= \frac{mZ\alpha}{n}$.  \hspace{10pt}For  $a = a_n$,  Eq. 99 reduces
to the equation for the eigenstates
\be
\sqrt{\rho(a_n)}|n l m\ra = |n l m; a_n)  .
\ee
Alternatively, if we hold $a$ constant, then $Z\alpha$  has the spectrum
\be
(Z\alpha)_n = \frac{na}{m}
\ee
with the corresponding eigenstates of $(Z\alpha)^{-1}$ being 
\be
|n l m; a )   .
\ee
The relationship between the usual energy eigenstates of the
energy $|n l m\ra $ and the eigenstates $| n l m; a)$ of $(Z\alpha)^{-1}$  is  
\be
|n \textit{l} m\ra = \frac{1}{\sqrt{\rho(a_n)}}|n \textit{l}m ; a_n)  ,
\ee
which requires that both sets of states have the same quantum numbers.  Note that the magnitude $(a|K(a)|a)$ is proportional to $\la 1/ar \ra_n=(1/a_n)(1/n^2a_0)$ where $a_0$ is the Bohr radius for the ground state, and so is positive and bounded. The kernel K(a) is real and symmetric in $p$ and $r$ and manifestly Hermitean. Since the kernel K in Eq. 98 is bounded, definite, and Hermitian with respect to the eigenstates $|n \textit{l}m ; a_n)$ \cite{mandf}  the set of normalized eigenstates
\be
|n l m; a) \hspace{12pt} n= 0, 1, 2....;\hspace{12pt} l = 0, 1,..n-2, n -1;\hspace{12pt} m= -l, -l+1, ...l-1, l.
\ee
where
\be
\left( \frac{1}{n} -K(a) \right)\hspace{5pt} |nlm ; a) =0
\ee
is a complete orthonormal basis for the hydrogenlike atom:
\be
\left(n l m  ;  a  | n^{\prime}  l^{\prime} m^{\prime} ; a\right)=\delta_{n n`} \delta_{l l`} \delta_{m m`}
\ee
\be
\sum_{nlm} |n l m ; a)(n l m ; a| = 1   .
\ee
There are several important points to notice with regard
to these eigenstates of the inverse of the coupling constant:\vspace{12
pt}

(1) \textbf{Because of the boundedness of K, there is no continuum portion in the eigenvalue spectrum of $(Z\alpha)^{-1}$, the eigenvalues are discrete}. Since K is a positive definite Hermitian 
operator, all eigenvalues are positive, real numbers. This feature leads to a unified treatment of all states of the hydrogenlike atom as opposed to the treatment in terms of energy
eigenstates in which we must consider separately the bound states and the continuum of scattering states.\vspace{12pt}

2) It follows from Eq. 104 that \textbf{the quantum numbers, multiplicities, and degeneracies of these states $|nlm;a)$ are precisely the same as
those of the usual bound energy eigenstates}. For example,
there are $n^2$ eigenstates of $(Z\alpha)^{-1}$ 
which have the
principal quantum number equal to $n$ or $(Z\alpha)$ equal to $\frac{na}{m}$. \vspace{12pt}

3) \textbf{A single value of the RMS momentum $a$ or the energy $E=\frac{-a^2}{2m}$ applies to all the states in our complete
basis}, as opposed to the usual energy eigenstates where
each nondegenerate state has a different value of $a$. We
have made this explicit by including $a$ in the notation
for the states:	$| nlm; a)$. Sometimes we will write the
states as $|nlm)$, provided the value of $a$ has been
specified. This behavior in which a single value of $a$
applies to all states will prove to be very useful. In essence it permits us to generalize from statements applicable in a subspace of Hilbert space with energy $E_n$ or energy parameter $a_n$ to the entire Hilbert space.\vspace{7pt}

(4)	\textbf{By a suitable scale change or dilation we may give the quantity $a$ any positive value we desire}. This is
effected by the unitary operator
\be
D(\lambda)=e^{i \frac{1}{2}(\bm{p} \cdot \bm{r}+\bm{r} \cdot \bm{p}) \lambda}
\ee
which transforms the canonical variables 
\be
D(\lambda) \bm{p} D^{-1}(\lambda)=e^{-\lambda} \bm{p} \hspace{13pt} D(\lambda) \bm{r} D^{-1}(\lambda)=e^{\lambda} \bm{r}
\hspace{13pt}
\ee
and the kernel K(a)
\be
D({\lambda})K(a)D^{-1}(\lambda)=K(a e^{\lambda})  .
\ee
and the eigenvalue equation
\be
\left(\frac{1}{n} - K(ae^{\lambda})\right)D(\lambda)|nlm; a) = 0.
\ee
Therefore the states transform as
\be
D(\lambda)|n l m; a) = |n l m; ae^{\lambda})  .
\ee
These transformed states form a new basis corresponding to the new value $e^{\lambda}a$ of the RMS momentum.

The relationship between the energy eigenstates and the $(Z\alpha)^{-1}$ eigenstates can be written using the dilation operator:
\be
|nl m\rangle=\frac{1}{\sqrt{\rho\left(a_{n}\right)}} D\left(\lambda_{n}\right) | n \ell m ; a)\hspace{13pt}
\text{where} \hspace{5pt}
e^{\lambda_n} = \frac{a_n}{a}  .
\ee
The usual energy eigenstates $|nlm \ra$ are obtained from the eigenstates of $(Z\alpha)^{-1}$ by first performing a scale change to insure
that the energy parameter $a$ has the value $a_n$ and then
multiplying by a factor $\rho^{-1/2}$.	The need for the 
scale change is apparent from dimensional considerations:
from the $(Z\alpha)^{-1}$ eigenvalue equation we see that the
eigenfunctions are functions of $p/a$ or $ar$ while from the energy eigenvalue equation the eigenfunctions are functions of $p/a_n$ and $a_nr$.
The factor $\rho^{-1/2}$ was required in order to convert
Schrodinger's equation to one involving a bounded Hermitean
operator.

Using the eigenstates of $(Z\alpha)^{-1}$ as our basis allows us to analyze the mathematical and physical structure of the hydrogenlike atom in the easiest and clearest way. 

\subsection[Another set of Eigenstates of Z alpha -1]{Another Set of Eigenstates of $(Z\alpha)^{-1}$}
We may transform Schrodinger's equation Eq. 95 to an eigenvalue equation for $(Z\alpha)^{-1}$ that differs from Eq. 96 by similar methods:
\be
\left( \frac{1}{n} -K_1(a)\right )\hspace{5pt} |nlm ; a) =0
\ee
where
\be
K_{1}(a)=\frac{1}{\sqrt{a r}} \frac{2 a^{2}}{p^{2}+a^{2}} \frac{1}{\sqrt{a r}}\hspace{25pt}\rho(a)=n/ar \hspace{13pt}\sqrt{\rho(a_n)}|n l m \ra = |n l m; a_n) 
\ee
This kernel, like K(a), is a bounded, positive definite, Hermitian operator so the eigenstates form a complete basis.  The relationship of these basis states to the energy eigenstates is the same as that of the previously discussed eigenstates of $(Z\alpha)^{-1}$ Eq. 114 but with $\rho(a) = n/ar$. The $n$ insures that the two sets of eigenstates have consistent normalization, which may be checked by means of the virial theorem.  The $n$ cancels out when similarity transforming from the basis of energy eigenstates to the basis of $(Z\alpha)^{-1}$. Note that classically, both kernels equal $1/ar_c$.

The first set of basis states of $(Z\alpha)^{-1}$ with $\rho(a)=\frac{p^2 + a^2}{2a^2}$ is more convenient to use when working in momentum space and the second set with $\rho(a)=n/ar$ is more convenient in configuration space.

Other researchers have used other approaches to secure a bounded kernel for the Schrodinger hydrogen atom, for example, by multiplying the equation from the left by $r$ to regularize it\cite{wulf2}. However, the methods used have not symmetrized the kernels to make them Hermitian, nor are all the generators of the corresponding groups Hermitian, and they have to redefine the inner product \cite{frons2}\cite{wulf2}.

\subsection[Transformation of A and L to the New Basis States]{Transformation of $\textbf{A}$ and $\textbf{L}$ to the New Basis States}
We must transform $\bm{A}$ as given in Eq. 89 and $\bm{L}=\bm{r}\times\bm{p}$ when we change our basis states from eigenstates
of the energy to eigenstates of the inverse coupling constant. The correct transformation may be derived by requiring that the transformed generators produce the same linear
combination of new states as the original generators produced of the old states. Thus since
\be
\bm{A}|nlm\rangle = \sum_{l',m'} |n l' m' \rangle A_{l'm'}^{lm}
\ee
where the coefficients $A_{l'm'}^{lm}$
are the matrix elements of $\bm{A}$,
we require that the transformed generator $\bm{a}$ satisfies the equation
\be
\bm{a}|nlm) = \sum_{l'm'} |nl'm') A_{l'm'}^{lm}  .
\ee
In other words, since the Runge-Lenz vector $\bm{A}$ is a symmetry operator of the original energy eigenstates, $\bm{a}$ will be a symmetry operator of the new states with precisely the same properties and matrix elements. Since $\bm{A}$ is Hermitian, $\bm{a}$ is Hermitian.

To obtain a differential expression for $\bm{a}$ acting on the new states we need to transform the generator using Eq. 114: \be 
\bm{a} = D^{-1}(\lambda_n)\left(\sqrt{\rho(a_n)}\bm{A}\frac{1}{\sqrt{\rho(a_n)}} \right)D(\lambda_n)  .
\ee
The effect of the scale change on the quantity in large parenthesis is to replace $a_n$ everywhere by $a$. By explicit calculation we find 
\be
\bm{a}=\frac{1}{2 a} \left(\frac{\bm{r} p^{2}+p^{2} \bm{r}}{2}-\bm{r} \cdot \bm{p} \bm{p}-\bm{p} \bm{p} \cdot \bm{r}\right)-\frac{a \bm{r}}{2}
\ee
for $\rho(a)= (p^2 + a^2)/2a^2 $. And we obtain
\be
\bm{a}=\frac{1}{2 a}\left(\frac{{\bm{r}} p^{2}+p^{2} \bm{r}}{2}-\bm{r} \cdot \bm{p} \bm{p}-\bm{p} \bm{p} \cdot \bm{r}-\frac{\bm{r}}{4 r^{2}}\right)-\frac{a \bm{r}}{2}
\ee
for $\rho(a)=n/ar $.

Both of these expressions for $\bm{a}$ are manifestly Hermitian.  In addition since there is no dependence on the principal quantum number these expressions are valid in the entire
Hilbert space, and not just in a subspace spanned by the degenerate states, as was the case when we used the energy
eigenstates as a basis (Eq. 89).

The angular momentum operator is invariant under scale changes and it commutes with scalar operators.
Therefore $\bm{L}$ is invariant under the similarity transformation  $\frac{1}{\sqrt{\rho(a_n)}}D(\lambda_n)$ and the expression for the angular
momentum operator with respect to the eigenstates of
$(Z\alpha)^{-1}$ is the same as the expression with respect to the
eigenstates of the energy.

\subsection[The <U'| Representation]{The $\la U'|$ Representation}
The $U'$ coordinates provide the natural representation for the investigation of the symmetries of the
hydrogenlike atom in quantum mechanics, as in classical
mechanics\cite{prime}.  Therefore we briefly consider the relevant
features of this representation and, in particular, its
relationship to the momentum representation.
The eigenstate $<U'|$ of $U_b, b =$ 1,2,3,4,  is defined
by
\be
\la U'|\hspace{3pt} U_b = U_{b}'\hspace{3pt}\la U'|
\ee
These states are complete on the unit hypersphere in four
dimensions:
\be
\int | U'\ra \la U' |\hspace{3pt} d^3\Omega ' = 1 
\ee
where $\Omega$ refers to the angles $(\theta_4, \theta, \phi )$ defined in Eq. 54.  The $U$ variables are defined in terms of the momentum variables and the quantity $a$ in Eq. 52. Therefore the momentum and the U operators commute
\be
[p_i, U_b] = 0
\ee
and the state $\la U'|$ is proportional to a momentum eigenstate $\la p'|$:
\be
\la U'| = \la p'|\hspace{3pt} \sqrt{ J(p)}
\ee
where the momentum eigenstate is defined by $\la p'|\bm{p} = \bm{p'} \la p|$ and
\be
\int d^3p' |p'\ra \la p'| =1  .
\ee 
The function $J(p')$ may be determined by equating the completeness conditions and substituting Eq. 125:   
\be
1= \int d^3p' |p'\ra \la p'|=\int d^3\Omega' |U'\ra \la U'| =    
 \int d^3 \Omega' J(p') |p'\ra \la p'|
\ee
which leads to the identification
of the differential quantities
\be
d^3p' = d^3 \Omega' J(p')
\ee
demonstrating that $J(p')$ is the Jacobian of the transformation from the p- to the U- space.  Noting that on the unit sphere
\be
U_4^2 = 1 - U_iU_i
\ee
we can compute the Jacobian
\be
J(p') = \left[\frac{p'^2 + a^2}{2a}\right]^3  .
\ee
Therefore from Eq. 125 we have the important result
\be
\la U'| = \la p'|\left[\frac{p^2 + a^2}{2a}\right]^{3/2}  .
\ee
We can use this result to compute the action of $\bm{r}$ on $\la U'|$ in terms of the differential operators.  Using the equation
\be
\la p' | \bm{r} = i \bm{\nabla}_{p'} \la p'|
\ee
we obtain
\be
\la U'| \bm{r} = \left(i \bm{\nabla} _{p'} -\frac{3i\bm{p'}}{p'^2 + a^2} \right) \la U'|   .
\ee

\subsubsection[Action of A and L on <U'|]{Action of $\textbf{a}$ and $\textbf{L}$ on $\la U'|$}
Using Eq. 133 for the action of $\bm{r}$ on $\la U'|$ and using the expression Eq 120 for $\bm{a}$, we immediately find that when acting on $\la U'|$, $\bm{a}$
has the differential representation
\be
\bm{a}`=\frac{i}{2 a}\left((p`^{2}-a^{2}) \bm{\nabla}_{p`}-2 \bm{p}` \bm{p}` \cdot \bm{\nabla}_{\bm{p}`}\right)
\ee
where
\be
\la U'|\bm{a} = \bm{a}' \la U' |   .
\ee
We can also write $\bm{a}'$ in terms of the $U'$ variables by using the relationship Eq. 52 between the $p$ and $U$ variables:
\be
\bm{a}' = U_4' i \bm{\nabla}_{U'} - \bm{U}' i \partial/\partial_4'
\ee
where the spatial part of the four vector $U'$ is $\bm{U}= (U_1, U_2, U_ 3)$ and $U_4$ is the fourth component.  This is the differential representation of a rotation operator mixing the spatial and the fourth components  of $U_a'$. When acting on the state $\la U'|$ , clearly $e^{i\bm{a}\cdot \bm{\nu}}$ generates a four-dimensional rotation that produces a new eigenstate $\la U"|$. To derive the form of the finite transformation explicitly we compute
\be
[a'_j, U'_j] = i U'_4 \delta_{ij}\hspace{14pt} [a'_j, U'_4] = - i U'_i   .
\ee
For a finite transformation $a^{i\bm{a}\cdot\bm{n}\nu}$ with $\bm{n}^2 = 1$, we have
\be
\begin{aligned} \bm{U}^{\prime \prime} &=e^{i \bm{a}` \cdot \bm{n}\nu} \bm{U}`e^{-i \bm{a}` \cdot \bm{n} \nu} \\ &=\bm{U}` -\bm{n} \bm{n} \cdot \bm{U}` +\bm{n} \bm{n} \cdot \bm{U}` \cos {\nu}-\bm{n} U_4` \sin{\nu} \end{aligned}
\ee
and
\be
\begin{aligned}U_4" &= e^{i \bm{a}` \cdot \bm{n}\nu} {U_4}`e^{-i \bm{a}` \cdot \bm{n} \nu}\\
&= U_4` \cos{\nu} + \bm{n}\cdot\bm{U}` \sin{ \nu}\end{aligned}  
\ee
These equations of transformation are like those for a
Lorentz transformation of a four-vector (r, it). We can illustrate the equations for $e^{ia_2\nu}$ (cf Eq. 58) which mixes the 2 and 4 components of $U'$:
\be
\begin{aligned}
U_1"&= U_1' \hspace{85pt}  U_3" = U_3'\\ 
U_2"&= U_2' \cos\nu - U_4' \sin \nu \hspace{18pt} U_4" = U_2' \sin \nu +U_4' \cos \nu \end{aligned} .
\ee
When $\bm{L}$ acts on $\la U'|$ it has the
differential representation
\be
\bm{L}' = \bm{U}'\times i \bm{\nabla}_{U'}
\ee
This result follows directly since $U_i'$ equals $p_i'$ times
a factor that is a scalar under rotations in three dimensions. When $e^{i\bm{L} \cdot\bm\omega}$ acts on $\la U'|$ it produces a new state $\la U"|$ , where the spatial components of $U'$ have
been rotated to produce $U"$.

In summary we see that $U'$ is a four-vector under
rotations generated by $\bm{a}'$ and $\bm{L}'$. Therefore the
states $\la U'|$ provide a vector representation of the group of rotations in four dimensions SO(4), with the generators $\bm{a}$ and $\bm{L}$.

\section{Wave Functions for the Hydrogenlike Atom}
In this section we analyze the wave functions of the hydrogenlike atom, working primarily in the $\la U'|$
representation and using eigenstates of the inverse of the coupling constant $(Z\alpha)^{_-1}$ for the basis states. In this representation
the wave functions are spherical harmonics in four dimensions. We derive the relationship of the usual
energy eigenfunctions in momentum space to the spherical harmonics and discuss the classical limits in momentum and configuration space.

\subsection{Transformation Properties of the Wave Functions under the Symmetry Operations}
We can show that the wave functions $Y_{nlm}(U')$ in
the $\la U'|$ representation with respect to the eigenstates of $(Z\alpha)^{-1}$
\be
Y_{nlm} (U') \equiv \la U'| nlm)
\ee
transform as four-dimensional spherical harmonics under
the four-dimensional rotations generated by the Runge-Lenz vector $\bm{a}$ and the angular momentum $\bm{L}$. We note
that the quantity $a$ is implicit in both the bra and the
ket in Eq. 142.  For our basis states we employ the set of $(Z\alpha)^{-1}$ eigenstates $|nlm)$ of the inverse coupling constant that are convenient for momentum space calculations ($\rho=\frac{p^2 + a^2}{2a^2}$).  We choose these states rather than those convenient for configuration space calculations because the $\la U'|$ eigenstates are proportional to the $\la p'|$ eigenstates.

If we transform our system by the unitary operator $e^{i\theta}$ where $\theta = \bm{L}\cdot \bm{\omega} + \bm{a}\cdot\bm{\nu}$, then the wave function in the new system is
\be
Y'_{nlm}(U') = \la U'| e^{i \theta}|nlm).  \ee
There are two ways in which we may interpret this transformation, corresponding to what have been called the
active and the passive interpretations. In the passive interpretation we let $e^{i \theta}$ act on the coordinate eigenstate $\la U'|$ .  As we have seen in Section 3.6, this
produces a new eigenstate $\la U"|$, where the four-vector $U"$ is obtained by a four-dimensional rotation of $U'$ (Eq. 138-139).
Thus we have
\be
Y_{nlm}'(U') = \la U" | nlm) = Y_{nlm}(U")  .
\ee
In the active interpretation we let $e^{i \theta}$ act on the
basis state $|nlm)$. Since $\bm{L}$ and $\bm{a}$ are symmetry
operators of the system, transforming degenerate states into each other, it follows that $e^{i \theta} |nlm )$ must be a linear combination of states with principal quantum number
equal to n. Therefore we have
\be
Y_{nlm}'(U') = \sum_{l'm'}\la U'|R_{nl'm'}^{nlm}|nl'm')\hspace{5pt}=\hspace{5pt}\sum_{l'm'}R_{nl'm'}^{nlm} Y_{nl'm'}(U')  .
\ee
The wave functions for degenerate states with a given $n$ transform
irreducibly among themselves under the four-dimensional rotations, forming a basis for an irreducible representation of SO(4) of dimensions $n^2$ . Equating the results of the two different interpretations gives
\be
Y_{nlm}(U") = \sum_{l'm'}R_{nl'm'}^{nlm} Y_{nl'm'}(U') .
\ee
The transformation properties Eq. 146 of $Y_{nlm}$ are precisely analogous to those of the three-dimensional spherical
harmonic functions. It follows that the $Y_{njm}$ are four-dimensional spherical harmonics\cite{band1}\cite{mandf}.

\subsection[Differential equation for the Four Dimensional Spherical Harmonics Ynlm(U')]{Differential Equation for the Four Dimensional Spherical Harmonics $Y_{nlm}(U')$}
The differential equation for the $Y_{nlm}(U')$ may be obtained from the equation
\be
(\bm{L'}^2 + \bm{a'}^2)Y_{nlm}(U') = (n^2 -1)Y_{nlm}(U')
\ee
which follows from $C_2=n^2 - 1$ and the definition of $C_2$, Eq. 81.  Substituting in the differential expressions Eq. 134 and 141 for $\bm{a'}$ and $\bm{L'}$ we find that $\bm{L'}^2 + \bm{a'}^2$  equals $\nabla^2_{U'} -(\bm {U'} \cdot \bm{\nabla}_{U'})^2$, which is the angular part of the Laplacian operator in four dimensions (cf in three dimensions, $\bm{L}^2/r^2 = p^2 - p_r^2 ).$ 
Thus Eq. 147 is the differential equation for four-dimensional spherical harmonics with the degree of homogeneity equal to  
$n-1$, which means $n^2$ such functions exist, in agreement with the know degree of degeneracy.

\subsection{Energy Eigenfunctions in Momentum Space}
We want to determine the relationship between the usual energy eigenfunctions in momentum space $\psi_{nlm}(p')\equiv \la p'|nlm\ra$ (with $a=a_n$)  and the four-dimensional spherical harmonic eigenfunctions $Y_{nlm}(U';a) = \la U' |nlm; a)$.

We choose the RMS momentum $a$ to have the value $a_n$.  If we use the expression Eq. 131 for $\la U'|$ in terms of $\la p'|$ 
\be
\la U'| = \la p'| \left(\frac {p^2 + a_n^2}{2a_n}\right)^{3/2}
\ee
and the expression Eq. 104 for the eigenstates of $(Z\alpha)^{-1}$ in terms of the energy eigenstates
\be
|nlm; a_n) = \sqrt{\frac{p^2 + a_n^2}{2a_n^2}}|nlm\ra
\ee
we find the desired result
\be
Y_{nlm}(U'; a_n) = \left(\frac{p^2 + a_n^2}{2a_n}\right)^2\frac{1}{\sqrt{a_n}}\psi _{nlm}(p') .
\ee
The usual method of deriving this relationship between the wave function in momentum space and the corresponding spherical harmonics in four dimensions involves transforming the Schrodinger wave equation to an
integral equation in momentum space\cite{fock}\cite{band1}. As in the classical case, we first replace $\bm{p}$ by $\bm{p}/a$ and perform a stereographic
projection from the hyperplane corresponding to the three-
dimensional momentum space to a unit hypersphere in a
four-dimensional space. The resulting integral equation
manifests a four-dimensional invariance. When the wave
functions are normalized as in Eq. 150, the solutions are
spherical harmonics in four dimensions. As another alternative
to this procedure, we can Fourier transform the configuration space wave functions directly \cite{morse2}.
\subsection {Explicit Form for the Spherical Harmonics}
The spherical harmonics in four dimensions can be expressed as\cite{bateman}:
\be
Y_{nlm}(\Omega) = N_1(n,l)(\sin \theta _4)^l C_{n -1 -l}^{l+1}(\cos \theta_4) \cdot\left[N_2(l,m)(\sin \theta)^m C_{l-m}^{m 
+ 1/2}(\cos \theta)\frac {e^{im \phi}}{\sqrt{2\pi}}\right]  .
\ee
The factor in brackets is equal to $Y_l^m(\theta, \phi)$, the usual spherical harmonic in three-dimensions\cite{bateman}. The Gegenbauer polynomials $C_n^\lambda$ of degree $n$ and order $\lambda$ are defined in terms of a generating function:
\be
\frac{1}{(1 - 2tx + t^2)^{\lambda}}=\sum_{n=0} t^n C_n^{\lambda} .
\ee
$N_1(n,l)$ and $N_2(l,m)$ are chosen to normalize the $Y_{nlm}$ on the surface of the unit sphere:
\be
\int|Y_{nlm}(\Omega)|^2 d^3\Omega_U = 1
\ee
where $d^3\Omega_U = \sin^2{\theta_4} \sin{\theta}d\theta d \phi$.  We find
\be
N_1(n,l)=\sqrt{\frac{2^{2l+1}}{\pi}\frac{n(n - l - 1)!(l!)^2}{(n+l)!}}
\ee
\be
N_2(l,M)=\sqrt{\frac{2^{2m}}{\pi}(l+\frac{1}{2})\frac{(l - m)!}{(l + m)!}[\Gamma (m + \frac{1}{2})]^2}    \hspace{5pt} .
\ee
In the next section we discuss the asymptotic behavior of $Y_{nlm}$ for large quantum numbers and compare it to the classical results of Section 3.

\subsection{Wave Functions in the Classical Limit}

\subsubsection{Rydberg Atoms}
Advances in quantum optics, such as the development of ultra short laser pulses, microwave spectroscopy, and atom interferometry, have opened new possibilities for experiments with atoms and Rydberg states, meaning hydrogenlike atoms in states with very large principal quantum numbers and correspondingly large diameter electron orbits. The pulsed electromagnetic fields can be used to modify the behavior of the orbital electrons. Semi-classical electron wave packets in hydrogenlike atoms were first generated in 1988 by ultrashort laser pulses, and today are often generated by unipolar terahertz pulses\cite{mako}\cite{bell}\cite{berry}. Over the last few decades there has been interest in the classical limit of the hydrogenlike atom for $n$ very large, Rydberg states, for a number of reasons\cite {eberH}: 1. Rydberg states are at the border between bound states and the continuum, and any process which leads to excited bound states, ions or free electrons usually leads to the production of Rydberg states. This includes, for example, photo-ionization or the application of microwave fields. The very large cross section for scattering is unique.  2. Rydberg states can be used to model atoms with a higher atomic number that have an excited valence electron that orbits beyond the core. 3. In Rydberg states, the application of electric and magnetic fields breaks the symmetry of the atom and allows the study of different phenomena, including the transition from classical to quantum chaos\cite{laksh}. 4. Rydberg atoms can be used to study coherent transient excitation and relaxation, for example the response to short laser pulses creating coherent quantum wave packets that behave like a classical particle.

The square of the wave function for a given quantum state gives a probability distribution for the electron that is independent of time. If we want to describe the movement of an electron in a semiclassical state, with a large radius, going around the nucleus with a classical time dependence, then we need to form a wave packet.  
 The wave packet is built as a superposition of many wave functions with a band of principal quantum numbers. 

 A variety of theoretical methods have been used to derive expressions for the hydrogen atom wave functions and wave packets for highly excited states. There is general agreement on the wave functions for large n, and that the wave functions display the expected classical behavior, elliptical orbits in configurations space, and great circles in the four dimensional momentum space\cite{kay}\cite{lena}\cite{bhau}\cite{mcan}.

 Researchers have proposed a variety of wave packets to describe Rydberg states. There are general similarities in the wave packets that describe electrons going in circular or elliptical orbits with a classical time dependence for some characteristic number of orbits, and it is maintained that the quantum mechanical wave packets provide results that agree with the classical results\cite{brow}\cite{bell}\cite{berry}\cite{kay} \cite{lena}\cite{bhau}\cite{mcan}\cite{pita}\cite{nauen}.  Most of the approaches exploit the SO(4) or SO(4,2) symmetry of the hydrogen atom which is used to rotate a circular orbital to an elliptical orbit. The starting orbital is often taken as a coherent state, which is usually considered a classical like state.  The most familiar example of a coherent state is for a one dimensional harmonic oscillator characterized by creation and annihilation operators $a^{\dagger}$ and $a$.  The coherent state $|\alpha\rangle$ is a superposition of energy eigenstates that is an eigenstate of $a$ where $a|\alpha\ra = \alpha |\alpha\ra$ for a complex $\alpha$. This coherent state will execute harmonic motion like a classical particle \cite{leon}.
 To obtain a coherent state for the hydrogenlike atom, eigenstates of the operator than lowers the principal quantum number n (which will be discussed in Section 7.4) have been used\cite{girar}, as well as lowering operators based on the equivalence of the four dimensional harmonic oscillator representation of the hydrogen atom\cite{lena}\cite{bhau}\cite{saty}. 
 
 In either case this coherent eigenstate is characterized by a complex eigenvalue, which needs to be specified. Several constraints have been used to obtain the classical wave packet that presumably obeys Kepler's Laws, such as requiring that the orbit lie in a plane so $\la z \ra =0$ for the orbital, or that $\la r - r_{classical} \ra$ be a minimum, or that some minimum uncertainty relationship is obtained.  In addition, there are issues regarding the approximations used, in particular, those that relate to time.  For times characteristic of the classical hydrogen atom, the wavepackets act like a classical system.  For longer times, the wavepacket spreads in the azimuthal direction and after some number of classical revolutions of order 10 to 100 the spread is $2\pi$ so the electron is uniformly spread over the entire orbit. The spread arises because the component wave functions forming the wavepacket have different momenta.  In two derivations, still longer times were considered, and recoherence was predicted to occur after about $n/3$ (where $n$ is the approximate principal quantum number) revolutions, although there is some difference in the predicted amount of recoherence\cite{eberH}\cite{mcan}. Because of the conservation of $\bm{L}$ and $\bm{A}$ the spread of the wave packets is inhibited except in the azimuthal direction.
 
 Brown took a different approach to develop a wavepacket for a circular orbit \cite{brow1}.  He first developed the asymptotic wave function for large $n$ and then optimized the coefficients in a Gaussian superposition to minimize the spread in $\phi$, obtaining a predicted characteristic decoherence time of about 10 minutes, considerable longer than any other predicted decoherence time. 

 Other authors have explored the problem from the perspective of classical physics and the correspondence principle\cite{kay}\cite{pita}\cite{liu}\cite{zver}.  Results from the different methods are similar with the basic conclusion that the wave functions are peaked on the corresponding classical Kepler trajectories: "atomic elliptic states sew the wave flesh on the classical bones "\cite{bell}. 
 
With the variety of experimental methods used to generate Rydberg states, a variety of Rydberg wave packets are created, and it is not clear which theoretical model, if any, is preferred\cite{eberH}. We take a very simple approach to forming a wavepacket and simply use a Gaussian weight for the different frequency components. This does not give an intentionally optimized wave packet but it is a much simpler approach and the result has all the expected classical behavior that is very similar to that obtained from much more complicated derivations. We start with a circular orbit and then do a SO(4) rotation to secure an elliptical orbit. We show that it has the classical period of rotation.   

\subsubsection{Wave functions in the Semi-classical Limit}

We need to derive the semiclassical limit of the wave functions that correspond to circular orbits in configuration space.  For this case, $\sin{\nu}$, which we interpret as the expectation value of the eccentricity, vanishes. We derive expressions for the wave functions in momentum space and then form a wavepacket.  To obtain corresponding expressions for elliptical orbits, we perform a rotation by $e^{i\bm{a} \cdot \nu}$ which does not alter the energy but changes the eccentricity and the angular momentum.

\vspace{7pt}
\underline{Case 1: Circular orbits, $\sin{ \nu} = e = 0$}

We derive the asymptotic form of $Y_{nlm}$ for large quantum numbers, where for simplicity we choose
the quantum numbers $n-1=l=m$ corresponding to a circular orbit in the 1-2 plane. From Eq.151 we see we encounter Gegenbauer polynomials of the form $C_0^{\lambda}$, which, by Eq. 152, are unity. For a very large $l$, $\sin^{l}{\theta}$ will have a very strong peak at $\theta = \pi/2$ so we make the expansion\cite{brown}
\be
\sin \theta=\sin \left(\frac{\pi}{2}+\left(\theta-\frac{\pi}{2}\right)\right)=1-\frac{1}{2}\left(\theta-\frac{\pi}{2}\right)^{2}+\cdots \hspace{14pt}  \approx e^{-(1/2)(\theta - \pi/2)^2}
\ee
to obtain
\be
\sin^{l}{\theta} \approx e^{-(1/2)l(\theta - \pi/2)^2}  .
\ee
The asymptotic forms for $N_1$ and $N_2$ can be computed using the properties of $\Gamma$\ functions: 
\be
\begin{array}{l}\quad \Gamma(2 z)=\left(\frac{1}{\sqrt{2 \pi}}\right) 2^{2 z-\frac{1}{2}} \Gamma(z) \Gamma\left(z-\frac{1}{2}\right) \vspace{7pt}\\ \lim _{z \rightarrow \infty} \Gamma(a z+b) \simeq \sqrt{2 \pi} e^{-a z}(a z)^{a z+b-\frac{1}{2}}.\end{array}  
\ee
We finally obtain\cite{note11}
\be
 Y_{n, n-1, n-1}(\Omega)= \sqrt{\frac{n}{2 \pi^{2}}} e^{-\frac{1}{2} n\left(\theta_{4}-\frac{\pi}{2}\right)^{2}}  \cdot e^{-\frac{1}{2} n\left(\theta-\frac{\pi}{2}\right)^{2}} e^{i(n-1) \phi}  .
\ee
which gives the probability density
\be
\left|Y_{n, n-1, n-1}(\Omega)\right|^{2}= \frac{n}{\left(2 \pi^{2}\right)} e^{-n\left(\theta_{4}-\frac{\pi}{2}\right)^{2}} \cdot  e^{-n\left(\theta-\frac{\pi}{2}\right)^{2}}  .
\ee
We have Gaussian probability distributions in $\theta_4$ and $\theta$ about the
value $\pi/2$.	The distributions are quite narrow with widths $\Delta\theta_4 \approx \Delta\theta \approx 1/\sqrt{n}$ and the spherical harmonic essentially describes a circle $(\theta_4=\theta = \pi/2)$ on the unit sphere in the 1-2 plane. As $n$ becomes very large, $U_4 =\cos \theta_4 \approx (r-r_c)/r$  (Eqs. 54 and Eq. 72) and $U_3=\sin\theta_4 \cos\theta$, which is proportional to $p_3$, both go to zero as $1/\sqrt{n}$ and the distribution approaches the great circle $U_1^2 + U_2^2 = 1$ that we found in Section 3.6
for a classical particle moving in a circular orbit in the
1-2 plane in configuration space. Note that this state is a stationary state with a constant probability density.  To get the classical time dependence we need to form a wavepacket.

\vspace{5pt}
\underline{Forming a Wavepacket}

We form a time dependent wavepacket for circular orbits by  superposing circular energy eigenstates:
\be
\chi (\Omega, t)=\sum_{n} {e}^{i t E_{n}} Y_{n, n-1, n-1} {A}_{n-N}
\ee
where $A_{n-N}$ is an amplitude peaked about $n=N>>1$.  For $n>>1$ we expand $E_n$ about $E_N$:
\be
{E}_{n}={E}_{{N}}+\left.\frac{\partial {E}}{\partial {n}}\right|_{{N}} {s}+\left.\frac{\partial^{2} {E}}{\partial {n}^{2}}\right|_{{N}} {s}^{2}+\ldots
\ee
where $s=n-N$.  From the equation for the energy levels, $E=-m{(Z\alpha})^2/(2n^2)$ we compute
\be
\frac{\partial E_n}{\partial n} \Big{|}_N =\frac{m(Z\alpha)^2}{N^3}=\sqrt{\frac{-8E_N^3}{m(Z\alpha)^2}}.
\ee
In agreement with the Bohr Correspondence Principle the right-hand side of this equation is just the classical
frequency $\omega_{cl}$	as given in Eq. 38. For the second order derivative we have 
\be
\frac{\partial^2E}{\partial n^2}\Big{|}_N = -\frac{3}{N}\omega_{cl} \equiv \beta
\ee
which gives 
\be
\chi(\Omega, t)=e^{-i t E_{N}} e^{i \phi(N-1)} \sum_{s=-N+1}^{\infty} e^{-i\left(\omega_{cl} s t-\left(\frac{\beta}{2}\right) s^{2} t-s \phi\right)} \cdot A_{s} | Y_{N+s, N+s-1, N+s-1}| .
\ee
We choose a Gaussian form for $A_s$
\be
A_s=\frac{1}{\sqrt{2\pi N}}e^{-s^2/(2N)}.
\ee
Brown used $A_s= C e^{-s^2 3\omega_{cl}t/N}$ which minimizes the diffusion in $\phi$ at time $t$ \cite{brow}. Since $| Y_{N+s,N+s-1,N+s-1}|$ varies slowly with $s$ for $N>>1$ we can take it outside the summation in Eq. 165. We
now replace the sum by an integral over $s$. Since $A_s$ is peaked about $N$ we can integrate from $s = -\infty$ to 
$s = +\infty$. We perform the integral by completing the square
in the usual way. The final result for the probability amplitude for a circular orbital wave packet is
\be
\begin{aligned}|x(\Omega, t)|^{2}=&\left|Y_{N, N-1, N-1}\right|^{2}\left(1+\beta^{2} t^{2} N^{2}\right)^{-1} \\ \cdot & \exp \left[-\left(\phi-\omega_{cl} t\right)^{2} \frac{N}{1+(\beta t N)^{2}}\right] \end{aligned}  .
\ee
This represents a Gaussian distribution in $\phi$ that is centered about the classical value $\phi=\omega_{cl}t$, meaning that the wavepacket is traveling in the classical trajectory with the classical time dependence . The width of the $\phi$ distribution is 
\be
\Delta\phi = (N)^{(-1/2)}(1 + \beta^2t^2N^2)^{1/2} =(N)^{(-1/2)}(1 + 9\omega_{cl}^2t^2)^{1/2}.
\ee
The distribution in $\phi$ at $t=0$ is very narrow, proportional to $1/\sqrt{N}$, but after several orbits $\Delta \phi$ is increasing linearly with time.

The distributions in $\theta_4$ and $\theta$ are Gaussian and centered
about $\pi/2$ in each case as for the circular wave function (cf. Eq. 160) with widths equal to $(N)^{-1/2}$. The
spreading of these distributions in time is inhibited
because of the conservation of angular momentum and energy. The detailed behavior of the widths depend on our use of the Gaussian distribution.  Other distributions will give different widths, although the general behavior is expected to be similar.

As a numerical example, consider a   hydrogen atom which is in the semiclassical region when the orbital diameter is about $1$ cm.  The corresponding principal quantum number is about $10^4$, the mean velocity is about $2.2 x 10^4$ cm/sec and the period about $1.5 x 10^{-4}$ sec.  After about 34 revolutions or $5 x 10^{-3}$ sec, the spread in $\phi$ is about $2 \pi$, meaning the electron is spread uniformly about the entire circular orbit. This characteristic spreading time can be compared to $1.6 x 10^{-3}$ sec for a fully optimized wave packets formed from coherent SO(4,2) states \cite{mcan}\cite{nand}. In order to make predictions about significantly longer times, we would need to retain more terms in the expansion Eq. 162 of $E_n$. 
\newpage

\underline{Case 2: Elliptical orbits $\sin\nu = e \neq 0$}

We can obtain the classical limit of the wave function for elliptical orbits 
by first writing our asymptotic form Eq. 159 for $Y_{n, n-1,n-1}$ in terms of the $U$ variables instead of the angular variables by using definitions Eq. 52, and setting $a=a_n$.  Retaining only the lowest order terms in $(\theta_4 - \pi/2)$ and $(\theta - \pi/2)$, we find 
 \be
Y_{n, n-1, n-1}(\hat{U})=\left(\frac{n}{2 \pi^{2}}\right)^{\frac{1}{2}} e^{i(n-1) \tan ^{-1}\left(\frac{U_{2}}{U_{1}}\right)} \cdot e^{-\frac{1}{2}n\left(U_{4}\right)^{2}} e^{-\frac{1}{2}n\left(U_{3}\right)^{2}}
\ee
For large n, this represents a circular orbit in the 1-2 plane.  We now perform a rotation by $A_2 \nu$.  which will change the eccentricity to $\sin{\nu}$, and change the angular momentum, but will not change the energy or the orbital plane. Using Eq. 140 to express the old coordinates in terms of the new coordinates, we find to lowest order
\be
Y'_{n, n-1,n-1}(U)=\left(\frac{n}{2 \pi^{2}}\right)^{1/2} e^{i(n-1) \tan ^{-1}\left(\frac{U_{2}}{U_{1} \cos v}\right)}  \cdot e^{-\frac{1}{2}n\left\{U_{2} \sin \nu -U_{4} \cos \nu\right\}^{2}} \cdot e^{-\frac{1}{2}n\left(U_{3}\right)^{2}}.
\ee
In Section 3.6 we found that the vanishing of the term in braces $0 =U_2 \sin\nu-U_4 \cos\nu $ specifies the classical great hypercircle orbit (Eq.56) corresponding to  an ellipse in configuration space with eccentricity $e = \sin \nu$ and lying in the 1—2
plane. The probability density $|Y'_{n,n-1,n-1}(U)|^2$ vanishes except within a hypertorus with a narrow cross section of radius approximately $\frac{1}{\sqrt{n}}$ which is centered about the
classical distribution. Since the width $\frac{1}{\sqrt{n}}$ of the distribution is constant in U space, it will not be constant when projected onto p space 
 
In terms of the original momentum space variables, the asymptotic spherical harmonic is 
\be
\begin{array}{l}\qquad \begin{array}{rl}Y_{n, n-1, n-1}(\bm{p})=\left(\frac{n}{2 \pi^{2}}\right)^{\frac{1}{2}} e^{i(n-1) \tan ^{-1}\left(\frac{p_{2}}{p_{1} \cos \nu}\right)} \\ \cdot \exp \left\{-\left(\frac{n}{2}\right)\left[p_{1}^{2}+\left(p_{2}-a \tan \nu\right)^{2}-\frac{a^{2}}{\cos ^{2} \nu}\right]^{2}\left(\frac{\cos \nu}{p^{2}+a^{2}}\right)\right\}^{2}\vspace{4pt} & \\ \cdot \exp \left\{-\left(\frac{n}{2}\right)\left(\frac{2 p_{3} a}{p^{2}+a^{2}}\right)\right\}^{2}\Big{|} a=a_{n}\end{array}\end{array}.
\ee
The expression in brackets corresponds to the momentum space classical
orbit equation we found previously (Eq. 49). As we
expect, $p_3$ is Gaussian about zero since the classical orbit is in the 1—2 plane.  We can simplify the
expressions for the widths by observing that to lowest order we can use Eq. 48, which implies
$p^2 + a^2 = 2a^2 + 2ap_2\tan\nu$
in the exponentials. The widths of both distributions therefore
increase linearly with $p_2$. We also note that since classically there exists a one-to-one correspondence
between each point of the trajectory in momentum space and
each point in configuration space, we may interpret the widths of the distributions using Eq. 32  $\frac{p^2 + a^2}{a^2}=\frac{2r_c}{r}$.  Accordingly the widths increase as the momentum increases or as the distance to the force center decreases (Fig. 8). 
\begin{figure}[ht] 
\centering
 \includegraphics[scale=0.8]{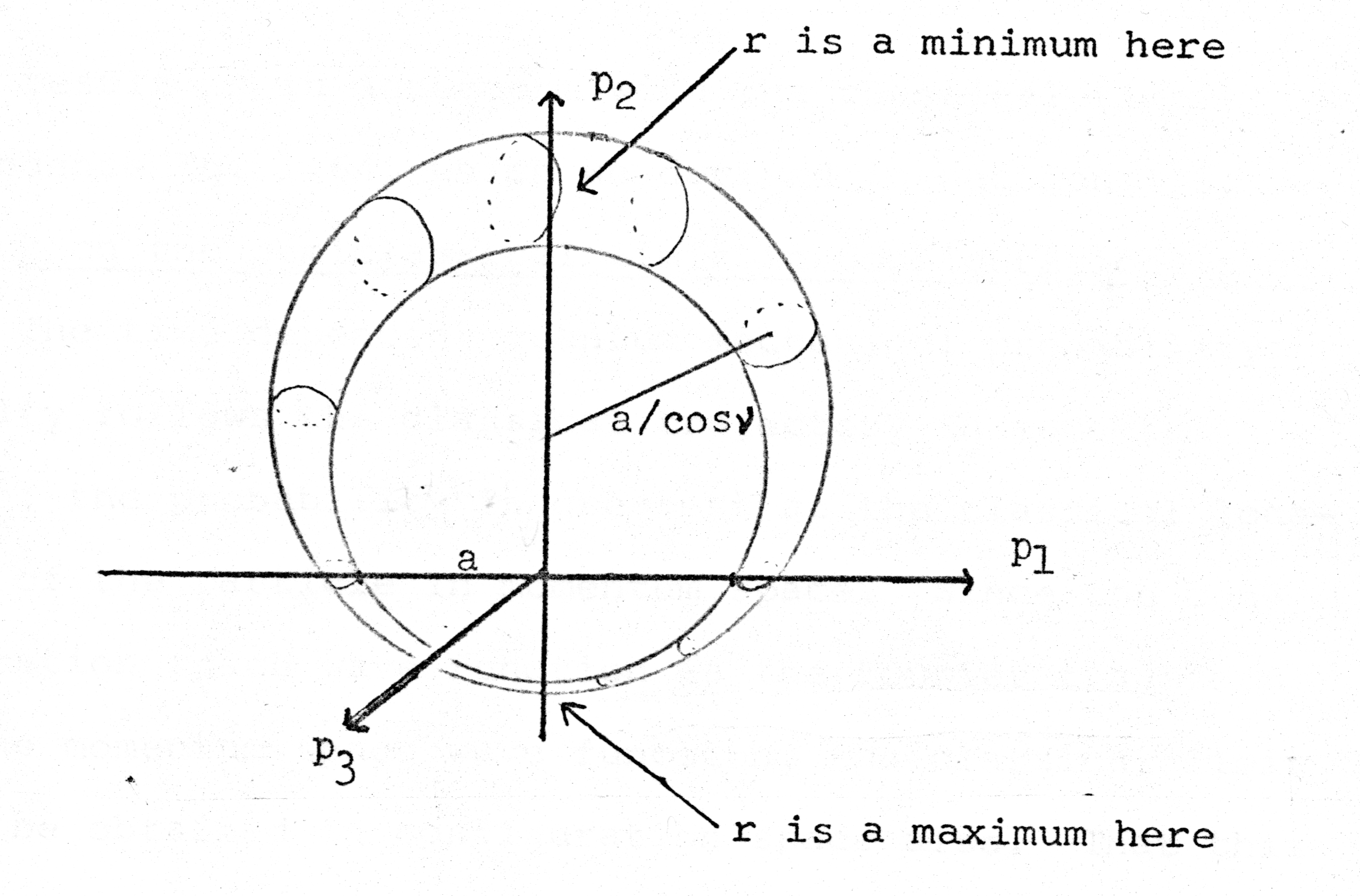}
 \caption{Wave function  probability distribution $|Y'_{n, n-1,n-1}(\bm{p})|^2$  in momentum space for large n, showing the variation in the width of the momentum distribution about the classical circular orbit.  The center of the distribution is at $p_2 = a \tan \nu$ The classical orbit is in the 1-2 plane.}
 \label{Fig8}
\end{figure}

\underline{Forming a Wave packet for Elliptical Motion}

We may form a time dependent wavepacket superposing the wave functions of Eq. 170. Care must be taken to include the first order dependence (through $a_n$) of $\tan^{-1}(U_2/U_1\cos\nu)$ on the principal quantum number when integrating over the Gaussian weight function. The result for the probability density is the
same as before (Eq. 167) except
$|Y'_{N,N-1,N-1}|^2$ (given in Eq. 170) replaces $|Y_{N,N-1,N-1}|^2$ and
\be
\omega_{{cl}} {t}=\tan ^{-1}\left(\frac{{U}_{2}}{{U}_{1} \cos \nu}\right)+\sin \nu \left({U}_{1}\right)
\ee
replaces $\omega_{cl}t=\phi$  .
The result is exactly the same as the classical time dependence Eq. 73. The spreading of the wave packet will be controlled by the same factor as for the circular wave packet.
\newpage

\underline{Remark on the Semiclassical Limit in Configuration Space}

The time dependent quantum mechanical probability
density follows the classical trajectory in momentum
space meaning that the probability is greatest at the classical location of the particle in momentum space. Since the configuration space wave function is the Fourier transform of the momentum space wave function, the classical limit
must be obtained in configuration space also. That this limit is obtained is made explicit by observing that the
momentum space probability density is large when 
\be
(U_2 \sin \nu - U_4 \cos \nu)^2 \approx 0 .
\ee
However from Section 3 we can show that  
\be
U_2 \sin\nu - U_4 \cos\nu = \cos\nu \Big(\frac{r - r_{classical}}{r_{classical}}\Big)
\ee
where  $r_{classical}$  is given by the classical orbit Eq. 33. Accordingly we see that the configurations space probability will be large when 
\be
\Big[\frac{r - r_{classical}}{r_{classical}}\Big]^2 \approx 0  .
 \ee

\subsection{Quantized Semiclassical Orbits}

It is convenient at times to have a semiclassical model for the orbitals of the hydrogenlike atom. Historically this was first done by Pauling and Wilson\cite{paul}. We can obtain a model by interpreting the classical formulae for the geometrical properties of the orbits as corresponding to the expectation values of the appropriate quantum mechanical expressions. 
Thus when the energy $E = -a^2/2m$ appears in a classical formula we employ the expression for $a$ for the quantized energy levels $a = \frac{1}{nr_0}$
where $r_0 = (mZ\alpha)^{-1}$ which is the radius 0.53 Angstrom of the ground state.  Similarly if $\bm{L}^2$ appears in a classical formula, we substitute $l(l+1)$ where $l$ is quantized $l$ = $0 , 1, 2..n-2, n-1;$ and $m$ , the component of $\bm{L}$ along the 3-axis is quantized: $m$= $-l, -l+1, -l+ 2,..l$.

\newpage
\underline{Orbits in Configuration Space}

Recalling Eq. 28 $r_c=\frac{mZ\alpha}{a^2}$ we see $ar_c=n$, which gives a semimajor axis of length $r_c = n^2r_0,$ where $r_0$ is the radius for the circular orbit of the ground state and $r_c$ is for a circular orbit for a state with principal quantum number $n$. The equations for the magnitude of $\bm{L}$ and $\bm{A}$ are 
\be
L=r_{c} a \cos v=n \cos v=\sqrt{l(l+1)} 
\ee
\be
A=r_{c} a \sin v=n \sin v=\sqrt{n^{2}-l(l+1)}
\ee
This gives an eccentricity $\sin\nu$ equal to
\be
e=\sin v=\sqrt{1-\frac{l(l+1)}{n^{2}}}
\ee
and a semi-major axis equal to
\be
b=r_c\sin{\nu}=n\sqrt{l(l+1)}.
\ee 
Note that the expression for $e$ is limited in its meaning.  For an $s$ state, it always gives e=1, and for states with $l=n-1$ it  give $e=\sqrt{1/n}$, not the classically expected 0 for a circular orbit.

\vspace{8pt}
\underline{Orbits in U-Space}

The corresponding great hypercircle orbits $(\nu,\Theta)$
in U-space are described by giving the quantized angle $\nu$,
between the three-dimensional hyperplane of the orbit and the 4-axis, and the quantized angle $\Theta$, between the hyperplane of the orbit and the 3-axis:
\be
\cos\nu=\sqrt{\frac{l(l+1)}{n^2}}
\ee
\be
\cos\Theta = \sqrt{\frac{m^2}{l(l+1)}}     .
\ee
Note the similarity in these two equations, suggesting that $m$ relates to $l$ the same way that $l$ relates to $n$, suggesting a generalization of the usual vector model of the atom which only describes the precession of $\bm{L}$ about the $z $ axis.

The results for orbits in configuration and momentum space illustrate some interesting features:

1.	The equation $ar = n$ illustrates that the characteristic dimensions of an orbit in configuration space and the corresponding orbit in momentum space are
inversely proportional, as expected since they are related by a Fourier transform, consistent with the Heisenberg Uncertainty Principle. 

2.	If $l = 0$ then no classical state exists. The orbit in configuration space degenerates into a line
passing through the origin while the corresponding circular orbit in momentum space attains an infinite radius
and an infinite displacement from the origin. Although
this seems peculiar from the pure classical viewpoint, quantum mechanically it follows since for S states there
is a nonvanishing probability of finding the electron within the nucleus.

In order to interpret these statements about quantized semiclassical elliptical orbits we observe that for quantum mechanical state of the hydrogenlike atom with definite $n,l, m$, the probability density is (1) independent of $\phi_r$ or $\phi_p$ and (2) it does not confine the electron to some orbital plane.  Since the quantum mechanical distribution for such a state specifies no preferred direction in the 1—2 plane, we must imagine this distribution as corresponding in some way to an average over all possible orientations of the semiclassical elliptical orbit. This interpretation is supported by the fact that the region within which the quantum mechanical radial distribution function differs largely from zero is included between the values of $r$ corresponding to the
semiclassical turning points $r_c (1 \pm \sin \nu )$.

\subsection{Four-dimensional Vector Model of the Atom}
In configuration space or momentum space the angle between the classical plane of the orbit and the 3-axis is $\Theta$  which is usually interpreted 
in terms
of the vector model of the atom in which we imagine $\bm{L}$
to be a vector of magnitude $\sqrt{l(l+1)}$ precessing about
the 3 axis, with $m$ as the component along the 3-axis.  This precession may be linked to the $\phi_r$ independence
of the probability and the absence of an orbital plane as
mentioned at the end of the preceding section. The precession constitutes a classical mechanism which yields the
desired average over all possible orientations of the semiclassical elliptical orbit. Since the angle $\Theta$ is restricted to have only certain discrete values one can
say that there is a quantization of space. 

The expression for $\cos\nu =\sqrt{l(l+1)/n^2}$ is quite analogous to that for $\Theta$ and so suggests a generalization of the vector model of the atom to four dimensions. The projection of the four-dimensional
vector model onto the physical three-dimensional subspace
must give the usual vector model. We can achieve this by imagining that a four-dimensional vector of length $n$,
where $n$ is the principal quantum number, is precessing in such a way that its 
3 and 4 components are constants,
while the 1 and 2 components vary periodically. The projection onto the 1—2—3 hyperplane is a vector of constant magnitude $\sqrt{l(l+1)}$ precessing about the 3-axis. The
component along the 3—axis is $m$. The component along the 4-axis is $A=\sqrt{n^2 - l(l+1)}$ the magnitude of the vector $\bm{A}$. The vectors $\bm{L}$ and $\bm{A}$ are perpendicular to each other.  Thus the precessing $n$ vector makes a constant angle $\Theta$   with the 3-axis and a
constant angle $\pi/2 - \nu$  with the 4-axis. Since both angles
are restricted to certain values, we may say that we have a quantization of four-dimensional space.

\section{The Spectrum Generating Group SO(4,1) for the Hydrogenlike Atom}

We consider the Schrodinger hydrogen atom and its unitary "noninvariance" or spectrum generating operators $e^{iD_i\beta_i}$ where $D_i$ is a generator and $\beta_i$ is a real parameter, using eigenstates of $(Z\alpha)^{-1}$ for our basis of our representation. These
operators transform an eigenstate of the kernel K with a definite value of the coupling constant (or principal
quantum number) into a linear combination of eigenstates with different values of the coupling constant (or different
principal quantum numbers), and different $l$ and $m$. Unlike the invariance generators $\bm{L}$  and $\bm{A}$, the noninvariance generators clearly do not generally commute with the kernel K, $[D_i, K] \neq 0$ so they change the principal quantum number.

The set of all invariance and noninvariance operators
forms a group with which we may generate all eigenstates in our complete set from a given eigenstate. We show that this group, called the spectrum generating
group of the hydrogenlike atom, is SO(4,1), the group of
orthogonal transformations in a 5-dimensional space with a
metric $g_{AB} = (-1, 1, 1, 1, 1),$  where $ A,B = 0, 1, 2, 3, 4.$  The
complete set of eigenstates of $(Z\alpha)^{-1}$ for the hydrogenlike atom forms a  unitary, irreducible, infinite-dimensional representation of SO(4,1) which, we shall find, can be decomposed into an infinite sum of irreducible representations of
SO(4), each corresponding to the degeneracy group for a
particular principal quantum number. A unitary representation means all generators are unitary operators.  An irreducible representation does not contain lower dimensional representations of the same group. In Section 6.3 we
discuss the isomorphism between the spectrum generating group SO(4,l) and the group of conformal transformations in
momentum space.  An isomorphism means the groups have the same structure and can be mapped into each other.

\subsection{Motivation for Introducing
the Spectrum Generating Group Group SO(4,1)}

 We have examined the group structure for the degenerate eigenstates of $(Z\alpha)^{-1}$ for the Schrodinger hydrogenlike atom: the $n^2$ degenerate states form an irreducible representation of SO(4). The next question we might
ask is: Do all or some of the states with different principal quantum numbers form an irreducible representation of some larger group which is reducible into SO(4)
subgroups? If such a group exists then it clearly is not
an invariance group of the kernel K (Eq. 106). If we want our noninvariance group to include just some of the
states then it will be a compact group, since unitary representations of compact groups can be finite dimensional.
If we want to include all states then it will be a noncompact group since there are an infinite number of eigenstates of $(Z\alpha)^{-1}$ and all unitary representations of
noncompact groups are infinite dimensional \cite{lie}.

We can find
a compact noninvariance group for the first N levels of the coupling constant
, $n = 1, 2, ... N$.	The dimensionality of our representation is
\be
\sum_{n=1}^{N} n^{2}=\frac{N(N+1)(2 N+1)}{6}     .
\ee 
Mathematical analysis of the group SO(5) shows that this is the dimensionality of
the irreducible symmetrical
tensor representation of SO(5) given by the tensor with 5 upper indices $T^{abc...}$ where $a,b,..= 1,2,3, 4$ or $ 5$ \cite{wigner}. Reducing this representation of SO(5) into its SO(4) components
gives 
\be
\begin{aligned}(\text { symm } \cdot \text { tensor } N)_{S O(5)}=&(0,0) \oplus\left(\frac{1}{2}, \frac{1}{2}\right) \oplus \ldots \quad \oplus\left(\frac{N-1}{2}, \frac{N-1}{2}\right) \\=&(\text { symm } \cdot \text { tensor } n=1)_{S O(4)}{\oplus} \\ &(\text { symm } \cdot \text { tensor } n=2)_{S O(4)}\oplus \\ &  \ldots \oplus \quad(\text { symm } \cdot \text { tensor } n=N)_{S O(4)} \end{aligned}
\ee
which is precisely the structure of the first N levels of
a hydrogenlike atom. If we want to include all levels then we guess that the appropriate noncompact group is SO(4,1),
whose maximal compact subgroup is SO(4). Thus we conjecture that all states form a representation of SO(4,1).

Consider the Lie algebra of O(4,1) and the general structure of its generators in terms of the canonical variables. The algebra of O(n) has $\frac{n(n-1)}{2}$ generators so to extend the algebra of O(n) to O(n+1) takes n generators, which can be taken as the components of a n-vector. To extend the Lie algebra from O(4) to O(5) or O(4,1) we can choose the additional generators $G_a$ to be components of a four-vector G under O(4):
\be
[S_{ab}, G_{c}] = i(G_b \delta_{ac} - G_a \delta_{bc})\hspace{14pt}a,b,c= 1,2,3,4   .
\ee
If we apply Jacobi's identity to $S_{ab}$, $G_a$, and $G_{b}$ and use Eq. 184 we find
\be
[S_{ab}, [G_a, G_b]] = 0  .
\ee
We require that the Lie algebra closes, so $[G_a, G_b]$ must be a linear combination of the generators, clearly proportional to $S_{ab}$ and we choose the normalization such that 
\be
[G_a, G_b] = -iS_{ab}   .
\ee
If we define
\be
G_4 = S_{40} = S ; \hspace{10pt} G_i = S_{i0} = B_i 
 \ee
and recall Eq. 79
$$ L_i=e_{ijk}S_{jk} \hspace{10pt}A_i=S_{i4}
$$
then the additional commutation relations that realize SO(4,1) may be written in terms of $\bm{L}, \bm{A}, \bm{B}$, and $S$:
\be
\begin{aligned}\left[{L}_{{i}}, {B}_{{j}}\right] &={i} \epsilon_{{ijk}} {B}_{\mathrm{k}} &\left[{L}_{i}\hspace{3pt}, {S}\right]\hspace{2pt} &=0 \\\left[{S}\hspace{3pt}, {A}_{{j}}\right] &={i} {B}_{{j}} &\left[{S}\hspace{2pt},{B}_{{j}}\right] &={iA}_{{j}} \\\left[{A}_{{j}}, {B}_{{k}}\right] &={i} \delta_{{jk}} {S} &\left[{B}_{{i}}, {B}_{{j}}\right] &=-{i} \epsilon_{{ijk}} {L}_{{k}}. \end{aligned}
\ee
The top two commutators show that $\bm{B}$
transforms as a three-vector under O(3) rotations and that $S$ is a scalar under rotations. Alternatively we can write the commutation relations in terms of the generators $S_{AB}$ , $A,B = 0,1,2,3,4$ :
\be
\left[S_{A B}, S_{C D}\right]=i\left(g_{A C} S_{B D}+g_{B D} S_{A C}-g_{A D} S_{B C}-g_{B C} S_{A D}\right)
\ee
where $g_{00} = -1, g_{aa} = 1$.

The commutators above follow directly from the mathematical theory of SO(4,1), but the theory does not tell us what these generators represent, just their commutation properties. We now investigate the general features of the representations of SO(4,1) provided by the hydrogenlike atom and how to represent the generators in terms of the canonical variables.

\subsection{Casimir Operators}
The two Casimir operators of SO(4,1) are\cite{bacr1}
\be
Q_{2}=-\frac{1}{2} S_{A B} S^{A B}=S^{2}+\bm{B}^{2}-\bm{A}^{2}-\bm{L}^{2}
\ee
and
\be
Q_{4}=-w_{A}{w}^{A}=(S \bm{L}-\bm{A} x \bm{B})^{2}-\frac{1}{4}[\bm{L} \cdot(\bm{A}+\bm{B})-(\bm{A}+\bm{B}) \cdot \bm{L}]^{2}
\ee
where $w_A = \frac{1}{8}\epsilon_{ABCDE}S^{BC}S^{DE}$ .

For SO(4), we recall that for the SO(4) representations the structure
of the generators in terms of the canonical variables led
to the vanishing of one Casimir operator
$ C_1=\bm{L}\cdot\bm{A}$ and consequently 
the appearance of only symmetrical tensor representations. We will find $Q_4$ vanishes for analogous reasons.

If $\bm{B}$ is a pseudovector, it is proportional to $\bm{L}$,
which is the only independent pseudovector that can be constructed
from the dynamical variables. The coefficient of proportionality, a scalar, $X$ need not commute with $H$:
\be
\bm{B}=X\bm{L} \hspace{10pt}[X,\bm{L}]=0\hspace{10pt}[X,H]\ne 0
\ee
Since $[B_i,B_j]= -ie_{ijk}L_k$ it follows that $X^2 = -1$ and
$\bm{B}$ would therefore be a constant multiple of $\bm{L}$ and not
an independent generator. Thus $\bm{B}$ must be a vector and expressible as
\be
\bm{B} = f\bm{r} + h\bm{p}
\ee
where f and h are scalar functions of $r,p^2,$ and $\bm{r}\cdot\bm{p}$.  Accordingly we find
\be
 \bm{B}\cdot\bm{L} = \bm{L}\cdot\bm{B}=0
\ee 
Further since $\bm{B}$ is a vector and $\bm{A}$ is a vector, $\bm{A}\times\bm{B}$ is a pseudovector and
therefore is proportional to $\bm{L}$ :
\be
\bm{A}\times\bm{B} = Y\bm{L}\hspace{7pt} ,\hspace{13pt} [Y,\bm{L}] = 0
\ee
For this equation to be consistent with the SO(4,1) commutation relations we find 
$Y=S$ and therefore
\be
\bm{A}\times\bm{B} = S \bm{L}  .
\ee
It follows from $\bm{L}\cdot \bm{A}=0$ and from Eqs. 194 and 196 that for the SO(4,1) representations realized by the hydrogenlike atom 
\be
Q_4=0   .
\ee
As with the SO(4) symmetry, the dynamics of the hydrogen atom require that only certain representations of SO(4,1)
appear. From the mathematical theory of irreducible infinite dimensional unitary representations of SO(4,1)
we have the following results:
\be
\begin{array}{l}\text{Class I: } \quad Q_{4}=0; \quad Q_{2} \text { real, }>0 \vspace{5pt}\\ \begin{aligned}& \text{SU(2) } \times \operatorname{SU}(2) \text { content: } \\ \hspace{10pt} (Q)^{I}=&(0,0) \oplus\left(\frac{1}{2}, \frac{1}{2}\right) \oplus(1,1) \oplus \ldots \end{aligned}\end{array}
\ee
\be
\begin{array}{l}\text { Class II: } Q_{4}=0, Q_{2}=-(s-1)(s+2), s=\text { integer }>0 \vspace{5pt} \\ \qquad \qquad \begin{aligned}  & \hspace{8pt} S U(2) \times S U(2) \text { content: } \vspace{4pt} \\ \quad (Q)^{I I}=&\left(\frac{s}{2}, \frac{s}{2}\right) \oplus\left(\frac{s+1}{2}, \frac{s+1}{2}\right) \oplus \cdots \end{aligned}\end{array}
\ee
The class I representations are realized by the complete set of eigenstates of $(Z\alpha)^{-1}$ for the hydrogenlike atom. Note, however, that we have an infinite number of such class I representations since $Q_2$	may have any positive real value. We shall find that for $Q_2= 2$ we may extend our group from SO(4,l) to SO(4,2). The class II representations are realized by the eigenstates of $(Z\alpha)^{-1}$
with principal quantum numbers from $n = s + 1$ to $n$
becomes infinite.  The first $s$ levels could, if we desire, be described by SO(5).

In this section we have analyzed the group structure and the representations using the complete set of eigenstates of $(Z\alpha)^{-1}$ for our basis. We might ask: What if we used energy eigenstates instead as a basis for the representations? From Section 4.3 we know that the quantum numbers and multiplicities of the $(Z\alpha)^{-1}$ eigenstates are precisely the same as those of the bound energy eigenstates.
Thus with the energy eigenstates as our basis, we would reach the same conclusions about the group structure as before but we would be including only the bound states in our representations and we would be ignoring all scattering states.

\subsection{Relationship of the Dynamical Group SO(4,1) to the Conformal Group in Momentum Space}

We can give a more complete analysis of the hydrogenlike atom in terms of SO(4,1) by considering the relationship between the four-dimensional rotations of the
four-vector $U'_{a}$ with $a=l,2,3,4,$  which we discussed in Section
4.6, and the group of conformal transformations in momentum
space. Conformal transformations preserve the angles between directed curves, but not necessarily lengths. The rotations generated by the Runge-Lenz vector $\bm{a}$ and the angular momentum $\bm{L}$ leave the scalar product
$U_a V^a$ of four-vectors invariant and therefore are conformal
transformations. The stereographic projection we employed
is also a conformal transformation. Since the product of
two conformal transformations is itself a conformal transformation, we must conclude that $\bm{a}$ generates a conformal
transformation of the momentum three-vector $\bm{p}$. 

In order to express the most general conformal transformation we must introduce two additional operators
that correspond to the operators $\bm{B}$ and $S$ introduced
in Section 6.1. By employing the isomorphism between the
generators $\bm{L}$, $\bm{a}$, $\bm{B}$, and $S$ of SO(4,1) and the generators
of conformal transformations in momentum space we can immediately obtain expressions for the additional generators
$\bm{B}$ and $S$ in terms of the canonical variables, which is our objective.  We need these additional generators to complete our SO(4,1) group for the hydrogen atom.

To derive the isomorphism we use the most convenient representation, namely that based on eigenstates of $(Z\alpha)^{-1}$ convenient for momentum space calculations ($\rho = \frac{(p^2 + a^2)}{2a^2}$). Once established, the isomorphism becomes a group theoretical statement and is independent of the particular representation.

\newpage
\underline{The Conformal Group in Momentum Space}

An arbitrary infinitesimal conformal transformation in momentum three-space may be written as
\be
\delta p_{j}=\delta a_{j}+\delta \omega_{j k} p_{k}+\delta \rho p_{j}+\left(p^{2} \delta c_{j}-2 p_{j} \bm{p} \cdot \delta \bm{c}\right)
\ee
where $\delta \omega_{jk} = - \delta \omega_{kj}$.

The terms in $\delta p_j$ arise as follows:
\be
\begin{array}{l}\delta a_{j} \text { translation generated by } \bm{R} \cdot \delta \bm{a} \\ \delta \omega_{j k} \text { rotation generated by } \bm{J} \cdot \delta \bm{\omega}, J_{ij}=\epsilon_{i j k} J_{k} \\ \delta \rho \text { dilation generated by } D \delta \rho \\ \delta {c}_{j} \text { special conformal transformation generated by } \bm{K} \cdot \delta \bm{c}\end{array}
\ee
This is a ten parameter group with the generators $(\bm{R}, \bm{J}, D, \bm{K})$ which obey the following commutation relations:
\be
\begin{array}{ll}{\left[{D}, {R}_{{j}}\right] \cdot={i} {R}_{{j}}} & {\left[{D}, {J}_{{i}}\right]=0} \\ {\left[{D}, {K}_{{j}}\right]=-{i} {K}_{{j}}} & {\left[{R}_{{i}}, {J}_{{k}}\right]=i \epsilon_{{i} j {k}} {R}_{{k}}} \\ {\left[{K}_{{n}}, {R}_{{m}}\right]=2 {i} \epsilon_{{nmr}} {J}_{{r}}-2 {i} \delta_{{mn}}}D & {\left[{J}_{{i}}, {J}_{{k}}\right]={i} \epsilon_{{i} {km}} \mathrm{J}_{\mathrm{m}}} \\ {\left[{R}_{{i}}, {R}_{{j}}\right]=0} & [{K}_{{i}}, {J}_{{k}}] ={i} \epsilon_{{i} {km}} {K}_{{m}} \\ {\left[{K}_{{j}}, {K}_{{j}}\right]=0} & \end{array}
\ee
There is an isomorphism between the algebra of the generators of conformal transformations and the dynamical noninvariance algebra of SO(4,1) of the hydrogen atom.  Since $J_i$ is the generator of spatial rotations we make the association $L_i=J_i$. Comparing the differential change in $p_i$ from a transformation generated by $\bm{A}\cdot\delta \bm{\nu}$ (in the representation with $\rho=\frac{(p^2+A^2)}{2a^2})$
 \be
\begin{aligned} \delta p_{i} &=i\left[\bm{a} \cdot \delta \bm{\nu}, p_{i}\right] \\ &=-\frac{1}{2 a}\left[\left(p^{2}-a^{2}\right) \delta\nu_{i}-2 \bm{p} \cdot \delta \bm{\nu} p_{i}\right] \end{aligned}
\ee
to the differential change in $p_i$ from a conformal transformation leads to the association
\be
a_{i}=\frac{1}{2}\left(\frac{K_i}{a}-a R_{i}\right)   .
\ee
To confirm the identification we can use the commutation relations of the conformal group to show that the O(4) algebra of $\bm{L}$ and $\bm{a}$ corresponds precisely to that of
$\bm{J}$ and $\frac{1}{2}(\frac{\bm{K}}{a}-a\bm{R})$.
This result alone suggests that our
SO(4) degeneracy group should be considered as a subgroup of the larger group SO(4,1).  It suggests introducing the operators 
\be
\bm{B} = \frac{1}{2}\left(\frac{\bm{K}}{a} + a\bm{R}\right)\hspace{20pt}  S=D  .
\ee
The commutation relations of $S$ and $\bm{B}$ which follow from Equations 205 and the commutation
relations Equation 202 are identical to the commutation relations
given for $S$ and $\bm{B}$ in Section 6.1. Thus by considering
the $\bm{a}$ and $\bm{L}$ transformations in momentum space as conformal transformations, we were led to introduce the generators $\bm{B}$ and $S$ and obtain the dynamical algebra SO(4,l).  Further, we are led to the expressions for these generators in terms of the canonical variables.

By comparing the expression for $\bm{a}$ in terms of the conformal generators with our known expressions for $\bm{a}$,
Eq. 120 or Eq. 121, we obtain expressions for	$K_i$ and $R_i$ in terms of the canonical variables. If we use the  eigenstates convenient for configuration space calculations ($\rho = n/ar)$ we make the identifications
\be
\begin{array}{l}\bm{K}=\frac{1} {2}(\bm{r} p^{2}+p^{2} \bm{r})-\bm{r} \cdot \bm{p} \bm{p}-\bm{p} \bm{p} \cdot \bm{r}-\frac{\bm{r}}{4 r^{2}} \vspace{3pt} \\ \bm{R}=\bm{r}  .\end{array}
\ee
Substituting these results in the equation for	$\bm{B}$ we find
\be
\bm{B}=\frac{1}{2 a}\left(\frac{p^{2} \bm{r}+\bm{r}{p}^{2}}{2}-\bm{r} \cdot \bm{p} \bm{p}-\bm{p} \bm{p} \cdot \bm{r}-\frac{\bm{r}}{4 r^{2}}\right)+\frac{a \bm{r}}{2}
\ee
which is a manifestly Hermitean operator valid throughout Hilbert space. Note that
\be
\bm{B} - \bm{a} = a \bm{r} .
\ee
To compute $D$ we substitute the expressions
for $\bm{K}$ and $\bm{R}$ into the commutation relation
$$D=\frac{i}{2}[K_i, R_i]$$
obtaining the result
\be
S=\frac{1}{2}(\bm{p}\cdot\bm{r} + \bm{r}\cdot\bm{p}) = D
\ee
which is identical to the generator of the scale change transformation $D(\lambda)$ defined in Eq. 109 in Section 4.3 .

The significance of the generator $D=S$ of the scale change in terms of SO(4,1) is apparent if we compute 
\be
e^{i \lambda D}\left(a \bm{R}\pm\frac{\bm{K}}{a}\right) e^{-i \lambda D}=a'\bm{R}\pm\frac{\bm{K}}{a'}
\ee
where $a'=e^{\lambda}a.$

The unitary transformation $e^{i\lambda D}$	may be viewed as generating an inner automorphism of SO(4,1) which is an equivalent representation of SO(4) that is characterized by a different value of the quantity $a$ or the energy. In other words under the scale change $e^{i\lambda D}$ the basis states for our representation of SO(4,1), $|n l m; a)$, transform to a new set, $|n l m; e^{\lambda}a)$ in agreement with
our discussion in Section 4.3.

Since the algebra of our generators closes we may also view $e^{i\lambda D}$
as transforming a given generator into a
linear combination of the generators. With the definitions of $\bm{a}$ and $\bm{B}$ (Eq. 204 and 205) we can easily show
that Eq. 210, with the upper sign, may also be written
\be
 e^{i \lambda D} \bm{B} e^{-i \lambda D}= \bm{B}\cosh \lambda +\bm{a} \sinh \lambda  .
\ee

\section{The Group SO(4,2)}

\subsection {Motivation for Introducing SO(4,2)}
We would like to express Schrodinger's equation as an algebraic equation in the generators of some group\cite{frons3}\cite{frons2}. As we are unable to do this with our SO(4,l) generators
$S_{AB}$ we again expand the group. To guide us we recall
that to expand SO(3) to SO(4) we added a three-vector of generators $\bm{A}$, and to expand SO(4) to SO(4,1) we
added a four-vector of generators $(S,\bm{B})$. In both cases
this type of expansion produced a set of generators
convenient for the study of the hydrogenlike atom. We
guess that the appropriate expansion of SO(4,1) is obtained by adding a five-vector (under SO(4,1)) of
generators $\Gamma_{A}$ to obtain SO(4,2)\cite{frons3}\cite{frons2}. We can provide additional motivation for this choice by considering Schrodinger's equation. The generators in terms of which
we want to express this equation must be scalars under $L_i$ rotations. Also we know $S = S_{40}$ (Eq. 187) generates scale changes of Schrodinger's equation. The fact that $S_{40}$
mixes the zero and four components of a five-vector
suggests that Schrodinger's equation may be expressed in 
terms of the components $\Gamma_0$ and $\Gamma_4$ which are scalars
under $L_i$, of the five vector $\Gamma_{A}$.  Since $\Gamma_{A}$ is a five-vector under SO(4,1), it must satisfy the equation
\be
\left[S_{A B}, \Gamma_{C}\right]=i\left(\Gamma_{B} g_{A C}-\Gamma_{A} g_{BC}\right)  .
\ee
The spatial components of $\Gamma_A$ which are $(\Gamma_1, \Gamma_2, \Gamma_3)=\bm{\Gamma}$ transform as a vector under rotations generated by $\bm{L}$. 

To construct the Lie algebra of SO(4,2)  we require that the set of operators $\{\Gamma_A, S_{AB}; A,B = 0 ,1,2,3,4\}$ must close under the operations of commutation.  By applying Jacobi's identity to $\Gamma_A, \Gamma_B,$ and $S_{AB}$ , and requiring that $\Gamma_{A}$ and $\Gamma_{B}$ do not commute,  we find $$[S_{AB},[\Gamma_{A},\Gamma_{B}]]=0 \hspace{10pt} A,B=0,1,2,3,4  .$$

Since we require that our Lie algebra closes, the
commutator of $\Gamma_A$ and $\Gamma_{B}$ must be proportional to $S_{AB}$.  We normalize $\Gamma$ so
\be
[\Gamma_A, \Gamma_B] = -iS_{AB} \hspace{10pt} A,B=0,1,2,3,4  .
\ee

If we define 
\be
S_{{A}5} = \Gamma_{{A}}=-S_{5A} \hspace{10pt} A=0,1,2,3,4   .
\ee
and recall
$$A_i=S_{i4}\hspace{10pt}B_i=S_{io}\hspace{10pt} L_i=e_{ijk}S_{jk}\hspace{10pt}S=S_{40}$$
then we may unite all the commutations relations of $\Gamma_{{A}}$ and $S_{AB}$ in the single equation  :
\be
[S_{\cal{AB}}, S_{\cal{CD}}]= i(g_{\cal{AC}}S_{\cal{BD}} + g_{\cal{BD}}S_{\cal{AC}}- g_{\cal{AD}}S_{\cal{BC}}-g_{\cal{BC}}S_{\cal{AD}})
\ee
where $\cal{A},\cal{B},..$ = $0, 1,2,3,4,5$ and $g_{00}=g_{55}=-1$; $g_{aa}=1, a=1, 2,3,4$.

These are the commutation relation for the Lie algebra of SO(4,2).  In terms of $\bm{A},\bm{B}, \bm{L},S$ and $\Gamma_A$ the additional commutation relations for the noncommuting generators are\cite{bednar}:
\be
\begin{array}{lll}{\left[{B}_{i}, \hspace{2pt}\Gamma_{j}\right]={i} \Gamma_{0} \delta_{i j}} \hspace{4pt}& {\left[{\Gamma}_{i}, \Gamma_{j}\right]=-{i} \epsilon_{i j k} {L}_{k}} \\ {\left[{A}_{i}, \Gamma_{j}\right]={i} \Gamma_{4} \delta_{i j}} & {\left[\Gamma_{i}, \Gamma_{0}\right]=-{i} {B}_{i}} \\ {\left[{L}_{i},\hspace{3pt} \Gamma_{j}\right]={i} \epsilon_{i j k} \Gamma_{k}} & {\left[\Gamma_{i}, \Gamma_{4}\right]=-{i} {A}_{i}} \\ & {\left[\Gamma_{4}, \Gamma_{0}\right]=-{i} S} \\ {\left[\hspace{2pt}{B}_{i}, \Gamma_{0}\right]={i} \Gamma_{i}} & {\left[{A}_{i}, \Gamma_{4}\right]=-{i} \Gamma_{i}} \\ {\left[\hspace{3pt}{S}, \hspace{3pt}\Gamma_{0}\hspace{2pt}\right]={i} \Gamma_{4}} & [\hspace{3pt}S, \hspace{2pt}\Gamma_{4}]=\hspace{3pt}{i} \Gamma_{{0}}\end{array}
\ee
\subsection{Casimir Operators}

The Lie algebra of SO(4,2) is rank three so it has three Casimir operators $W_2$, $W_3$, and $W_4$\cite{kyri2}:
 \be
 W_2 = -\frac{1}{2}S_{\cal{AB}}S^{\cal{AB}}=Q_2+\Gamma_{A}\Gamma^{{A}}
 \ee
where $Q_2$ is the nonvanishing SO(4,1) Casimir operator Eq. 190 and
\be
 W_3=\epsilon^{\cal{A}BCDEF}S_{\cal{AB}}S_{\cal{CD}}S_{\cal{EF}}
\ee 
\be 
 W_4=S_{\cal{AB}}S^{\cal{BC}}S_{\cal{CD}}S^{\cal{DA}}  .
\ee 
\newpage
\underline{Computation of $W_3$}

 We can show that $W_3 = 0$
 from dynamical considerations similar to those used in the discussion of SO(4,1) Casimir operators. The only terms that can be included in $W_3$ are scalars formed from products of three generators with different indices
\begin{align}
  &\bm{B}\cdot\bm{A}\times\bm{\Gamma}, &  &\bm{A}\cdot\bm{\Gamma}\times\bm{B}, & &\bm{\Gamma}\cdot\bm{B}\times\bm{A}\\
  &\Gamma_{4}\bm{L}\cdot\bm{B},        &
  &\Gamma_0\bm{L}\cdot\bm{A},          &
  & S\bm{\Gamma}\cdot\bm{L}
\end{align}
 It is interesting that these terms are actually all pseudoscalars. Terms like $\bm{B}\cdot\bm{A}\times \bm{L}$ are simply not possible because of the structure of $W_3$. We know that $\bm{\Gamma}=(\Gamma_1, \Gamma_2, \Gamma_3)$ must not be pseudovector, otherwise it would be proportional to $\bm{L}$.  Since it is a vector, it must equal a linear combination of $\bm{r}$ and $\bm{p}$.  Therefore we conclude
 \be
 \bm{\Gamma}\cdot\bm{L}=\bm{L}\cdot\bm{\Gamma}=0 .
 \ee
Since $\bm{\Gamma}$ and $\bm{B}$ are both vectors and $\bm{L}$ is the only pseudovector we have $\bm{\Gamma}\times\bm{B}=\Lambda \bm{L}$  .  In order to determine the scalar $\Lambda$
we evaluate the commutators 
\be
 [B_k,(\bm{\Gamma} \times \bm{B})_k], \hspace{13pt}\text{and      }[\Gamma_{k}, (\bm{\Gamma}\times \bm{B})_k]
 \ee
 and find
 \be
  \bm{\Gamma} \times \bm{B} = \Gamma_0 \bm{L} = -\bm{B}\times \bm{\Gamma}    .         \ee                                  The analogous equations for $\bm{A}$ and $\bm{\Gamma}$, and for $\bm{A}$ and $\bm{B},$ are                           
  \begin{align}
    \bm{\Gamma}\times \bm{A}&= -\Gamma_4 \bm{L}&=-\bm{A}\times\bm{\Gamma}\\  
 \bm{A}\times\bm{B}&=\hspace{8pt}S\bm{L} \hspace{5pt}&=-\hspace{3pt}\bm{B}\times \bm{A} .
 \end{align}
 
From Eqs. 225 and 226 we see that because of the dynamical structure of the generators each of the quantities in the first line of Eq. 220 is proportional to the
quantity directly below in Eq. 221. We have also
shown that (Eqs. 194, 222)
\be \bm{L}\cdot \bm{B} = \bm{L}\cdot\bm{A}=\bm{\Gamma}\cdot\bm{L}=0  . \ee 
Accordingly each scalar in our list vanishes and
\be
 W_3=0.
\ee

 \underline{Computation of $W_2$}

 In order to compute $W_2$ we need to evaluate 
\be
 \Gamma^2 \equiv \Gamma_A \Gamma^A =\Gamma_4^2 + \Gamma_i \Gamma^i - \Gamma_0^2  .
 \ee 
From the structure of $W_2$ as shown in Eq. 217, we
see that $\Gamma^2$ must be a number since $W_2$ and $Q_2$ are
both Casimir operators and therefore numbers for a
particular representation. Accordingly we have
\be
[\Gamma^2, \Gamma_A] = 0   .
\ee
From this equation we can deduce a lemma allowing us to easily evaluate $W_2$ and	$W_4$ in terms of the number $\Gamma^2$. Using Eq. 230 and the definition of $S_{AB}$ Eq. 213 we find
$$
\Gamma^AS_{AB} + S_{AB}\Gamma^A = 0  .  
$$
Contracting Eq. 212 with $g_{AC}$ gives
$$ S_{AB} \Gamma^A -\Gamma^A S_{AB} = 4i\Gamma_B .$$ 
Consequently it must follow that
\be
S_{AB}\Gamma^A = 2 i \Gamma_B = -\Gamma^A S_{AB}  .
\ee
We are now able to evaluate the quantity
$$ 
 S_{AB}S^{B}_{\hspace{3pt}C}=i S_{AB}[\Gamma^B, \Gamma_C]\\=i(S_{AB}\Gamma^B\Gamma_C-S_{AB}\Gamma_C\Gamma^B)  .$$  
 Using Eq. 212 for the commutator of $S_{AB}$ with $\Gamma_C$ and Eq. 231 for the contraction $S_{AB}\Gamma^B$ we prove the lemma
 \begin{equation}
     S_{AB}S^B_{\hspace{3pt}C} = 2 i S_{CA} -\Gamma_A \Gamma_C + \Gamma^2 g_{AC}   .
 \end{equation}
The value of the SO(4,1) Casimir operator $Q_2 =\frac{1}{2}g^{AC}S_{AB}S^B_{\hspace{3pt}C}$ follows directly from the lemma:
\begin{equation}
    Q_2=2\Gamma^2  .
\end{equation}
So we have from Eq. 217
\begin{equation}
  W_2=3\Gamma^2  .
\end{equation}

\underline{Computation of $W_4$}

The Casimir operator $W_4$ can be written as
\be
W_4 =S_{\mathrm{A}\cal{B}}S^{\cal{B}\mathrm{C}}S_{C\cal{D}}S^{\cal{D}\mathrm{A}}+S_{\mathrm{A}\cal{B}}S^{\cal{B}\mathrm{5}}S_{5\cal{D}}S^{\cal{D}\mathrm{A}}+S_{5\cal{B}}S^{\cal{B}\mathrm{C}}S_{\mathrm{C}\cal{D}}S^{\cal{D}\mathrm{5}}+S_{5\cal{B}}S^{\cal{B}\mathrm{5}}S_{5\cal{D}}S^{\cal{D}\mathrm{5}}  .
 \ee
 where $\cal{B},\cal{D}$  $= 0,1,2,3,4,5$  and $\mathrm{A,C}=0,1,2,3,4$.

 In order to evaluate $W_4$ in terms of $\Gamma^2$ we compute $S_{\mathrm{A}\cal{B}}S^{\cal{B}\mathnormal{C}}$ . Recalling $\Gamma_{\mathrm{A}}=S_{\mathrm{A}5}=-S_{5\mathrm{A}}$ we see
 \begin{equation}
 S_{\mathrm{A}\cal{B}}S^{\cal{B}\mathrm{C}} =\Gamma_{\mathrm{A}}\Gamma^{\mathrm{C}} + S_{\mathrm{A}\mathnormal{B}}S^{\mathnormal{B}\mathrm{C}}  
 \end{equation}
 where only $\cal{B}$ goes from 0 to 5.
 Substituting the lemma Eq. 232, we find 
\begin{equation}
 S_{\mathrm{A}\cal{B}}S^{\cal{B}\mathrm{C}}=2 i S^C_{\hspace{3pt}\mathrm{A}}+\Gamma^2 g_A^{\hspace{3pt}\mathrm{C}}   .
\end{equation} 
 From Eq. 231, it follows that
 \be
 S_{\mathnormal{5}\cal{B}}S^{\cal{B} \mathrm{C}}=2 i \Gamma^{\mathrm{C}} .
 \ee
 Substituting Eqs. 233, 234, 237, and 238 into Eq. 235 for $W_4$ we find
 \be
 W_4=6(\Gamma^2)^2 - 24\Gamma^2  .
 \ee
 The fact that the nonvanishing Casimir operators ($Q_2, W_2$, and $W_4$) for
SO(4,1) and SO(4,2) are given in terms of $\Gamma^2$ implies that the representation of
SO(4,2) determines the particular representation of
SO(4,l) appropriate to the hydrogenlike atom. In turn the value of $\Gamma^2$ is determined by the structure of the
$\Gamma$s in terms of the canonical variables.  In Section 7.4
we derive these structures and find that
$$\Gamma^2 = 1.$$
 Therefore the quadratic SO(4,1) Casimir operator $Q_2$ has the value
 $$Q_2=2$$
 and the SO(4,2) Casimir operators have the values:
 $$W_2 = 3 \hspace{15pt} W_3=0 \hspace{15pt}W_4= -18. $$
 The researchers that have published different representations of SO(4,2) based on the hydrogen atom that give their Casimir operators all have $W_2=3$ (or its equivalent) and $W_3$=0 \cite{baru4}\cite{kibl1} \cite{wulf2}, however, two authors have representations with $W_4=0$\cite{baru4}\cite{wulf2} and one \cite{kibl1} has $W_4=-12$, compared to our value of -18. 

 From the mathematical theory of representations it follows that our representations of SO(4,1) and SO(4,2) are both unitary and irreducible.  This means there is no subset of basis vectors that transform among themselves as either SO(4,1) or as SO(4,2). 
\vspace{8pt} 
 
 \subsection{Some Group Theoretical Results}
 
 In this section we derive the transformation properties of the generators of SO(4,2) and then a novel contraction formula that will prove useful for situations in which we want to employ perturbation theory, for example, in our calculation of the radiative shift for the hydrogen atom in Section 8.  We will work primarily with the SO(4,2) generators expressed as the combination of the SO(4,1) generators $S_{AB}$ and the five-vector $\Gamma$, with $g_{AB} = (-1, 1, 1, 1, 1),$  where $ A,B = 0, 1, 2, 3, 4.$

\vspace{8pt}
\underline{Transformation Properties of the Generators}

We can evaluate quantities like
\be
^{AB}\Gamma_{B}(\theta)\equiv e^{iS_{A B}\theta}\Gamma_{B} e^{-i S_{A B} \theta}\hspace{10pt}\text{no sum over A or B}
\ee
by expanding the exponentials in an infinite series and then using the commutation relations Eqs. 212 and 213 of the
generators $S_{AB}$ and $\Gamma_A$, $\Gamma_B$ repeatedly. However it is easier to solve the differential equations satisfied by $^{AB}\Gamma_B$ and to use the appropriate boundary conditions.  Differentiating Eq. 240 and using the commutation relations, we obtain the equations
\be
\frac{d}{d\theta}\hspace{3pt}  {^{AB}\Gamma}_B = - g_{BB}\hspace{3pt} {^{AB}\Gamma}_A \hspace{13pt} \frac{d^2}{d\theta^2}\hspace{3pt}  {^{AB}\Gamma}_B = - g_{AA}g_{BB}\hspace{3pt} {^{AB}\Gamma}_B 
\ee
which have the solution
\be
{^{AB}\Gamma}_{B}=\Gamma_{B} \cos \sqrt{g_{A A} g_{B B}} \theta+\frac{g_{B B}}{\sqrt{g_{A A} g_{B B}}} \Gamma_{A} \sin \sqrt{g_{A A} g_{B B}} \theta   .
\ee 
 Using a similar procedure we find
 \be
 e^{i\Gamma_{A}{\theta}} S_{A B} e^{-i \Gamma_{A}{\theta}}=S_{A B} \cosh \sqrt{g_{A A}} \theta +\sqrt{g_{A A}}\Gamma_{B} \sinh \sqrt{g_{A A}} \theta
 \ee
 \be
 e^{i\Gamma_{A}{\theta}} \Gamma_{B} e^{-i \Gamma_{A}{\theta}}=\Gamma_{ B} \cosh \sqrt{g_{A A}} \theta +\frac{1}{\sqrt{g_{A A}}}S_{AB} \sinh \sqrt{g_{A A}} \theta
 \ee
 where no summation over $A$
or $B$ is implied.

These formulae, Eq. 242-244, give the SO(4,2) transformation properties of the SO(4,2) generators.
\vspace{5pt}

\underline{The Contraction Formula}

 If we multiply Eq. 244 from the right by $e^{i\Gamma_{A}\theta}$ 
 and then contract from the left with $\Gamma_{B}$ we obtain
 \be
\sum_{B} \Gamma^B e^{i \Gamma_A \theta}\Gamma_B = \left[(1-g_{AA} \Gamma_A^2)\cosh{\sqrt{g_{AA}}\theta}+\frac{2i\Gamma_A}{\sqrt{g_{AA}}}\sinh{\sqrt{g_{AA}}\theta}\right]e^{i\Gamma_A \theta}+g_{AA}\Gamma^2_A e^{i\Gamma_{A} \theta}
 \ee
 where we have used $\Gamma^2=1$ and Eq. 231.  Expanding the hyperbolic functions in terms of exponentials and collecting terms gives
 \be
 \sum_B \Gamma^B e^{i\Gamma_A \theta}\Gamma_B = \frac{1}{2}(1 + \frac{i\Gamma_A}{\sqrt{g_{AA}}})^2 e^{i(\Gamma_A-i\sqrt{g_{AA}})\theta}+\frac{1}{2}(1 - \frac{i\Gamma_A}{\sqrt{g_{AA}}})^2 e^{i(\Gamma_A+i\sqrt{g_{AA}})\theta}+g_{AA}\Gamma_A^2 e^{i\Gamma_A \theta}   .
 \ee
 A Fourier decomposition of a function $\Gamma_A$ may be written
 \be
 f(\Gamma_A) = \frac{1}{2\pi}\int d\theta \hspace{2pt}h(\theta) e^{i\Gamma_A \theta}  .                                           
 \ee
 Consequently we have
 \be
 \sum_B \Gamma^B f(\Gamma_A)\Gamma_B = \frac{1}{2}(1 + \frac{i\Gamma_A}{\sqrt{g_{AA}}})^2 f(\Gamma_A-i\sqrt{g_{AA}})+\frac{1}{2}(1 - \frac{i\Gamma_A}{\sqrt{g_{AA}}})^2 f(\Gamma_A+i\sqrt{g_{AA}})+g_{AA}\Gamma_A^2 f(\Gamma_A)   .
\ee 
By performing a suitable rotation we can generalize this formula from functions of $\Gamma_A$ to functions of $\Gamma_A n^A$ where $n_A n^A = \pm 1.$ For $n^2=-1$ we start with $\Gamma_A = \Gamma_0$ and rotate to obtain a very general result
\be 
 \sum_{B} \Gamma_Bf(n\Gamma) \Gamma^B = \frac{1}{2}(n\Gamma +1)^2 f(n\Gamma +1) + \frac{1}{2}(n\Gamma -1)^2 f(n\Gamma -1) - (n\Gamma)^2 f(n\Gamma)   .
\ee 
We will have occasion to apply this formula for the special case 
\be 
 f(n\Gamma) = \frac{1}{\Gamma n - \nu}    .
 \ee
Using the representation
\be
\frac{1}{\Gamma n - \nu}  =\int_0^\infty ds e^{\nu s} e^{-\Gamma n s}
\ee
we obtain the result
\be
\Gamma_A \frac{1}{\Gamma n - \nu} \Gamma^A  =-2 \nu \int_0^\infty ds\hspace{2pt} e^{\nu s}\frac{d}{ds}(\sinh^2 \frac{s}{2} \hspace{2pt}e^{-\Gamma n\hspace{2pt} s})
\ee
which is in a form convenient for perturbation calculations.

\vspace{8pt}
\underline{Derivation of the $\Gamma_A$ in terms of the Canonical Variables}

For our basis states we shall use eigenstates of
$(Z\alpha)^{-1}$ convenient for configuration space calculations
($\rho = na/r$). We choose these states rather than those convenient for momentum space calculations because they
lead to simpler expressions for the $\Gamma_A$ in terms of the
canonical variables, although the expression for $\bm{a}$ is slightly more complicated. Thus our states obey Eq. 115
\be
\left[\frac{1}{K_1(a)} - n\right]|n l m)=0   .
\ee
We know that $K_1^{-1}$ must commute with the generators of the SO(4) symmetry group $\bm(a)_i=S_{i4}$ and $S_{ij} = \epsilon _{ijk}L_k$.  This suggests that we choose 
\be
\Gamma_0 = [K_1(a)]^{-1} = \sqrt{ar} \frac{p^2+a^2}{2a^2}\sqrt{ar} = \frac{1}{2}\left(\frac{\sqrt{r}p^2\sqrt{r}}{a}+ar\right)
\ee
so that 
\be
\left(\Gamma_0 - n\right)|nlm)=0   .
\ee
This last equation is the Schrodinger equation expressed in our language of SO(4,2): our states $|nlm)$ are eigenstates of $\Gamma_0$ with eigenvalue $n$.

To find $\Gamma_4$,  we calculate $\Gamma_4=-i[S, \Gamma_0]$, using Eq. 209 for $S$, 
\be
\Gamma_4=\sqrt{ar} \frac{p^2-a^2}{2a^2}\sqrt{ar} = \frac{1}{2}\left(\frac{\sqrt{r}p^2\sqrt{r}}{a}-ar\right)  .
\ee
Sometimes it is convenient to use the linear combinations
\be
\Gamma_0 - \Gamma_4 = ar  \hspace{30pt} \Gamma_0 + \Gamma_4 = \frac{\sqrt{r}p^2\sqrt{r}}{a}
\ee
which can be used to express the dipole transition operator \cite{baru0}. 
We can find $\Gamma_i$ from Eq. 216,  $\Gamma_i = -i[B_i, \Gamma_0]$
\be
\Gamma_i = \sqrt{r}p_i \sqrt{r}
\ee
which we might have guessed initially since $[rp_i,rp_j]\sim L_k$. 
Every component of $\Gamma_A$ is Hermitean, consequently the
generators $S_{AB}$ given by the commutators Eq. 213 are also Hermitian. We
may verify explicitly that these expressions for $\Gamma_A$ lead
to a consistent representation of all generators in the
SO(4,2) Lie algebra.

Under a scale change generated by $S$,    $\Gamma_i$ is invariant and $\Gamma_4$ and $\Gamma_0$
transform in the same manner as $\bm{a}$ and $\bm{B}$ (Eq. 210):
they retain their form but $a$ is transformed into $e^{\lambda}a$:
\be
e^{i\lambda S} \left\{ \frac{\Gamma_0}{\Gamma_4} \right\} e^{-i\lambda S}=\frac{1}{2} \left( \frac{\sqrt{r}p^2\sqrt{r}}{e^{\lambda} a} \pm e^{\lambda} a r \right)
\ee

The scale change generates an inner automorphism of
SO(4,2) characterized by a different value of the parameter $a$.

\subsection{Subgroups of SO(4,2)}
The two most significant subgroups are generated by\cite{bednar}:
\vspace{5pt}

1. $L_i, a_i$ or $S_{jk}, S_{i4}$, forming an SO(4) subgroup. These generators commute with $\Gamma_0$ and therefore constitute the degeneracy group for states of energy
$-a^2/(2m)$ and fixed principal quantum number $n$ (or fixed coupling constant $na/m$). The Casimir operator for this subgroup is
\be
\bm{a}^2 + \bm{L}^2 = n^2 -1 =\Gamma_0^2 -1  .
\ee
We discussed this subgroup in Section 4.2 in terms of $\bm{L}$ and $\bm{A}$ and the states $|nlm\ra$.  The same results are obtained with the generators $\bm{L}$ and $\bm{a}$ with the states $|n l m)$. For example, we have the raising and lowering operators for $m$ and $l$ (Eqs 93, 94). With the definition 
\be
L_{\pm}=L_{1}\pm i L_2
\ee
it follows that
\be
[L_3, L_{\pm}] = \pm L_{\pm}\hspace{20pt}
\ee
which gives
\be
L_{\pm} |nlm ) = \sqrt{(l(l+1)-m(m\pm 1)}|n l\hspace{3pt} m\pm 1 )\hspace{15pt}L_3|nlm)=m|nlm) .
\ee
for $l \ge 1$.
 In analogy to $L_{\pm}$ one can define 
\be
a_{\pm}=a_1 \pm i a_2  
\ee
which obey the relations
\be
[a_3, a_{\pm}] = \pm L_3 \hspace{15pt} [L_3, a_{\pm}] = \pm a_{\pm}
\ee
and 
\begin{align}
a_{\pm}|nlm)=&\mp \left(\frac{(n^2-(l+1)^2)(l+2 \pm m)(l+1 \pm m)}{4(l+1)^2-1}\right)^{\frac{1}{2}}|n \hspace{3pt}l+1 \hspace{3pt} m \pm 1 )\\  &\pm \left(\frac{(n^2 - l^2)(l \mp m)(l - 1  \mp m)}{4 l^2 -1}\right )^{\frac{1}{2}}|n \hspace{3pt}l-1 \hspace{3pt}m \pm 1)
\end{align}
for $l\ge 1$.
The action of $a_{\pm}$ is not directly analogous to that of $L_{\pm}$ because we are using $|nlm)$ as basis states.  If we used $|n a_3 l_3=m )$ as basis states, the action would be similar.
An operator that changes only the angular momentum is $a_3$
\be
a_3|n l m\ra=\left(\frac{(n^2 - (l+1)^2)((l+1)^2-m^2)}{4(l+1)^2-1}\right)^{\frac{1}{2}} |n\hspace{2pt} l+1 \hspace{2pt}m\ra+\left(\frac{(n^2 -l^2)(l^2-m^2)}{4l^2 -1}\right)^{\frac{1}{2}}|n \hspace{3pt}l-1\hspace{3pt}m\ra   .
\ee
for $l\ge 1$. Since $a_3$ commutes with $L_3$ and $\Gamma_0$, it does not change $n$ or $m$. 
\vspace{5pt}

2. $\Gamma_4$, $S=S_{40}$, $\Gamma_0$, forming a SO(2,1) subgroup.  These operators commute with $\bm{L}$ but not with $\Gamma_0$, hence then can change $n$ but not $\bm{L}$ or $m$.  The Casimir operator for this subgroup is
\be
\Gamma_0^2 -\Gamma_4^2 -S^2 = \bm{L}^2 = l(l+1)  .
\ee
We can define the operators\cite{bednar}
\be
j_1 = \Gamma_4 \hspace{15pt} j_2 = S \hspace{15pt} j_3 = \Gamma_0
\ee
with commutators
\be
[j_1, j_2 ]= -i  j_3 \hspace{8pt}[j_2,j_3]= i j_1 \hspace{10pt} [j_3, j_1]= i j_2
\ee
We can define the raising and lowering operators
\be
j_{\pm}=j_1 \pm i j_2 = \Gamma_4 \pm iS 
\ee
which obey the commutation relations
\be
[j_{\pm}, j_3] = \mp j_{\pm}  .
\ee
We find (in analogy to Eq. 263) 
\be
\Gamma_0|nlm)=n|nlm) \hspace{15pt} (\Gamma_4 \pm i S)|n l m)=\sqrt{n(n\pm 1)-l(l+1)}|n \pm 1\hspace{3pt} l m)
\ee
We can express the action of $\Gamma_0 - \Gamma_4 = ar$
on our states
\be
ar|nlm) = \frac{1}{2} \bigg{(} (n)(n-l)-l(l+1) \bigg{)} ^{\frac{1}{2}}|n-1 \hspace{3pt}l m)\hspace{5pt} +n|n\hspace{3pt}lm)\hspace{5pt} +\frac{1}{2} \bigg{(} (n)(n+l)-l(l+1) \bigg{)} ^{\frac{1}{2}}|n+1 \hspace{3pt}l m)
\ee
As mentioned previously, the operator $S$ generates scale changes as shown in Eq. 259, where the value of $a$ is changed.  We can also express the action of $S$ equivalently as  transforming $\Gamma_0$ into $\Gamma_4$
\be
e^{i S\lambda}\Gamma_0 e^{-i S \lambda} = \Gamma_0 \cosh{\lambda} - \Gamma_4 \sinh {\lambda}\hspace{20pt} e^{i S\lambda}\Gamma_4 e^{-i S \lambda} = \Gamma_4 \cosh{\lambda} - \Gamma_0 \sinh {\lambda}  .
\ee

\subsection{Time Dependence of SO(4,2) Generators}
For a generator to be a constant it must commute with the Hamiltonian as discussed in Section 2.1. Since the SO(4,2) group is the non-invariance or spectrum generating group, the additional generators do not all commute with the Hamiltonian.  It is notable that as far as we know only one paper considers the time dependence of the generators of non-invariance groups in general and one considers SO(4,2) specifically \cite{doth2time}\cite{mcan}. Our results certainly clarify and make explicit the time dependence, and  show it is just a particular aspect of the SO(4,2) transformations.  In our representation $| nlm ; a)$, the Hamiltonian has been transformed into $\Gamma_0$ and the Schrodinger energy eigenvalue equation has become $\Gamma_0 |nlm)=n|nlm).$  Accordingly, all the generators that commute with $\Gamma_0$ are constants of the motion, which includes $\bm{a}$, $\bm{L}$.   
The other operators, $\bm{B}, \bm{\Gamma}, S, \Gamma_4$ have a time dependence given by Eqs. 243 and 244, for example  
\be
S(t)=e^{iHt}S(0)E^{-iHt}=e^{i\Gamma_0 t}Se^{-i\Gamma_0 t}=S\cos t + \Gamma_4 \sin t  .
\ee
\be
\Gamma_4(t)=e^{iHt}\Gamma_{4}(0)E^{-iHt}=e^{i\Gamma_0 t}\Gamma_4e^{-i \Gamma_0 t}=\Gamma_4 \cos t -  S \sin t.
\ee
Consequently, terms like $j_{\pm}$ have a simple exponential time dependence
\be
j_{\pm}(t)=j_{\pm}(0) e^{\pm it}.
\ee
Similarly $\bm{\Gamma}\pm i \bm{B}$ has an exponential time dependence.

\subsection{Expressing the Schrodinger Equation in Terms of the
Generators of SO(4,2)
}

We can write the Schrodinger equation for the
energy eigenstate $E_n$ of a particle in a Coulomb potential in terms of the SO(4,2) generators, which correspond to the energy $-a^2/2m$, by making a scale change. From Section 4.3, Eq. 114, the relationship between the Schrodinger energy eigenstate $|nlm\ra$ and the eigenstate of $(Z\alpha)^{-1}$ is:
\be
|n l m; a)= e^{-iS \lambda _n} \sqrt{\rho(a_n)}|n l m\ra
\ee
where
\be
e^{\lambda _n} = \frac{a_n}{a} \hspace{15pt}\rho(a_n)=\frac{n}{a_n r}  .
\ee
Substituting Eq. 280 into the eigenvalue equation Eq. 255 for $|nlm;a)$ and employing the transformation Eq. 276, we find the usual Schrodinger equation can be expressed in SO(4,2) terms as
\be
(\Gamma n -n)\sqrt{\rho(a_n)}|n l m\ra =0
\ee
where 
\be
\Gamma n \equiv \Gamma_A n^A=\Gamma_0 n^0 +\Gamma_i n^i + \Gamma_4 n^4  \hspace{1pt}
\ee
\be\hspace{2pt}n^o = \cosh {\lambda_n}= \frac{a^2 + a_n^2}{2 a a_n}\hspace{3pt},\hspace{12pt} n^i =0, \hspace{12pt}n^4 = -\sinh {\lambda_n} = \frac{a^2-a_n^2}{2 a a_n}
\ee
and $n_A n^A = n_4^2 - n_0^2 = -1$.

Eq. 282 expresses Schrodinger's equation for an ordinary energy eigenstate $|nlm \ra$ with energy $E_N=-a_n^2/2m $ in the language of SO(4,2).  It shows the relationship between these energy eigenstates and the basis states of $(Z\alpha)^{-1}$ used for the SO(4,2) representation, 

\section{SO(4,2) Calculation of the Radiative Shift for the Schrodinger Hydrogen Atom}

In the 1930's it was generally believed that the Dirac equation predicted the energy levels of the hydrogen atom with excellent accuracy, but there were some questions about the prediction that the energy levels for a given principal quantum number and given total angular momentum were independent of the orbital angular momentum.  To finally resolve this issue, in 1947 Willis Lamb and his student Robert Retherford at Columbia University in New York City employed rf spectroscopy and exploited the metastability of the hydrogen $2s_{1/2}$ level in a beautiful experiment and determined that the $2s_{1/2}$ and $2p_{1/2}$ levels were not degenerate and that the energy difference between the levels was about {1050 MHz}, or 1 part in $10^6$ of the $2s_{1/2}$ level \cite{lam1}\cite{lam2}. Shortly thereafter
Hans Bethe \cite{bet1} published a ground breaking nonrelativistic quantum theoretical
calculation of the shift assuming it was due to the interaction of the electron with the ground state electromagnetic field of the quantum vacuum field. This radiative shift accounted for about 96\% of the measured shift. The insight that one needed to include the interaction of the atom with the vacuum fluctuations and how one could actually do it ushered in the modern world of quantum electrodynamics\cite{maclayrad}. Here we compute in the non-relativistic dipole approximation and to first order in the radiation field, as did Bethe, the radiative shift but we use group theoretical methods based on the SO(4,2) symmetry of the non-relativistic hydrogen atom as developed in this paper.  Bethe's calculation required the numerical sum over intermediate states to obtain the average value of the energy of the states contributing to the shift. In our calculation, we do not use intermediate states, and we derive an integral equivalent to Bethe's log, and more generally derive the shift for all levels in terms of a double integral.

An expression for the radiative shift $\Delta_{NL}$ for energy level $E_{N}$ of a hydrogen atom in a state $|NL\ra$  can be readily obtained using second order perturbation theory (to first order in $\alpha$ the radiation field)  \cite{bethe}\cite{bands}\cite{milonni}\cite{eide}
\be
\Delta_{NL} = \frac{2\alpha c}{3 \pi m^2}\sum_n ^s \int_0^{\omega_c} d\omega \frac{(E_n-E_N)\la N L|p_i|n \ra \la n|p_i|N L \ra}{E_n - E_N + \omega -i\epsilon}  ,
\ee
where $\omega_C$ is a cutoff frequency for the integration which we will take as $\omega_c = m$.

This expression, which is the same as Bethe's,  has been derived by inserting a complete set of states $|n\ra \la n|$, a step which we eliminate with our group theoretical approach:
\be
\Delta_{NL}=\frac{2\alpha }{3\pi m^2} \int_0^{\omega_c} d\omega  \la NL|p_i \frac{H-E_N}{H-(E_N-\omega) - i\epsilon}p_i|NL \ra 
\ee
If we add and subtract $\omega$ from the numerator we find the real part of the shift is 
\be
Re\Delta_{NL}=\frac{2\alpha}{3\pi m^2}Re\int_0^{\omega_c} d\omega [\la NL|p^2 |NL \ra - \omega \Omega_{NL}]
\ee
where
\be
\Omega_{NL}=\la NL| p_i \frac{1}{H-E_N +\omega -i\epsilon} p_i |NL \ra
\ee
and
\be
H=\frac{p^2}{2m}-\frac{Z\alpha}{r}.
\ee
The imaginary part of the shift gives the width of the level\cite{maclayrad}.

The matrix element $\Omega_{NL}$ can be converted to a matrix element of a function of the generators $\Gamma_A$ taken between
eigenstates $|nlm)$ of  $(Z\alpha)^{-1}$. To do this we insert
factors of $1 = \sqrt{r}\frac{1}{\sqrt{r}}$ and use the definitions of the $\Gamma_A$ in terms of the canonical variables, Eqs. 254, 256, 258. Letting the parameter $a$ take the value $a_N$ we obtain the result
\be
\Omega_{NL}=\frac{m\nu}{N^2}(NL|\Gamma_i\frac{}{\Gamma n(\xi) - \nu}\Gamma_i|NL)
\ee
where
\be
n^0(\xi)=\frac{2+\xi}{2\sqrt{1+\xi}}=\cosh \phi \hspace{20pt}n^i = 0 \hspace{20pt} n^4(\xi)= -\frac{\xi}{2\sqrt{1+\xi}}=-\sinh \phi  
\ee
and 
\be
\xi=\frac{\omega}{|E_N|} \hspace{20pt}\nu=\frac{N}{\sqrt{1+\xi}}=Ne^{-\phi}.
\ee
From the definitions we see $\phi = \frac{1}{2}ln(1 + \xi) > 0$ and $n_A(\xi)n^A(\xi) = -1$.  The quantity
$$\nu =\frac{mZ\alpha}{\sqrt{-2m(E_N-\omega)}}$$ may be considered the effective principal quantum number for a state of energy $E_N - \omega$.  The contraction over $i$ in $\Omega_{NL}$ may be evaluated using the group theoretical formula Eq. 252:
\be
\begin{aligned} \Omega_{N L} &=-2 \frac{m \nu^{2}}{N^{2}} \int_{0}^{\infty} d s e^{\nu s} \frac{d}{d s}\left(\sinh ^{2} \frac{s}{2} M_{N L}(s)\right) \\  & \vspace{5pt} -m \frac{\nu}{N^{2}} (N L|\Gamma_{4} \frac{1}{\Gamma n(\xi)-\nu} \Gamma_{4} | N L) + m\frac{\nu}{N^2} (N L| \Gamma_0 \frac{1}{\Gamma n(\xi) - \nu} \Gamma_0|NL) \end{aligned}
\ee
where
\be
M_{NL}(s) = (NL| e^{-\Gamma n(\xi)\hspace{1pt} s} |NL)  .
\ee
In order to evaluate the last-two terms in	$\Omega_{NL}$ we can
express the action of $\Gamma_4$ on our states in terms of
$\Gamma n(\xi) - \nu$. Substituting the equation
\be
\Gamma_0|NL)=N|NL)
\ee
into the expression for $\Gamma n(\xi) - \nu$, with $n(\xi)$ given 
by Eq. 291, gives
\be
\Gamma_{4}=N-\left(\frac{1}{\sinh \phi}\right)\left(\Gamma n(\xi)-\nu\right)
\ee
when acting on the state |NL). If we substitute Eq. 296 into the expression for the $Re \Delta E_{NL}$ Eq. 287, using Eq. 293 for $\Omega_{NL}$, and simplify using the virial theorem
$$(NL|p^2|NL) = a_N^2$$
we find that the term in $p^2$ exactly cancels the last two
terms in $\Omega_{NL}$,  yielding the result
\be
Re \Delta E_{NL}=\frac{4 m \alpha(Z \alpha)^{4}}{3 \pi N^{4}} \int_{0}^{\phi _c} d \phi  \sinh \phi e^{\phi} \int_{0}^{\infty} d s\hspace{2pt} e^{\nu s} \frac{d}{d s}\left(\sinh ^{2} \frac{s}{2} M_{N L}(s)\right)
\ee
where
\be
\phi_c= \frac{1}{2}ln \left(1 + \frac{\omega_c}{|E_N|} \right)=\frac{1}{2}ln \left( 1+\frac{2N^2}{(Z\alpha)^2} \right)
\ee
and $\omega_c = m$.
\vspace{5pt}

\underline{Comparison to the Bethe Logarithm}

The first order non-relativistic radiative shift is commonly given in terms of the Bethe logarithm $\gamma(N,L)$ which is interpreted as the weighted average over all states, including scattering states, of $ln\frac{|E_n - E_N|}{\frac{1}{2}m(Z\alpha)^2}$ \cite{bands} :
\be
\begin{array}{l}\gamma(N, L) \sum_{n}^{S}\left(E_{n}-E_{N}\right)\left\langle N 0\left|p_{i}\right| n\right\rangle\left\langle n\left|p_{i}\right| N 0\right\rangle \\ \hspace{4pt} =\sum_{n}^{S}\left(E_{n}-E_{N}\right)\left\langle N L\left|p_{i}\right| n\right\rangle\left\langle n\left|p_{i}\right| N L\right\rangle \ln \frac{\left|E_{n}-E_{N}\right|}{\frac{1}{2} m(Z \alpha)^{2}}\end{array}  .
\ee
We use the dipole sum rule
\be
2 \sum_{n}^{s}\left(E_{n}-E_{N}\right)\left\langle N\left|p_{i}\right| n\right\rangle\left\langle n\left|p_{i}\right| N\right\rangle=\left\langle N\left|\nabla^{2} V\right| N\right\rangle
\ee
and apply it for the Coulomb potential $\nabla^{2} V(r)=4 \pi Z \alpha \delta(r)$. The use of the Bethe log allowed Bethe to take the logarithmic expression obtained from the frequency integration outside the summation over the states, and replace it with the average value. Only the S states contribute to the expectation value in Eq. 300, giving from Eq. 285 an expression for the shift
\be
\operatorname{Re} \Delta E_{N L}=\left[\frac{4 m}{3 \pi} \alpha(Z \alpha)^{4}\right] \frac{1}{N^{3}}\left\{\delta_{L 0} \ln \frac{2}{(Z \alpha)^{2}}-\gamma(N, L)\right\} .
\ee
Comparing the shift in terms of $M_{NL}$ Eq. 297 to the shift in terms of $\gamma (N,L)$ we find that the Bethe log is 
\be
\gamma(N, L)=-\int_{0}^{\phi_c} d \phi \sinh \phi\hspace{2pt} e^{\phi} \int_{0}^{\infty} d s \hspace{2pt} e^{\nu s} \frac{d}{d s}\left(\sinh ^{2} \frac{s}{2}  \hspace{3pt} M_{N L}(s)\right) +\delta_{L0} ln\frac{2}{(Z\alpha)^2}
\ee

\subsection{Generating Function for the Shifts}

We can derive a generating function for the shifts for any eigenstate characterized by $N$ and $L$ if we multiply Eq. 297 by $N^4 e^{_\beta  N}$ and sum
over all $N, N \ge L + 1$. To simplify the right side of the resulting equation we use the fact that
the O(2,1) algebra of $\Gamma_0$, $\Gamma_4$, and $S$ closes.  We can compute
the sum on the right hand side:
\be
\sum_{N={L}+1}^{\infty} e^{-\beta {N}} M_{NL}=\sum_{{N}={L}+1}^{\infty}( {NL} | e^{-\bm{j} \cdot \bm{\psi}}| {NL} ) .
\ee
where
\be
e^{\hspace{1pt}-\bm{j}\cdot \bm{\psi}} \equiv e^{-\beta \Gamma_0} e^{-s \Gamma n(\xi)}  .
\ee
We perform a $\bm{j}$ transformation (which leaves the trace invariant) such that
\be
e^{-\bm{j} \cdot \psi} \rightarrow e^{-j_3 \psi}=e^{-\Gamma_0 \psi}
\ee
so
\be
\sum_{N={L}+1}^{\infty} e^{-\beta {N}} M_{NL}=\sum_{{N}={L}+1}^{\infty}( {NL} | e^{-j_3 \psi}| {NL} ) =\sum_{N={L}+1}^{\infty} e^{-N \psi} 
\ee
\be
=\frac{e^{-\psi (L+1)}}{1-e^{-\psi}}   .
\ee
In order to find a particular $M_{NL}$ we must expand the right hand side
of the equation in powers of $e^{-\beta}$ and equate the coefficients to those on the left hand side. First we need an
equation for $e^{-\psi}$.	This can be obtained using the
isomorphism between $\bm{j}$ and the Pauli $\bm{\sigma}$ matrices:
\be
(\Gamma_4, S, \Gamma_0) \rightarrow (j_1,j_2,j_3) \rightarrow (\frac{i}{2}\sigma_1,\hspace{2pt} \frac{i}{2}\sigma_2,\hspace{2pt} \frac{1}{2} \sigma_3)
\ee
Using the formula 
\be
e^{\frac{i}{2}s\bm{n} \cdot \bm{\sigma}} =\cos{\frac{s}{2}}+i\bm{n} \cdot \bm{\sigma}\sin \frac{s}{2}
\ee
where $|\bm{n}|=1$, we find 
\be
\cosh \frac{\psi}{2}=\cosh \frac{\beta}{2} \cosh \frac{s}{2}+\sinh \frac{\beta}{2} \sinh \frac{s}{2} \cosh \phi   .
\ee
We can rewrite this equation in a form easier for expansion
\be
e^{+\frac{1}{2} \psi}=d e^{\frac{1}{2} \beta}+b e^{-\frac{1}{2} \beta}-e^{-\frac{1}{2} \psi}
\ee
where
\be
\begin{array}{l}d=\cosh \frac{s}{2}+\sinh \frac{s}{2} \cosh \phi \\ b=\cosh \frac{s}{2}-\sinh \frac{s}{2} \cosh \phi\end{array}   .
\ee
Let $\beta$ become very large and iterate the equation for $e^{-\frac{1}{2}\psi}
$ to obtain the result
\be
{e}^{-\psi}={Ae}^{-\beta}\left[1+{A}_{1} {e}^{-\beta}+{A}_{2} {e}^{-2 \beta}+\ldots\right]   
\ee
where
\be
\begin{aligned} A=A_{0} &=\frac{1}{d^{2}} \\ A_{1} &=-\left(\frac{2}{d}\right)\left(b-d^{-1}\right) \\ A_{2} &=3 d^{-2}\left(b-d^{-1}\right)^{2}-2^{-2}\left(b-d^{-1}\right) \\ & \vdots \end{aligned}
\ee
Note $b-d^{-1} = -d^{-1} \sinh^2 \frac{s}{2} \sinh^2 \phi$.

\subsection{The Shift Between Degenerate Levels}

Expressions for the energy shift between degenerate levels with the same value of N may be obtained directly from the generating function using Eqs. 297, 306 and 307. We find
$$
 \sum_{N=L+1} e^{-\beta N} N^{4} \operatorname{Re} \Delta E_{N L}-\sum_{N=L^{'}+1}^{\infty} e^{-\beta N} N^{4} \operatorname{Re} \Delta E_{N L^{'}}= $$
\be
\frac{4 m \alpha(Z \alpha)^{4}}{3 \pi} \int_{0}^{\phi_c} d \phi e^{\phi} \sinh \phi \int_{0}^{\infty} d s e^{\nu s} \frac{d}{d s}\left(\sinh^2 \frac{ s}{2} \frac{e^{-\psi(L+1)}-e^{-\psi\left(L^{\prime}+1\right)}}{1-e^{-\psi}}\right)   .
\ee

For an example, consider $L=1, L'=0$.  For the shifts between levels we obtain
$$
\sum_{N=2}^{\infty} e^{-\beta N} N^{4} \operatorname{Re}\left(\Delta E_{N O}-\Delta E_{N 1}\right)+\operatorname{Re} \Delta E_{10}\hspace{2pt} e^{-\beta} =
$$
\be
\frac{4 m \alpha(z \alpha)^{4}}{3 \pi} \int_{0}^{\phi_c} d \phi \hspace{2pt} e^{\phi} \sinh \phi \int_{0}^{\infty} d s e^{\nu s} \frac{d}{d s}\left(\sinh ^{2} \frac{s}{2} e^{-\psi}\right)
\ee
Substituting Eq. 313 for $e^{-\psi}$, using the coefficient $AA_{N-1}$ of $e^{-N\beta}$, gives
\be
Re(\Delta E_{N0}-\Delta E_{N1}) =
\frac{4m \alpha(Z \alpha)^{4}}{3 \pi N^{4}} \int_{0}^{\phi_c} d \phi e^{\phi} \sinh \phi \int_{0}^{\infty} d s e^{\nu s} \frac{d}{d s}\left(\sinh ^{2} \frac{s}{2} A A_{N-1}\right)  .
\ee
where $A$ and $A_{N-1}$ are given in Eq. 314 in terms of the variables of integration $s$ and $\phi$.

\vspace{5pt}
\underline{General Expression for $M_{NL}$}
\vspace{8pt}

Once we have a general expression for $M_{NL}$, we can use Eq. 297 to calculate the shift for any level $E_{NL}$.
We can obtain expressions for the values of $M_{NL}$ by
letting	$\beta$ become large, expanding the denominator in Eq. 307 and equating coefficients of powers of $e^{-\beta}$.
For large $\beta$, we have large $\psi$.   We have
$$
\frac{e^{-\psi(L+1)}}{1-e^{-\psi}}=\sum_{m=1}^{\infty} e^{-\psi(m+L)}
$$
and for large $\beta$ it follows from Eq. 313 that 
\be
\sum_{N=L+1}^{\infty} e^{-\beta N} M_{N L}=\sum_{m=1}^{\infty}\left[e^{-\beta} A\left(1+A_{1} e^{-\beta}+\ldots\right)\right]^{m+L}   .
\ee
Using the multinomial theorem\cite{mandf}, the right side of the equation becomes 
\be
\sum_{m=1}^{\infty} A^{m+L} \sum_{r, s, t, \ldots} \frac{(m+L) !}{r ! s ! t ! \ldots} {A_{1}}^{s} A_{2}^{t} \ldots e^{-\beta(m+L+s+2 t+\ldots)}  .
\ee
where $r + s + t + ...=m +L$.  

To obtain the expression for $M_{NL}$ we note 
$N$ is the coefficient of $\beta$ so
$N=m + L +s + 2t + ...=r + 2s + 3t + ...$
Accordingly we find
\be
M_{N L}=\sum_{r, s, t, \ldots} A^{(r+s+t+\ldots)} \frac{(r+s+t+\ldots) !}{r ! s ! t ! \ldots} A_{1}^{s} A_{2}^{t} \ldots
\ee
where $r + 2s + 3t + \ldots =N$ and $r + s + t + \ldots$ >L. 

By applying this formula we obtain the results:

\vspace{5pt}
N=1:
\be
M_{10} = A
\ee
\hspace{20pt}N=2:
\be
\begin{array}{l}
\qquad\qquad M_{20}=A^2 + AA_1 \\
\qquad\qquad M_{21}=A^2
\end{array}
\ee

\vspace{5pt}

\underline{Shifts for N=1 and N=2}
\vspace{7pt}

To illustrate these results we can calculate the shift for a given energy level using Eq. 297.  For 
$N=1$, we note from Eq. 321 that $M_{10}=A$, and from Eq. 314 that $A=1/d^2$.  We find that the real part of the radiative shift for the $1S$ ground state is

\be
Re \Delta {E}_{10}=\frac{4 m \alpha(Z \alpha)^{4}}{3 \pi} \int_{0}^{\phi_c} {d} \phi {e}^{\phi} \sinh \phi \int_{0}^{\infty} {ds e^{s{e^{-\phi}}}} \frac{d}{ds} \frac{1}{\left(\coth \frac{s}{2}+\cosh \phi \right)^{2}}
\ee

For the shift between two states Eq. 317 can be used. For the N=2 Lamb shift between 2S-2P states, the radiative to first order in $\alpha$\ is 
\be
Re(\Delta E_{20} - \Delta E_{21})= \frac{m \alpha(Z \alpha)^{4}}{6 \pi} \int_{0}^{\phi_c} d \phi e^{\phi} \sinh ^{3} \phi \int_{0}^{\infty} d s e^{2 s e^{-\phi}} \frac{d}{d s} \frac{1}{\left(\coth \frac{s}{2}+\cosh \phi\right)^{4}}
\ee

The s integral can be computed in terms of a Jacobi function of the second kind \cite{bateman}.

As one check on our group theoretical methods, we can compare our matrix elements $(1 0| e^{iS\phi}|n 0)$ with those of Huff \cite{huff}. To go from Eq. 304 to Eq. 305, we did a rotation $R(\phi)=e^{i\phi S}$ generated by $S$ that transformed $\Gamma n$ into $\Gamma_0$.  For $N, L = 1,0$ we have 
\be
M_{10} =(10|e^{-\Gamma n s}|10) = (10|R(\phi)e^{-\Gamma_0 s} R^{-1}(\phi)|10)= \frac{1}{(\cosh \frac{s}{2} + \sinh  \frac{s}{2} \cosh \phi )^2}
\ee
Expanding the hyperbolic functions, we get
$$
M_{10}=\frac{4 {e}^{-s}}{(1+\cosh \phi)^{2}}\left[1-e^{-s} \tanh ^{2} \frac{s}{2}\right]^{-2} 
$$
\be
=\frac{4}{(1+\cosh \phi)^{2}} \sum_{n=1}^{\infty} ne^{-ns }\left(\tanh ^{2} \frac{\phi}{2}\right)^{n-1}   .
\ee
We can also compute $M_{10}$ by inserting a complete set of states and using $\Gamma_0 |n 0) = n|n0)$ in Eq. 325.  Because the generator $S$ is a scalar, only states with $L=0, m=0$ can contribute:
\be
M_{10}=\sum_{n l m} e^{-n s} |n(10|R(\phi)|n 0)|^2  .
\ee  
Comparing this to Eq. 326, we make the identification
\be
|(10|R(\phi)| n 0)|^{2}=\frac{4 n}{(1+\cosh \phi)^{2}}\left(\tanh ^{2} \frac{\phi}{2}\right)^{n-1}  .
\ee
Huff computes this matrix element by analytically continuing the known $O(3)$ matrix element of $e^{iJ_y \phi}$ obtaining
\be
|\la10|R(\phi)| n 0\ra|^{2}=\frac{4 n}{\cosh ^{2} \phi-1}\left(\tanh ^{2} \frac{\phi}{2}\right)^{n} \cdot[_2F_1(0,-1;n;\frac{1}{2}(1 - \cosh \phi))]^2   .
\ee
By algebraic manipulation and using $_2F_1 = 1$ for the arguments here, we see his result agrees with our much more simply expressed result from group theory.
 
\section{Conclusion and Future Research}

Measuring and explaining the properties of the hydrogen atom has been central to the development of modern physics over the last century. One of the most useful and profound ways to understand its properties is through its symmetries, which we have explored in this paper, beginning with the symmetry of the Hamiltonian, which reflects the symmetry of the degenerate levels, then the larger non-invariance and spectrum-generating groups which include all the states. The successes in using symmetry to explore the hydrogen atom led to use of symmetry to understand and model other physical systems, particularly elementary particles.

The hydrogen atom will doubtless continue to be one of testing grounds for fundamental physics. Researchers are exploring the relationship between the hydrogen atom and  quantum information \cite{rau}, the effect of non-commuting canonical variables $[x_i,x_j]\ne0$ on energy levels \cite{cast}\cite{ala}\cite{gna}, muonic hydrogen spectra \cite{hag1}, and new physics using Rydberg states \cite{jon}\cite{cantu}.  The ultra high precision of the measurement of the energy levels has led to new understanding of low Z two body systems, including muonium, positronium, and tritium\cite{eide}. As mentioned in the introduction, measurements of levels shifts are currently being used to measure the radius of the proton\cite{beyer}. We can expect that further investigations of the hydrogen atom and hydrogenlike atoms will continue to reveal new vistas of physics and that symmetry considerations will play an important part.

\funding{This research received no external funding.}

\acknowledgments{I would especially like to thank Lowell S. Brown very much for the time and energy he has spent on my education as a physicist. I also thank Peter Milonni for his encouragement and his helpful comments and many insightful discussions and MDPI for the invitation to be the Guest Editor for this special issue of Symmetry on Symmetries in Quantum Mechanics.  }
\conflictsofinterest{The author declares no conflict of interest.}
\reftitle{References and Notes}



\begin{thebibliography}{999}

\bibitem{brow}Brown, L. Bound on Screening Corrections in Beta Decay, Phys. Rev. 135, B314(1964).

\bibitem{beyer}Beyer, A. et al, The Rydberg constant and proton size from atomic hydrogen, Science 358, 89 (2017).

\bibitem{moh}Mohr, P. D.B. Newell, and B.N. Taylor CODATA recommended values of the fundamental physical constants: 2014, Rev. Mod. Phys. 88, 035009 (2016). Arxiv:1507.07956 [physics.atom-ph].

\bibitem{rigden}Rigden, J.  Hydrogen, The Essential Element. Harvard University Press, 2002, Cambridge, MA, USA.

\bibitem{lamb}Lamb, W.; Retherford, R.  Fine Structure of the Hydrogen Atom by a Microwave Method. \emph{Phys. Rev.} \textbf{1947}, \emph{72},~241.

\bibitem{bethe}Bethe, H. The Electromagnetic Shift of Energy Levels. \emph{Phys. Rev.} \textbf{1947}, \emph{72}, 339.

\bibitem{maclayrad}Maclay, J. History and Some Aspects of the Lamb Shift. \emph{Physics} \textbf{2020}, \emph{ 2},105.

\bibitem{nother}Noether, E. (1918). "Invariante Variationsprobleme". Nachrichten von der Gesellschaft der Wissenschaften zu Göttingen. Mathematisch-Physikalische Klasse. 1918: 235–257.

\bibitem{lie}Hamermesh, M.,Group Theory, Adddison-Wesley Publishing Co., Reading, MA (1962).  

\bibitem{weyl}Weyl, H. The Theory of Groups and Quantum Mechanics, Second Edition, Dover Reprint, Dover Publications, NY NY.
Originally published in German in 1928.

\bibitem{wigner}Wigner, E., Group Theory and its Application to the Quantum Mechanics of Atomic Spectra, Academic Press, New York, New York, 1959.

\bibitem{barg}Bargmann, V., Zur Theorie des Wasserstffatoms, Z. Phys. 99, 576 (1936). 

\bibitem{lapl}Laplace, P. A Treatise of Celestial Mechanics(Dublin, 1827).

\bibitem{pauli}Pauli, W., Uber das Wasserstoffspektrum vom Standpunkt der neuen Quantummechanik, Z. Phys. 36, 336 (1926). Eng. trans.:Sources of Quantum Mechanics (North-Holland, Amsterdam, 1967, B. van der Waerden, Ed.).

\bibitem{maci}McIntosh, H., On Accidental Degeneracy in Classical and Quantum Mechanics,
Am. J. Phys. 27, 620 (1959).

\bibitem{hult}Hulthen, E., Über die quantenmechanische Herleitung der Balmerterme, Z. Phys. 86,21(1933). 

\bibitem{units}We employ natural Gaussian units so $\hbar=1$, $c=1$, and $\alpha = (e^2/\hbar c) \approx 1/137$. The notation for indices and vectors is $\mu, \nu, ..= 0,1,2,3;\hspace{4pt} i,j,.= 1,2,3;\hspace{4pt} p_{\mu}p^{\mu}=-p_0^2 + \bm{p}^2, \hspace{4pt}\bm{p}=(p_1, p_2,p_3), \hspace{4pt} g_{\mu \nu}=(-1,1,1,1)$.

\bibitem{fock}Fock, V., Zur Theorie des Wasserstoffatoms, Z. Phys. 98, 145 (1935). 

\bibitem{dirac}Dirac, P.,Quantum Mechanics, Oxford University Press, Oxford England, First Edition, 1930; Fourth Edition, 1958.

\bibitem{gell}Gell-Mann, M. Symmetries of Baryons and Mesons, Phys. Rev. 125, 1067(1962).

\bibitem{schw1}Schwinger, J.,Coulomb's Green's Function,  J. Math. Phys. 5, 1606 (1964).

\bibitem{neem}Ne'eman, Y. Algebraic Theory of Particle Physics, W.A. Benjamin, New York, New York (1967).

\bibitem{neem2}Ne'eman, Y., Derivation of strong interactions from a gauge invariance
Nuclear Physics 26, 222 (1961). 

\bibitem{gell2}Gell-Mann, M.,
A schematic model of baryons and mesons, Phys. Letters 8, 214 (1964).

\bibitem{eight}Gell-Mann, M and Y. Ne'eman, The Eightfold Way, W.A.Benjamin Co., N.Y.,N.Y., 1964. 

\bibitem{doth}Dothan, Y., M. Gell-Man, and Y. Ne'eman, Series of Hadron Energy Levels as Representations of Non-Compact Groups, Phys. Letters 17, 148 (1965).

\bibitem{nambu}Nambu, Y., Infinite-Component Wave Equations with Hydrogenlike Mass Spectra, Phys. Rev. 160, 1171(1967).

\bibitem{dyso1}Dyson, F., 1966,"Symmetry Groups in Nuclear and Particle Physics", W. A. Benjamin, New York, New York, (1966).

\bibitem{thom}Thomas, L., On the Unitary Representations of the Group of de Sitter Space, Ann. Math. 42, 113 (1941).

\bibitem{hari}Harish-Chandra, Representations of Semisimple Lie Groups II, Trans. Am. Math. Soc. 76, 26 (1954).

\bibitem{baru6}Barut, A., P. Budini and C. Fronsdal,Two examples of covariant theories with internal symmetries involving spin, Proc. Roy. Soc. (London) A291, 106 (1966).

\bibitem{malk}Malkin,I. and V. Man'ko, Symmetry of the Hydrogen Atom, Soviet Physics JETP Letters 2,146 (1966). 

\bibitem{baru0}Barut, A. and K. Kleinert, Transition Probabilities of the Hydrogen Atom from Noncompact Dynamical Groups, Phy. Rev. 156, 1541(1967). 

\bibitem{baru4}Barut, A. and H. Kleinert, Transition Form Factors in the H Atom, Phys. Rev. 160, 1149(1967). 

\bibitem{band1}Bander, M. and C. Itzykson, Group Theory and the Hydrogen Atom (I), Rev. Mod. Phys. 38, 330(1966). 

\bibitem{band2}Bander, M. and C. Itzykson, Group Theory and the Hydrogen Atom (II), Rev. Mod. Phys. 38, 330(1966).

\bibitem{frons3}Fronsdal, C., Infinite Multiplets and Local Fields, Phys. Rev. 156, 1653 (1967). 

\bibitem{frons2}Fronsdal, C., Infinite Multiplets and the Hydrogen Atom, Phys. Rev. 156, 1665 (1967). 

\bibitem{baru7}Barut, A. and C. Fronsdal, On Non-Compact Groups. II Representations of the 2+1 Lorentz Group,     Proc. Roy. Soc. (London) A287, 532 (1965). 

\bibitem{fron1}Fronsdal, C., Relativistic Lagrangian Field Theory for Composite Systems, Phys. Rev. 171, 1811(1968). 

\bibitem{prat}Pratt, R. and T. Jordan, Coulomb Group Theory for and Spin, Phys. Rev. 188, 2534(1969). 

\bibitem{frons8}Fronsdal, C. Relativistic and Realistic Classical Mechanics of Two Interacting Point Particles, Phys.Rev. D 4, 1689 (1971).

\bibitem{kyri2}Kyriakopoulos, R., Dynamical Groups and the Bethe-Salpeter Equation, Phys. Rev. 174, 1846 (1968). 

\bibitem{lieber}Lieber, M., O(4) Symmetry of the Hydrogen Atom and the Lamb Shift, Phys. Rev. 174, 2037 (1968).

\bibitem{huff}Huff, R., Simplified Calculation of Lamb Shift Using Algebraic Techniques, Phys. Rev. 186,1367(1969).

\bibitem{must1}Musto, R. Generators of SO(4,1) for the Quantum Mechanical Hydrogen Atom, Phys. Rev. 148, 1274(1966).

\bibitem{baruMag}Barut, A. and G. Bornzin, SO(4,2)-Formulation of the Symmetry Breaking in Relativistic Kepler Problems with of without Magnetic Charge, J. Math. Phys. 12,841 (1971).

\bibitem{baru9}Barut, A. and H. Kleinert, Current Operators and Majorana Equation for the Hydrogen Atom from Dynamical Groups, Phys. Rev. 157, 1180(1967). 

\bibitem{mack}Mack, G. and I. Todorov, Irreducibility of the Ladder representations when restricted to the Poincare Subgroup, J. Math Phys. 10, 2078 (1969).

\bibitem{deco}Decoster, A. Realization of the Symmetry Groups of the Nonrelativistic Hydrogen Atom, Nuovo Cimento, 68A, 105(1970). 

\bibitem{engl}Englefield, M., Group theory and the Coulomb Problem, Wiley-Interscience, New York, 1972.

\bibitem{bednar}Bednar, M. Algebraic Treatment of Quantum-Mechanical Models with Modified Coulomb Potentials, Ann. of Phys. 75, 305 (1973).

\bibitem{wulf7}Wulfman, C. and Y. Takahata, Noninvariance Groups in Molecular Quantum Mechanics, J. Chem. Phys. 47,4888(1967).

\bibitem{wybou}Wybourne, B. Symmetry Principles in Atomic Spectroscopy, J. de Physique 31, C4-33,(1970). From his book Wybourne, B.G., Symmetry Principles in Atomic Spectroscopy, John Wiley and Sons. Inc., New York, 1970.

\bibitem{mari}Mariwalla, K., Dynamical Symmetries in Mechanics, Physics Reports 20, 287 (1975). 

\bibitem{akyi}Akyildiz, Y., On the dynamical symmetries of the Kepler problem, J. Math. Phys. 21, 665(1980).

\bibitem{fron4}Fronsdal, C. and R. Huff, Two-Body Problem in Quantum Field Theory, Phys. Rev. D 3, 933(1971).

\bibitem{wulf1}Group Theory and Its Applications, ed. E. Loebl (Academic Press, New York, 1971).

\bibitem{baru2}Barut, A. and W. Rasmussen, The hydrogen atom as a relativistic elementary particle I. The wave equation and mass formulae, J. Phys. B6, 1695 (1973). 

\bibitem{baru3}Barut, A. and W. Rasmussen, The hydrogen atom as a relativistic elementary particle II. Relativistic scattering problems and photo-effect, J. Phys. B6, 1713 (1973).

\bibitem{barut1}Barut, A. and G. Bornzin, Unification of the external conformal symmetry group and the
internal conformal dynamical group, J. Math. Phys. 15, 1000(1974).

\bibitem{baru5}Barut, A., C. Schneider, and R. Wilson, Quantum theory of infinite component fields, J. Math. Phys. 20, 2244 (1979).

\bibitem{shib}Shibuya, T. and C. Wulfman, The Kepler Problem in Two-Dimensional Momentum Space, Am. J. Phys. 33, 570 (1965).

\bibitem{dahl}Dahl, J. Physical Interpretation of the Runge-Lenz Vector, Phys. Let. 27A, 62(1968). 

\bibitem{coll}Collas, P., Algebraic Solution of the Kepler Problem Using the Runge-Lenz Vector, Am. J. Phys. 38, 253 (1970).

\bibitem{rodg}Rodgers, H. Symmetry transformations of the classical Kepler problem, J. Math. Phys. 14, 1125(1973).

\bibitem{maju}Majumdar, S. and D. Basu, O(3,1) symmetry of the hydrogen atom, J. Phys.A : Math. Nuc. Gen.7, 787(1974).

\bibitem{stic}Stickforth,J., The classical Kepler problem in momentum space, Am.J. Phys.46, 74 (19  ).

\bibitem{ligo}Ligon, T. and M. Schaaf, On the Global Symmetry of the Classical Kepler Problem, Rep. Math. Phys. 9,281(1976). 

\bibitem{laks}Lakshmanan, M. and H. Hasegawa, On the canonical equivalence of the Kepler problem in coordinate and momentum space, J.Phys. Math. Gen. L889, (1984).

\bibitem{ross}O'Connell, R. and K. Jagannathan, Illustrating dynamical symmetries in classical mechanics: The Laplace-Runge-Lenz vector revisited, Am. J. Phys.71, 243 (2003). 

\bibitem{vale}Valent, G. The hydrogen atom in electric and magnetic fields: Pauli's 1926 article, Am. J. Phys. 71, 171 (2003).

\bibitem{more}Morehead, J. Visualizing the extra symmetry of the Kepler problem, Am. J. Phys. 73, 234 (2005).

\bibitem{lee}Huntington, L. and M. Nooijen, An SO(4) invariant Hamiltonian and the two‐body bound state. I: Coulomb interaction between two spinless particles, Intl. J. Quant. Chem. 109, 2885 (2009).

\bibitem{barut2}Barut, A. and A. Bohm, and y. Neeman, Dynamical Groups and Spectrum Generating Algebras, World Scientific, Singapore(1986).

\bibitem{grei}Greiner,W. and B Muller, Quantum Mechanics, Symmetries, Springer-Verlag, Berlin (1989).

\bibitem{gilm}Gilmore, R. Lie Groups, Lie Algegras and Some of their Applications (Dover Books on Mathmatics), Dover, Mineola, NY, NY (2005).

\bibitem{kibl1}Kibler, M. On the use of the group SO(4,2) in atomic and molecular physics, Molecular Physics 102, 1221(2004).

\bibitem{hamm}Hammond, I. and S. Chu, Irregular wavefunction behavior in dimagnetic Rydberg atoms:a dynamical SO(4,2) group study, Chem. Phys. Let. 182,63 (1991).  

\bibitem{lev}Lev, F., Symmetries in Foundation of Quantum Theory and Mathematics, Symmetry 12, 409 (2020).

\bibitem{wulf2}Wulfman, C., Dynamical Symmetry, World Scientific Publishing, Singapore, 2011.

\bibitem{john}Johnson, M. and B. A. Lippmann, “Relativistic Kepler problem,” Phys. Rev. 78, 329(1950).

\bibitem{bied1}Biedenharn, L. “Remarks on the relativistic Kepler problem,” Phys. Rev. 126, 845–851 (1962).

\bibitem{lanik}Lanik, J., The Reformulations of the Klein-Gordon and Dirac Equations for the Hydrogen Atom to Algebraic Forms, Czech. J. Phys. B19, 1540(1969).

\bibitem{stahl}Stahlhofen, A. Helv. Phys. Acta 70, 1141(1997).

\bibitem{chen4}Chen, J. and D. Deng and M. Hu, SO(4) symmetry in the relativistic hydrogen atom, Phys. Rev. A 77, 034102(2008).

\bibitem{khac}Khachidze, T. and A. Khelashvili, The hidden symmetry of the Coulomb problem in relativistic quantum mechanics: from Pauli to Dirac, Am. J. Phys. 74, 628(2006). 

\bibitem{zhan}Zhang, F.,B.Fu, and J. Chen, Dynamical symmetry of Dirac hydrogen atom with spin symmetry and its connection to Ginocchio's oscillator, Phys.Rev. A 78, 040101(R)(2008).

\bibitem{heine}Heine, V. Group theory in Quantum Mechanics, Dover Publications, New York New York, 1993. Originally published by Pergamon Press in 1960. 

\bibitem{noth}Noether, Emmy; Mort Tavel (translator) (1971). "Invariant Variation Problems". Transport Theory and Statistical Physics. 1 (3): 186–207.
arXiv:physics/0503066. doi:10.1080/00411457108231446. (Original in Gott. Nachr. 1918:235–257)

\bibitem{neue}Neuenschwander, Dwight E. (2010). Emmy Noether's Wonderful Theorem. Johns Hopkins University Press. ISBN 978-0-8018-9694-1.

\bibitem{hanc}Hanca, J.; Tulejab, S.; Hancova, M. (2004). "Symmetries and conservation laws: Consequences of Noether's theorem". American Journal of Physics. 72 (4): 428–35. 

\bibitem{byer}Byers, Nina (1998). "E. Noether's Discovery of the Deep Connection Between Symmetries and Conservation Laws". arXiv:physics/9807044. Citation is Proceedings of a Symposium on the Heritage of Emmy Noether, held on 2–4 December 1996, at the Bar-Ilan University, Israel, Appendix B.

\bibitem{history}The daughter of a mathematician, she wanted to be a mathematician, but since women were not allowed to take classes at the University of Erlingen, she audited courses. She did so well in the exams, that she received a degree and was allowed to enroll in the university and received a PhD in 1907. She remained at the university, unpaid, in an unofficial status, for 8 years. Then she went to the University at Gottengen, where she worked for 8 years with no pay or status before being appointed as Lecturer with a modest salary. She was invited in 1915 by Felix Klein and David Hilbert, two of the most famous mathematicians in the world at the time, to work with them and address issues in Einstein's theory of General Relativity about energy conservation. She discovered Nother's First Theorem (and a second theorem also). She remained there until 1933 when she, as a Jew, lost her job. At Einstein's suggestion, she went to Bryn Mawr College in Pennsylvania. She died from ovarian cysts two years later. 

\bibitem{sudar}Mukanda, N.,O'raifeartaigh, and Sudarshan, E. Characteristic Noninvariance Groups of Dynamical Systems, Phys. Rev. Let. 15, 1041 (1965).

\bibitem{kyri}Kyriakopoulos. E., Algebraic Equations for Bethe-Salpeter and Coulomb Green's Functions, J. Math. Phys. 13, 1729 (1972).

\bibitem{lipk}Lipkin, H.,Lie Groups for Pedestrians, Dover Publications, Mineola, New York (2001),

\bibitem{center}Were it not for this displacement of the force center, the observation that a rotated circle projects onto a plane as an ellipse would manifest the four-dimensional symmetry of the hydrogenlike atom directly in configuration space. The elliptical orbits could be viewed as projections of a rotated hypercircle onto a three-dimensional hyperplane. These considerations can be applied with some modification to the three-dimensional harmonic oscillator for which the force center and the center of the ellipse coincide. 

\bibitem{cordinate}This equation and any other equation written in this specific coordinate system can be generalized to an arbitrary coordinate system by noting that the Cartesian unit vectors may be written in a manner that is independent of the coordinate system: $\bm{i}=\frac{\bm{A}}{A}, \quad \bm{j}=\frac{\bm{L} \times \bm{A}}{LA}, \quad \bm{k}=\frac{\bm{L}}{L}
$.

\bibitem{brown}Brown, L. Unpublished lecture notes.

\bibitem{angle}We define the angle between a three-dimensional hyper-plane and a line as $\pi/2$ minus the angle between the line and the normal to the hyperplane.

\bibitem{later} It is desirable to first show that $\bm{A}$ (and of course $\bm{L}$) generate rotations of the hypersphere or $\hat{U}$. However, as we prefer to do the necessary calculations in terms of commutators rather than Poisson brackets, we defer these considerations to Section 4. There we show that the generator $L_i$ rotates $\hat{U}$ about the $i-4$ plane; the generator $A_1$ rotates $\hat{U}$ about the 2-3 plane, etc., thereby changing the orbit with respect to the 4-axis and changing the eccentricity.

\bibitem{petit}Bois, G., Tables of Indefinite Integrals (New York, Dover Pub1ications, Inc., 1961), p. 123.

\bibitem{doth2}Using Eq. 44 and \cite{cordinate}, Eq. 73 may be written as $cos^{-1}(\bm{U}\cdot \bm{A}/A)=\bm{p}\cdot \bm{r}/ar_c  +\omega_{cl}t$. This agrees with the time dependent function $\phi = \bm{p}\cdot{r}/a r_c  - \omega_{cl}t$ Eq. 70  defined in \cite{doth2time}.

\bibitem{doth2time}Dothan, Y., Finite-Dimensional Spectrum-Generating Algebras, Phys. Rev. D 2, 2944 (1970).

\bibitem{brownNon}Brown, L. Forces giving no orbit precession, Am. J. Phys. 46, 930 (1978).

\bibitem{bacry}Bacry, H. in Lectures in Theoretical Physics, Vol. IXA, edited by W.E. Brittin, A. O. Barut and Marcel Guenin (Gordon and Breach, New York, New York, 1967).

\bibitem{barutSHO}Barut, A. Dynamics of a Broken $SU_N$ Symmetry for the Oscillator,  Phys. Rev. 139, B1433(1965). 

\bibitem{boit}Boiteux, M. The Three-Dimensional Hydrogen Atom as a Restricted Four-Dimensional Harmonic Oscillator, Physica 65, 381(1972).

\bibitem{hugh}Hughes, J. The harmonic oscillator:values of the SU(3) invariants, J. Phys.A Math Gen 6,453(1973).

\bibitem{chen1}Chen, A., Hydrogen atom as a four-dimensional oscillator, Phys. Rev. A 22, 333(1980). 

\bibitem{chen2}Chen, A. Homomorphism between SO(4,2) and SU(2,2), Phys.Rev. A 23, 1653(1981).

\bibitem{kibl2}Kibler, M. and T. Negadi, Connection between the hydrogen atom and the harmonic oscillator: The zero-energy case,  Phys. Rev. A 29, 2891 (1984).

\bibitem{chen3}Chen, A. and M. Kibler, Connection between the hydrogen atom and the four-dimensional oscillator, Phys.Rev. A 31,3960(1985). 

\bibitem{gerr}Gerry, C., Coherent states and the Kepler-Coulomb problem, Phys.Rev. A, 33,6 (1986).

\bibitem{chen3a}Chen, A., Coulomb–Kepler problem and the harmonic oscillator, Am. J. Phys. 55, 250 (1987).

\bibitem{meer}van der Meer, J. The Kepler system as a reduced 4D oscillator, J. Geom. and Phys. 92, 181(2015). 

\bibitem{bacr1}Bacry, H., The de Sitter Group $L_{4,1}$ and the Bound States of the Hydrogen Atom, Nuovo  Cimento, 41, A222 (1966). 

\bibitem{bied2}Biedenharn, L., Wigner Coefficients for the $R_4$ Group and Some Applications, J. Math. Phys. 2, 433 (1961).

\bibitem{paul}Pauling, L. and and E. Bright Wilson, Introduction to Quantum Mechanics (New York, McGraw-Hill,
1935), p. 36 and p. 139.

\bibitem{shif}Shiff, L. , Quantum Mechanics, McGraw Hill, NY, NY (1955).

\bibitem{mandf}Morse, P. and H. Feshbach, Methods of Theoretical Physics, Vol. 1, McGraw-Hill, NY, NY, (1953).

\bibitem{prime}The primes indicates eigenvalues of operators, and unprimed quantities indicate abstract operators. The quantity $x'$ means the four-vector $(t',\vec{r'})$.

\bibitem{morse2}Morse, P. and H. Feshbach, Methods of Theoretical Physics, Vol. 2, McGraw-Hill, NY, NY, (1953).

\bibitem{bateman}Higher Transcendental Functions, Bateman Manuscript Project, Vol. 2, Ed. A. Erdeli, McGraw-Hill Book Co., NY, NY (1953).

\bibitem{mako}Makowski, A. and and P. Pepłowski, Zero-energy wave packets that follow classical orbits, Phys. Rev. A 86, 042117 (2012). 

\bibitem{bell}Bellomo, P. and C. Stroud, Jr., Classical evolution of quantum elliptical orbits, Phys. Rev. A 59, 2139 (1999).

\bibitem{berry}Berry, M. and K.E. Mount, Semiclassical approximations in wave mechanics, Rep. Prog. Phys. 35, 315 1972.

\bibitem{eberH}Eberly, J and C. Stroud, Chapters 14 (Rydberg Atoms) and Chapter 73 (Coherent Transients) in Springer Handbook of Atomic, Molecular, and Optical Physics, Ed. G. Drake, Springer Science and Business Media, N.Y.,N.Y. (2006).

\bibitem{laksh}Lakshmanan, M. and K. Ganesan, Rydberg atoms and molecules-Testing grounds for quantum manifestations of chaos, Current Science, 68, 38 (1995). 
                                        
\bibitem{kay}Kay, K., Exact Wave Functions for the Coulomb Problem from Classical Orbits, Phys. Rev. 25,5190 (1999). 

\bibitem{lena}Lena, C., D.Deland, and J. Gay, Wave functions of Atomic Elliptic States,  Europhys. Lett. 15, 697 (1991). 

\bibitem{bhau}Bhaumik, D., B. Dutta-Roy, and G. Ghosh, Classical limit of the hydrogen atom, J. Phys. A: Math. Gen. 19, 1355, (1986). 

\bibitem{mcan}McAnally, D. and A. Bracken, Quasiclassical states of the Coulomb system and SO(4, 2), J. Phys. A: Math. Gen. 23 2027 (1990).

\bibitem{pita}Pitak, A. and J. Mostowski, Classical limit of position and matrix elements for Rydberg atoms, Eur. J. Phys. 39, 025402(2018).

\bibitem{nauen}Nauenberg, M. Quantum wavepackets on Kepler elliptical orbits, Phys. Rev. A 40l, 1133 (1989).

\bibitem{leon}Leonhardt, U., Measuring the Quantum State of Light, Cambridge University Press, Cambridge, UK (1997).

\bibitem{girar}Barut, A., and L. Girardello,
New ``coherent'' states associated with non-compact groups , Commun. Math. Phys. 21, 41 (1971).           . 

\bibitem{saty}Satyanarayana, M. Squeezed coherent states of the hydrogen atom, J. Phys. A: Math. Gen. 19, 1973(1986). 

\bibitem{brow1}Brown L.S., Classical limit of the hydrogen atom, Am. J. Phys.41,525 (1973). 
                          
\bibitem{liu}Liu, Q. and B. Hu, The hydrogen atom's quantum-to-classical correspondence in Heisenberg's correspondence principle, J.Phys. A: Math. Gen. 34, 5713(2001).

\bibitem{zver}Zverev, V and B. Rubinstein, Dynamical symmetries and well-localized hydrogenic wave packets, Proc. of Inst. of Math. of NAS of Ukraine, Vol. 50, Part 2, 1018(2004)

\bibitem{nand}Nandi, S. and C. Shastry, Classical limit of the two-dimensional and three-dimensional hydrogen atom, J. Phys. A: Math. Gen. 22, 1005 (1989).

\bibitem{biden}Biedenharn, L. and N. Swamy, Remarks on the Relativistic Kepler Problem. II. Approximate Dirac-Coulomb Hamiltonian Possessing Two Vector Invariants, Phys. Rev. 133, B1353 (1964).

\bibitem{lam1}Lamb, W.; Retherford, R. Fine Structure of the Hydrogen Atom by a Microwave Method. \emph{Phys. Rev.} \textbf{1947}, \emph{72},~241.

\bibitem{lam2}{Lamb, W.}; Retherford, R. Fine Structure of the H Atom, Part I. \emph{Phys. Rev.} \textbf{1950}, \emph{79}, 549.

\bibitem{bet1}Bethe, H. The Electromagnetic Shift of Energy Levels. \emph{Phys. Rev.} \textbf{1947}, \emph{72}, 339.

\bibitem{bands}Bethe, H.; Salpeter, E. \emph{The Quantum Mechanics of One and Two Electron Atoms}; {Springer-Verlag,Berlin, Germany},~1957.

\bibitem{milonni}Milonni, P. \emph{The Quantum Vacuum}; Academic Press: San Diego, CA, USA, 1994.

\bibitem{eide}Eides, M., H. Grotch, and V. Shelyuto, Theory of Light Hydrogenic Bound States, Springer Tracts in Modern Physics 222, Springer, Berlin, Germany (2007).

\bibitem{baru1}Barut, A., Dynamical Groups, University of Canterbury Press, Christchurch, New Zealand, 1972. 

\bibitem{prax}Praxmeyer, L., Hydrogen atom in phase space: the Wigner representation, J. Phys. A: Math. Gen. 39, 14143(2006).

\bibitem{note11}The wave function in momentum space $\psi(p)$ is obtained by multiplying $Y_{nlm}$ by the normalizing factor $\frac{(a_n)^{3/2}}{(1 - U_4)^2}$, cf Eq. 150.


\bibitem{jent2}Jentschura, U. and P. Mohr, Calculation of hydrogenic Bethe logarithms for Rydberg States, Phys. Rev. A 72, 012110(2002).

\bibitem{jents3}Jentschura, U., E. LeBigot, J. Evers, P. Mohr, and C. Keitel, Relativistic and radiative shifts for Rydberg states, J. Phys. B: At. Mol. Opt. Phys. 38 S97(2005).

\bibitem{jents7}Jentschura, U. and P. Mohr, J. Tan, Fundamental constants and tests of theory in Rydberg states of one-electron ions, J. Phys. B: At. Mol. Opt. Phys. 43 074002 (2010).

\bibitem{rau}Rau, A. and G. Alber, Shared symmetries of the hydrogen atom and the two-bit system, J. Phys. B: At. Mol. Opt. 50, 242001 (2017).

\bibitem{cast}Castro, P. and R. Kullock, Physics of the $SO_p(4)$ Hydrogen Atom, Theo. and Math. Phys. 185, 1678(2015).

\bibitem{ala}Alavi, A. and N. Rezaei, Dirac equation, hydrogen atom spectrum and the Lamb shift in dynamical non-commutative spaces, Pramana-Journal of Physics 88 (5) (2017).

\bibitem{gna}Gnatenko, K. P., Y. S. Krynytskyi, and V. M. Tkachuk, Perturbation of the ns levels of the hydrogen atom in rotationally invariant noncommutative space, Modern Physics Letters A 30 (8) (2015). 

\bibitem{hag1}Haghighat, M. and M. Khorsandi, Hydrogen and muonic hydrogen atomic spectra in non-commutative space-time, European Physical Journal C 75 (1) (2015).

\bibitem{jon}Jones, Matthew, Robert M. Potvliege, and Michael Spannowsky, Probing new physics using Rydberg states of atomic hydrogen. Phys. Rev. Research 2, 013244 (2020)

\bibitem{cantu}Cantu, S.H., Venkatramani, A.V., Xu, W. et al. Repulsive photons in a quantum nonlinear medium. Nat. Phys. (2020). https://doi.org/10.1038/s41567-020-0917-6

\end{thebibliography}
\end{document}